\documentclass[twocolumn,iop]{emulateapj}
\usepackage{apjfonts}
\usepackage{color}

\usepackage{amsmath,graphicx,longtable,hyperref}
\usepackage{natbib,threeparttable,rotating}
\usepackage{ulem}
\usepackage{multirow}
\newcommand{\etal}{et al.}
\newcommand{\hbeta}{H{$\beta$}}
\def\CIV{C\,{\sc iv}}
\def\MgII{Mg\,{\sc ii}}
\newcommand{\code}{\texttt}

\begin{document}

\title{The Sloan Digital Sky Survey Reverberation Mapping Project: Comparison of Lag Measurement Methods with Simulated Observations}

\author{Jennifer I-Hsiu Li$^{1}$, Yue Shen$^{1,2,*}$, W.~N. Brandt$^{3,4,5}$, C.~J. Grier$^{3,4,6}$, P.~B. Hall$^{7}$, L.~C. Ho$^{8,9}$, Y. Homayouni$^{10}$, K. Horne$^{11}$, D.~P. Schneider$^{3,4}$, J.~R. Trump$^{10}$, D.~A. Starkey$^{11}$}

\altaffiltext{1}{Department of Astronomy, University of Illinois at Urbana-Champaign, Urbana, IL 61801, USA}
\altaffiltext{2}{National Center for Supercomputing Applications, University of Illinois at Urbana-Champaign, Urbana, IL 61801, USA}
\altaffiltext{*}{Alfred P. Sloan Research Fellow}
\altaffiltext{3}{Department of Astronomy \& Astrophysics, The Pennsylvania State University, University Park, PA, 16802, USA}
\altaffiltext{4}{Institute for Gravitation and the Cosmos, The Pennsylvania State University, University Park, PA 16802, USA}
\altaffiltext{5}{Department of Physics, 104 Davey Lab, The Pennsylvania State University, University Park, PA 16802, USA}
\altaffiltext{6}{Steward Observatory, The University of Arizona, 933 North Cherry Avenue, Tucson, AZ 85721, USA}
\altaffiltext{7}{Department of Physics and Astronomy, York University, Toronto, ON M3J 1P3, Canada}
\altaffiltext{8}{Kavli Institute for Astronomy and Astrophysics, Peking University, Beijing 100871, China}
\altaffiltext{9}{Department of Astronomy, School of Physics, Peking University, Beijing 100871, China}
\altaffiltext{10}{Department of Physics, University of Connecticut, 2152 Hillside Rd Unit 3046, Storrs, CT 06269, USA}
\altaffiltext{11}{SUPA Physics/Astronomy, Univ. of St. Andrews, St. Andrews KY16 9SS, Scotland, UK}

\shorttitle{SDSS-RM: Methodology of Lag Measurements}

\shortauthors{Li \etal}

\begin{abstract}
{We investigate the performance of different methodologies that measure the time lag between broad-line and continuum variations in reverberation mapping data using simulated light curves that probe a range of cadence, time baseline, and signal-to-noise ratio in the flux measurements. We compare three widely-adopted lag measuring methods: the Interpolated Cross-Correlation Function (ICCF), the $z$-transformed Discrete Correlation Function (ZDCF) and the MCMC code {\tt JAVELIN}, for mock data with qualities typical of multi-object spectroscopic reverberation mapping (MOS-RM) surveys that simultaneously monitor hundreds of quasars. We quantify the overall lag detection efficiency, the rate of false detections, and the quality of lag measurements for each of these methods and under different survey designs (e.g., observing cadence and depth) using mock quasar light curves. Overall {\tt JAVELIN} and ICCF outperform ZDCF in essentially all tests performed. Compared with ICCF, {\tt JAVELIN} produces higher quality lag measurements, is capable of measuring more lags with timescales shorter than the observing cadence, is less susceptible to seasonal gaps and S/N degradation in the light curves, and produces more accurate lag uncertainties. We measure the H$\beta$ broad-line region size-luminosity (R-L) relation with each method using the simulated light curves to assess the impact of selection effects of the design of MOS-RM surveys. The slope of the R-L relation measured by {\tt JAVELIN} is the least biased among the three methods, and is consistent across different survey designs. These results demonstrate a clear preference for {\tt JAVELIN} over the other two non-parametric methods for MOS-RM programs, particularly in the regime of limited light curve quality as expected from most MOS-RM programs. }

\end{abstract}


\section{Introduction}

Reverberation Mapping \cite[RM;][]{Blandford_1982,Peterson_2014} is the primary technique to measure super massive black hole (SMBH) masses. Unlike other mass estimators (for example, stellar kinematics), RM does not require high spatial resolution in order to resolve the sphere of influence of the central black hole (BH). Instead, RM monitors different parts of the electromagnetic spectrum (corresponding to different emitting regions of the AGN) and measures the timing of ``light echoes'' between different regions. Variable ionizing emission is emitted from an accretion disk surrounding the central black hole. As the UV/optical radiation from the accretion disk travels outward, it is reprocessed by various components of the AGN, for example, the broad-line region (BLR) and the dusty torus \cite[e.g.,][]{Peterson_1997}. RM measures the delay between signals at different wavelengths to probe the structure and kinematics of various regions of the AGN.

Most spectroscopic RM efforts have measured the time delay between the continuum emission (arising in the accretion disk) and the broad emission lines (produced by high-velocity gas clouds in the BLR) using optical spectra. Assuming the BLR is virialized, one can measure the black hole mass ($M_{\rm BH}$) using the BLR size ($R_{\rm BLR}$) inferred from the time delay and the BLR virial velocity determined from the width of a broad emission line ($\Delta V$) using the following equation:
\begin{equation}
M_{BH} = f \frac{R_{BLR} \Delta V^2}{G}\ ,
\end{equation}
where $f$ is a dimensionless scale factor of order unity, called the virial coefficient, that accounts for BLR geometry, kinematics, and inclination.

\begin{figure*}[t]
\centering
\includegraphics[width=\textwidth]{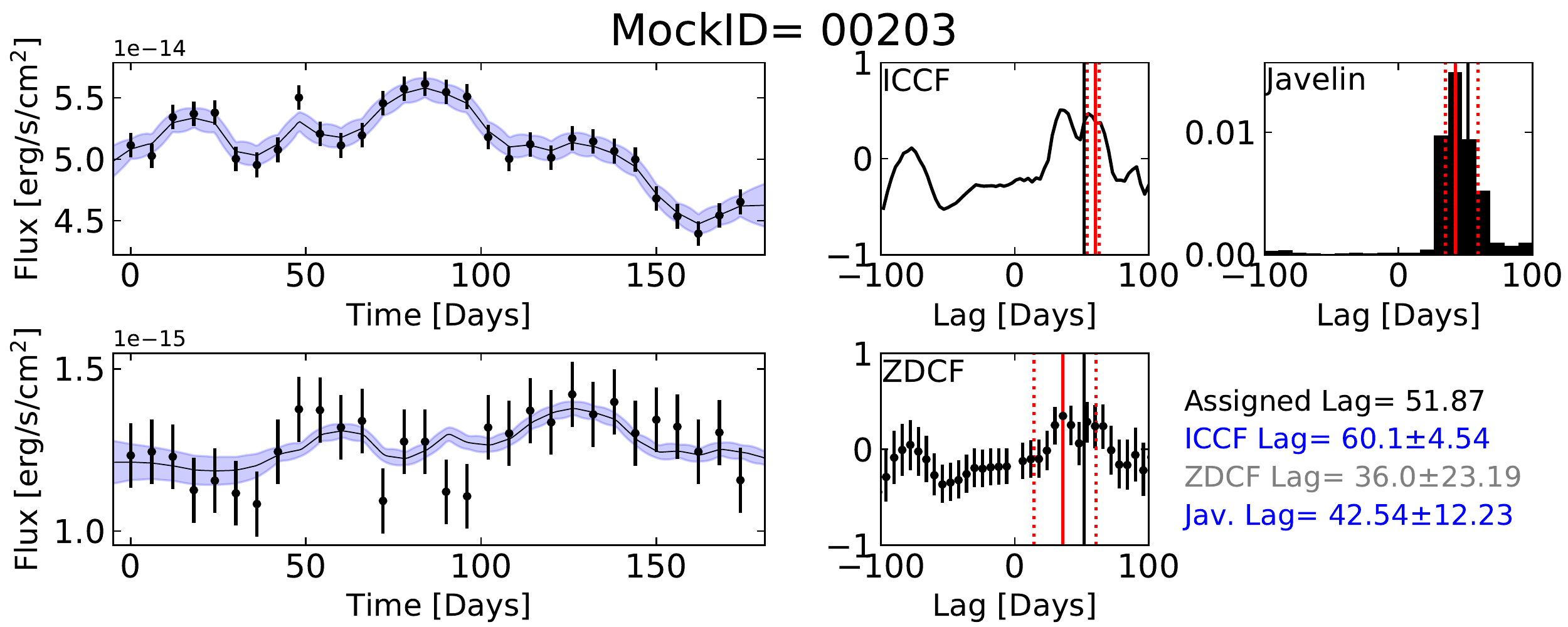}
\caption{An example of our simulated light curves and lag measurements using three different methods. Left: the simulated light curves (continuum in the top panel and emission line in the bottom panel) and predicted light curve models from {\tt JAVELIN} (shaded blue area). The right three panels display the ICCF, ZDCF, and the posterior distribution function from {\tt JAVELIN}. The black solid line marks the assigned lag of the mock quasar, and the red vertical lines indicate the measured lag (solid) and their uncertainties (dotted). In this case, the measured lags from ICCF and {\tt JAVELIN} are considered as true detections (see criteria in \S\ref{sec:criteria}), and the measured lag from ZDCF is not considered a detection.}
\label{fig:lc_example}
\end{figure*}

One of the most important results of past RM studies is the discovery of a correlation between the H$\beta$ BLR radius and the luminosity of the AGN \citep[the R-L relation, e.g.,][]{Laor_1998, Wandel_1999, Kaspi_2000, Kaspi_2005, Bentz_2006, Bentz_2009b, Bentz_2013}, which is the basis of the empirical single-epoch (SE) method \citep{Vestergaard_Peterson_2006, Shen_2013} for BH mass estimation that utilizes single-epoch spectroscopy. Using the measured R-L relation and assuming it applies to objects at different redshifts and luminosities, BH masses of broad-line quasars can be estimated with the luminosity and broad-line width measured from single-epoch spectra. Due to its simplicity, the SE method is widely used to estimate quasar BH masses \citep{Vestergaard_Peterson_2006, Kelly_2013}, although its reliability for emission lines other than H$\beta$ and in the high-redshift and high-luminosity regime remains to be tested. 

Traditional RM studies have focused only on the brightest sources with the highest variability and generally the strongest BLR lines in the local universe ($z<0.1$) to ensure successful measurements of time lags. So far our understanding of the BLR and the R-L relation is based on only $\sim$60 local AGN, which is a biased representation of he distant and luminous quasar population. The Sloan Digital Sky Survey Reverberation Mapping Project \citep[SDSS-RM, ][]{Shen_2015a} is a large-scale RM program that simultaneously monitors 849 uniformly-selected quasars over a broad range of $i$-band magnitude (15.0$<i<$21.7) and redshift (0.1$<z<$4.5), which greatly expands the AGN parameter space for which RM has been conducted. With its multiplex capability, SDSS-RM also dramatically improves the observing efficiency of RM, and thus can extend the redshift and luminosity range for which RM lag measurements are feasible. The first-season data from SDSS-RM has already produced lags for different emission lines in a luminosity-redshift regime largely unexplored by past RM studies  \citep{Shen_2016b, Li_2017, Grier_2017}, {and the multi-year data have started probing lags at even higher redshifts and luminosities \citep{Grier_2019}.}

With an industrial-scale MOS-RM program such as SDSS-RM, it is important to understand the interplay among the quasar sample, {variability characteristics}, survey design and observation sensitivity in order to evaluate/forecast the overall success and limitations of lag measurements. Lag detections strongly depend on the design of the monitoring program, including the cadence, total observation baseline, seasonal/weather gaps, and the signal-to-noise ratio (S/N) of the flux measurements. The complicated selection function induced by these various survey parameters may lead to preferential lag detections in a certain time range and may thus introduce potential selection biases when assessing any intrinsic correlations between lags and quasar properties (such as the R-L relation). In addition, the often poor S/N and lower-amplitude variability in quasars produce low-quality measurements or even false detections. Biases may also arise from different methods and assumptions used by a specific lag-measuring technique when applied to the typical survey-quality light curves produced by MOS-RM programs, as most of these techniques were originally developed using high-quality data from local AGN.

Detailed simulations of mock data are required to quantify the detection efficiency and quality of lag measurements for MOS-RM programs and assess the strengths and weaknesses of different lag-measuring techniques \citep[e.g.,][]{Peterson_1998,Shen_2015a,King_2015}. In this paper, we use a set of simulated observations of a uniform quasar sample (similar to the SDSS-RM sample after down-sampling) to conduct an investigation on a set of lag-measuring methods: the Interpolated Cross-Correlation Function \citep[ICCF, ][]{Gaskell_Peterson_1987}, $z$-Transformed Discrete Correlation Function \citep[ZDCF, ][]{Alexander_2013} and {\tt JAVELIN} \citep{Zu_2011}. {Although all three methods are widely used in the literature, there has not been a comprehensive comparison of their performance over a broad range of light curve properties. In some recent RM work \citep[e.g.,][]{Grier_2017, Homayouni_2018, Edelson_2019}, {\tt JAVELIN} and ICCF are found to yield consistent lag measurements, but {\tt JAVELIN} lag uncertainties are often smaller than those for ICCF.} The main purposes of this study are to inform current and upcoming MOS-RM programs and to understand selection biases introduced by the MOS-RM program design. This work expands our previous investigation \citep{Shen_2015a} that only focused on the traditional ICCF method to advise the design of the SDSS-RM program.

Section \ref{sec:data} describes the generation of our uniform mock quasar sample and its simulated continuum and broad-line light curves. Section \ref{sec:measurelags} presents the methods we use for measuring lags. We compare these different methods using results from the uniform sample in Section \ref{sec:results}, where we down-sample the uniform quasar sample to provide results that can be compared to realistic, flux-limited MOS-RM programs. Section \ref{sec:reallife} introduces a statistical approach to efficiently eliminate false detections from low-quality light curves and present the measurement results from this statistical approach. The implications for the observed R-L relation are discussed in Section \ref{sec:discussion}, and the results are summarized in Section \ref{sec:con}. Throughout this work, we adopt a $\Lambda$CDM cosmology with $\Omega_{\Lambda}=0.7$, $\Omega_{M}=0.3$, and $h=0.7$.


\section{Simulations}\label{sec:data}

A sample of 100,000 mock quasars and their associated light curve pairs were generated following the procedures described by \cite{Shen_2015a}. We first generate a quasar sample uniformly distributed over a grid of $i$-band magnitude (15$<M_{i}<$22) and redshift (0$<z<$5), and calculate the absolute $i$-band magnitudes ($M_{i}$) using K-corrections from \cite{Richards_2006}. The chosen $i$-band and redshift grids are similar to those selected for the SDSS-RM program. Using a power-law spectral index of 0.5 in $F_{\nu}$, we convert the absolute $i$-band magnitudes to monochromatic rest-frame continuum luminosities $L_{5100}$,  $L_{3000}$, and  $L_{1350}$, which correspond to the continuum wavelengths commonly adopted for use with \hbeta, \MgII\ , and \CIV\ reverberation mapping, respectively. To simplify the simulations, we consider RM for a single line in a given redshift interval: \hbeta\ for $z\leq$0.9, \MgII\ for 0.9$<z\leq$2.2 and \CIV\ for $z>$2.2. 

We assign equivalent widths of \hbeta, \MgII\ and \CIV\ as functions of the continuum luminosities of each mock quasar using empirical relations and dispersions measured from the SDSS DR7 quasar sample \citep{Shen_2011} and their corresponding broad-line luminosities. BH masses are assigned using the single-epoch mass estimator based on \hbeta\ and the model broad-line widths and continuum luminosities \citep{Vestergaard_Peterson_2006}. 

For the majority of this work, we focus on the single-season program simulation (with a duration of 180 days) and measure \hbeta\ lags with a few \MgII\ lags at intermediate redshifts. For the multi-season simulation (Section \ref{sec:multi-yr}), we use the same \hbeta\ R$_{\rm BLR}$-L relations for \MgII\ and \CIV\ because the actual R-L relations for the other lines are not as well-established as that for \hbeta. We adopt the average \hbeta\ R$_{\rm BLR}$-L relation at 5100\AA\ with a dispersion of 0.15 dex \citep{Bentz_2009b} to assign the expected BLR lags:
\begin{equation}
\log_{10}\big(\frac{\tau}{{\rm days}}\big) = -21.3 + (0.519)\times \log_{10}\big(\frac{\lambda L_{\lambda, 5100}}{{\rm erg\, s^{-1}}}\big)\ .
\end{equation}
Although only a handful of \MgII\ lags have been reported in the literature \citep[e.g.,][]{Reichert_1994, Dietrich_1995, Metzroth_2006,Shen_2016a, Czerny_2019}, previous studies have demonstrated that the lags for broad \MgII\ and \hbeta\ line widths are correlated \citep[e.g.,][]{Wang_2009,Shen_2011,Wang_2019}, and that \MgII\ may be used as a substitute for \hbeta\ at $z$ $>$1 \citep[e.g.,][]{McLure_2004, Shen_2012, Trakhtenbrot_2012}. The \CIV\ R-L relation at high redshift is currently constrained by only a handful of high redshift quasars with measured \CIV\ lags \citep[e.g.,][]{Kaspi_2007, Lira_2018, Grier_2019}. {In local low-luminosity AGN, \CIV\ lags are found to be smaller than \hbeta\ lags by a factor of $\sim 2$ \citep[e.g.,][]{Peterson_Wandel_1999,Peterson_Wandel_2000}. However, the discrepancies in different R-L relations will not affect our results, as the purpose of this study is to show how well each lag-measuring method recovers the assigned lags under different observing circumstances.}

For each mock quasar, we generate a continuum light curve with daily sampling, assuming that quasar continuum variability follows the Damped Random Walk (DRW) model \citep[a.k.a. the Ornstein-Uhlenbeck process or the first-order continuous autoregressive (CAR(1)) process, e.g.][]{Kelly_2009, Kelly_2011, Kozlowski_2010, Macleod_2010, Macleod_2012}. The DRW model describes a stochastic process with a damping timescale $\tau$ (the timescale for the time series to become uncorrelated) and a driving variability amplitude $\sigma$. {While the short ($<$day) and long ($>$years) timescales of quasar variability are not well-constrained by existing observations and may deviate from the DRW model \citep[e.g.,][]{Macleod_2010, Macleod_2012,Mushotzky_2011,Simm_2016, Guo_2017, Smith_2018}, our simulations will  focus on the timescales where the observed quasar variability can be approximately described by DRW models (1--1000 days). In Section \ref{sec:psd}, we further test the capabilities of each lag measuring method with simulated non-DRW light curves.}

The DRW parameters, $\tau$ and $\sigma$, can depend on the rest-frame color, luminosity and black hole mass of the quasar \citep{Macleod_2010, Macleod_2012}. {We assign the DRW parameters following the empirical Equation 6 from \cite{Macleod_2012} and using simulated quasar properties. 
However, \cite{Kozlowski_2017a, Kozlowski_2017b} reported that the scaling relations between DRW parameters and quasar properties in \cite{Macleod_2012} might be biased or might not exist. In this work, we only use the DRW parameters to produce realistic stochastic light curves to mimic quasar variability. Furthermore, we do not attempt to recover the DRW parameters during the fitting to mock light curves; instead, we fix the DRW parameters and only use the DRW model as a tool to interpolate light curves. } 

The daily-sampled DRW continuum light curve is constructed using the assigned DRW parameters, and the emission line light curve is generated by convolving the continuum light curve with a Gaussian transfer function with an offset equal to the assigned lag and a width of 1/10 of the assigned lag. The transfer function describes the emission line response to  the continuum variability and is related to the physical structure and kinematics of the BLR \citep{Blandford_1982}. The choice of the transfer function width is motivated by velocity-resolved lag observations \citep[e.g.,][]{Grier_2013, Skielboe_2015, Pancoast_2018}, but we have tested different transfer function widths and found that the results are insensitive to this detail \citep[e.g.,][]{Shen_2015a}. 

For each simulation set, we down-sample the full light curves to 30 epochs with a cadence of 6 days to mimic the first-year light curves from the SDSS-RM program. {Following the assumptions of \cite{Shen_2015a}, we adopt fiducial uncertainties of 10$^{-15}$ erg~s$^{-1}$~cm$^{-2}$ and 10$^{-16}$ erg~s$^{-1}$~cm$^{-2}$ for the continuum and line light curves, which are the typical flux uncertainties in the SDSS DR9 BOSS quasar catalog \citep{Paris_2012}, to represent the sensitivity of our simulated survey.} {The median relative uncertainties are $\sim$2\% for continuum fluxes and $\sim$10\% for line fluxes for the final down-sampled, flux-limited sample that mimics the SDSS-RM program (see Section \ref{sec:downsample} for details).} Finally, the fluxes are resampled in the down-sampled light curves by adding to the original flux a Gaussian random deviate with zero mean and a dispersion equal to the flux uncertainty. Figure \ref{fig:lc_example} (left panels) presents an example of our simulated light curves. 

\begin{figure}
\centering
\includegraphics[width=0.5\textwidth]{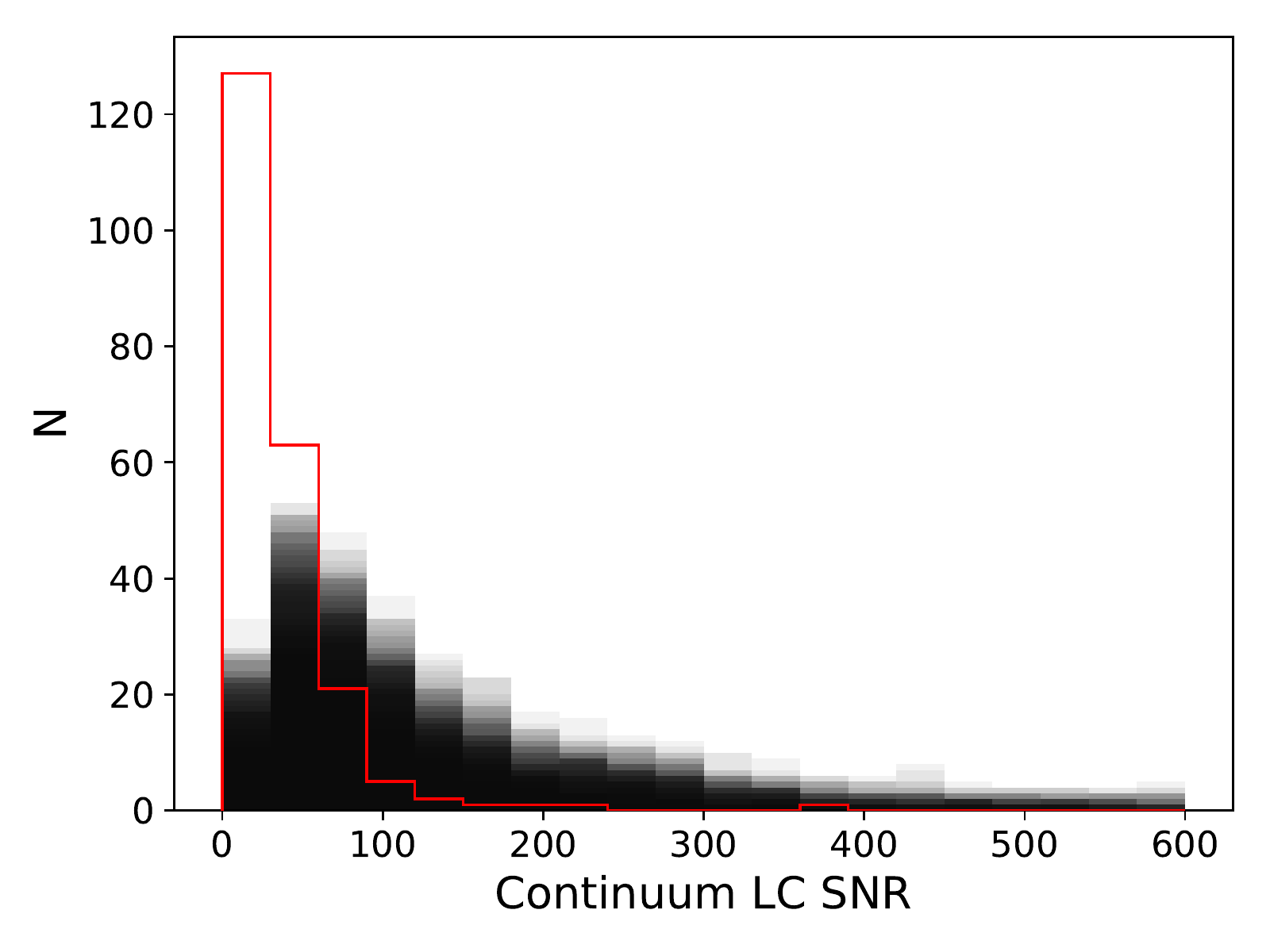}
\includegraphics[width=0.5\textwidth]{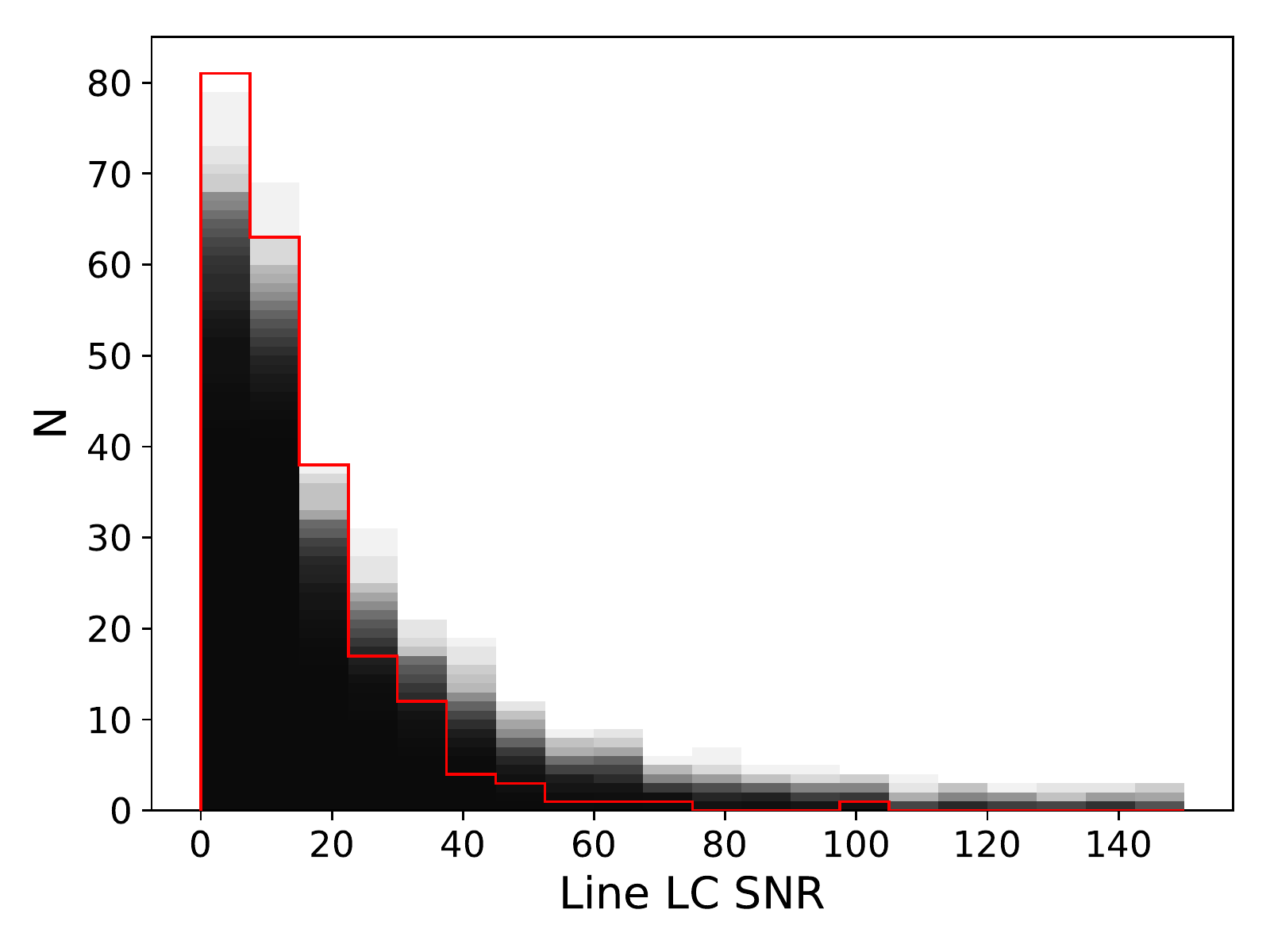}
\caption{S/N of the simulated light curves (black shaded histogram) compared to that of the \cite{Grier_2017} light curves (red open histogram). The S/N of the simulated light curves is represented with 50 realizations of randomly selected down-sampled subsets (see \S \ref{sec:downsample} for details of the down-sampling procedure) from the uniform sample to match the redshift and $i$-band magnitude distribution of the \cite{Grier_2017} sample.}
\label{fig:lc_snr}
\end{figure}

Compared to the light curves from actual SDSS-RM data used in \cite{Grier_2017}, the median S/N {(flux over flux uncertainty)} of the simulated continuum and line light curves are $\sim$3.5 and $\sim$1.5 times larger at similar $i$-magnitude and redshift (Figure \ref{fig:lc_snr}). {The continuum light curves in \cite{Grier_2017} include additional photometric monitoring data from the Steward Observatory Bok 2.3 m telescope and the 3.6 m Canada-France-Hawaii Telescope. An inter-calibration of the light curves was performed with the Continuum REprocessing AGN MCMC ({\tt CREAM}) software \citep{Starkey_2016}, which corrected for detector properties, telescope throughputs, and other properties specific to the individual telescopes. In addition, {\tt CREAM} applied a corrective term to the continuum and line light curve uncertainties to account for the inter-calibration and additional systematic uncertainties, which inflated the uncertainties by a factor of a few. In most of our simulations, we will not use the inflated uncertainties and will not discuss the effects of systematic flux uncertainties in individual light curves. Instead, we will discuss the effect of light curve S/N on lag detection using inflated uncertainties that include these corrections and systematics in Section \ref{sec:diss_err}.}

\section{Measuring Time Lags}\label{sec:measurelags}

We measure time lags with three methods commonly used in the literature: ICCF, ZDCF and {\tt JAVELIN}. ICCF measures the cross correlation between linearly interpolated light curves by assuming light curves are smooth between epochs. ZDCF does not use any interpolation and calculates the discrete cross correlation based solely on the observed data points only. Finally, {\tt JAVELIN} assumes the DRW model to describe the variations of the light curves and utilizes Markov chain Monte Carlo \citep[MCMC, e.g.,][]{EMCEE} to fit for the best time lag. While there are other methods available, i.e., non-parametric techiniques \citep{Skielboe_2015,Chelouche_2017}, Discrete Correlation Function \citep[DCF, ][]{Edelson_Krolik_1988} and \code{CREAM} \citep{Starkey_2016}, the three chosen methods are the most commonly used in analyzing the light curves and measuring time lags; we thus limit our study to these three. 

Below we describe each of the three methods in further detail. 

\subsection{Interpolated Cross-Correlation Function}\label{sec:ccf}

The most frequently used technique of measuring RM time lags is the ICCF method. ICCF calculates time lags by shifting and linearly interpolating the two light curves, calculating the cross-correlation coefficient $r$ at each given time lag ($\tau$) and finding the most likely time lag by locating the maximum $r$ over a grid of lag values. ICCF is designed for high-cadence observations (i.e., traditional RM with the aim for high success rate), and it is unclear to what extent ICCF can be applied to low-to-moderate quality light curve data from MOS-RM programs such as SDSS-RM.

In this work, we implement ICCF using the publicly available PyCCF code \citep{pyCCF} adapted from the original ICCF code written by B. Peterson \citep{Peterson_1998}. For a 180-day observing baseline, we compute the ICCF with a search range of $\pm$100 days to require that at least roughly half of the observations are included in the calculation of ICCF. {We tested different values ({0.1, 0.2, 0.5, 1.0, and 2.0 times of the light curve cadence}) for the $\tau$ grid spacing. The overall ICCF shape does not change drastically with different $\tau$ grid spacing; however, the ICCF may have spurious spikes or become over-smoothed when the grid density is too high or too low. We selected half of the light curve cadence to be the $\tau$ grid spacing, which yields reasonably smooth CCFs for our mock light curves.}

\begin{figure}
\centering
\includegraphics[width=9cm]{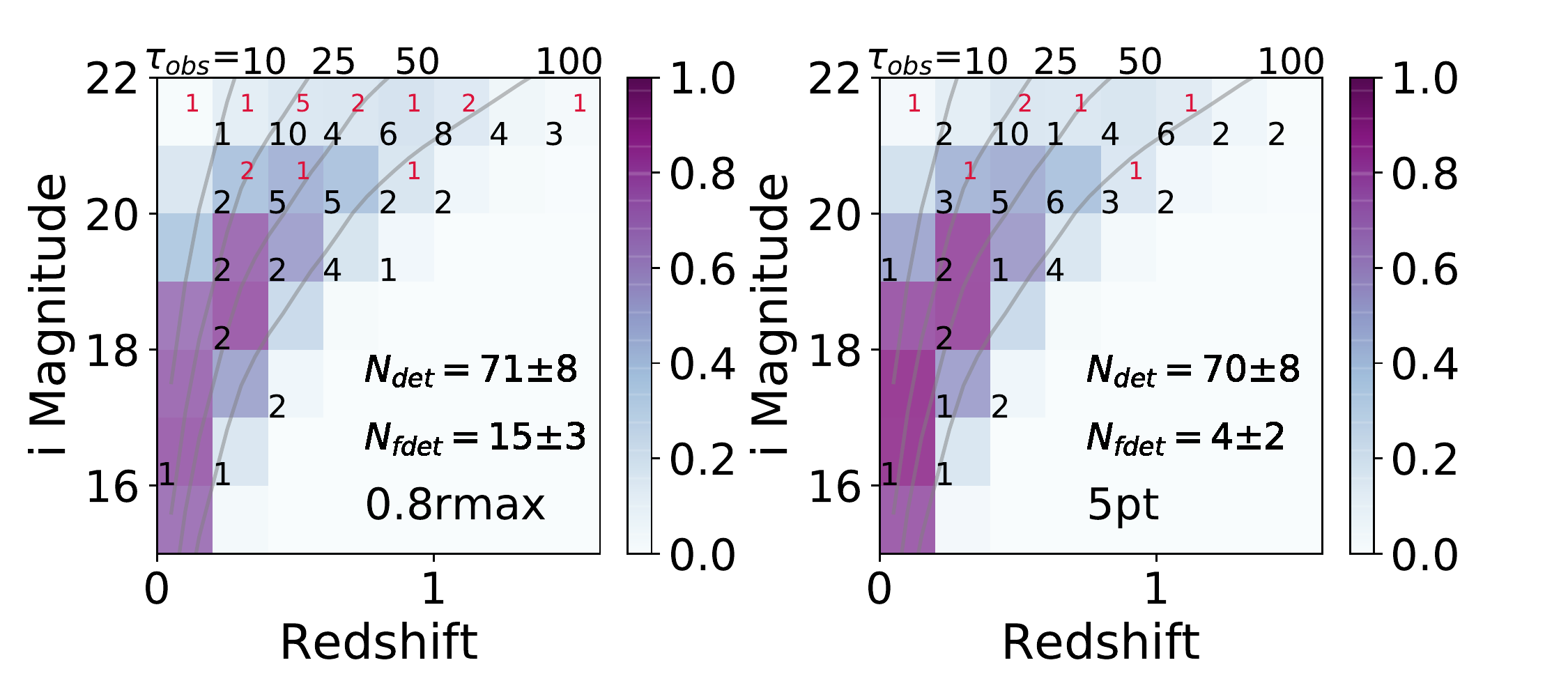}
\caption{Detection efficiency of the ICCF method for simulations with a cadence of 6 days and 30 epochs, using different ICCF centroid calculation schemes. Left panel: the centroid is calculated with all points above 0.8 of the maximum $r$; right panel: the centroid is calculated using 5 points centered on the peak. The colormap represents the detection efficiency and the numbers are the detection counts (true detections in black and false detections in red) of a single down-sampling realization. The total numbers of true and false detections shown in the lower-right corner are the median and uncertainties derived from 100 down-sampling realizations. The grey contours show the approximate constant lags from the R-L relation from \cite{Bentz_2009b}.}
\label{fig:peakdefinition}
\end{figure}

We adopt the traditional flux randomization/random subset sampling (FR/RSS) procedure \citep{Peterson_1998} to obtain the measured time lag and its uncertainties. For 1,000 Monte Carlo (MC) realizations, we randomize the flux measurements by their uncertainties and use a subset of light curve points (chosen at random with repetition) to calculate the CCF. The flux randomization accounts for the flux measurement uncertainties. By choosing random subsets of observations, we can avoid artificial lags introduced by the sampling characteristics of our observations or certain combinations of a few epochs. The centroid computed over five points centered around the ICCF peak is used as the measured time lag $\tau_{cent}$ in each realization. This approach is slightly different from the conventional method of calculating $\tau_{cent}$ from all the data points with $r>0.8\times r_{max}$. We found that with sparse light curves, CCFs occasionally have multiple strong peaks, which causes the centroid calculated in the conventional method to be biased. With the 5-point method, we are guaranteed to calculate a centroid from the local region of the strongest peak, ignoring the impact of aliased lags from sparse light curves. Figure \ref{fig:peakdefinition} demonstrates that the 5-point method eliminates the majority of false detections while retaining similar detection efficiency (defined as the fraction of objects with a detected lag, see Section \ref{sec:criteria} for our detection criteria). Finally, the cross correlation centroid distribution (CCCD), which is the distribution of the measured $\tau_{cent}$ in all MC realizations, is used to define the final lag and its measurement uncertainty, as described in detail in Section \ref{sec:aliases}.

\subsection{$z$-Transformed Discrete Correlation Function}\label{sec:zdcf}

The $z$-transformed discrete correlation function \citep[ZDCF,][]{Alexander_2013} is a modified version of the original DCF proposed by \cite{Edelson_Krolik_1988}. DCF analyzes the correlations in time series data with a conservative approach by merely calculating the cross correlation of the data points, without any interpolation. DCF calculations can avoid effects of correlated errors between continuum and line fluxes measured from the same spectrum, and yield more conservative uncertainties. However, DCF does not perform well for light curves with irregular or sparse cadences. 

ZDCF incorporates two improvements to the original DCF: the implementation of equal-population binning and the uncertainty calculations using the $z$-transform. For each given light curve pair, we calculate and sort the time differences between all data pairs from the two light curves. The ZDCF time lag grid is determined by requiring equal numbers of data pairs in each lag bin, i.e., the ZDCF time grid resolution is adaptive to the sampling of the light curves: when the sampling is denser, ZDCF has better resolution at certain time lags. Next, we calculate the correlation coefficient for the data pairs in each bin, and the uncertainty is calculated following \cite{Alexander_2013} using the $z$-transform method. The above procedure is repeated for 100 Monte Carlo realizations, where in each iteration the observed fluxes are randomly altered by the flux uncertainties. The final ZDCF is the {\it average} of the 100 Monte Carlo realizations. 

To determine peak position and its uncertainties, ZDCF calculates the maximum likelihood from the likelihood function instead of using the traditional FR/RSS method to prevent interpolation of data. We calculate the likelihood of point $i$ being the maximum in the final averaged ZDCF, which is approximately the product of the possibilities for point $i$ to be larger than any other point $j$ in the ZDCF \citep[see][for the complete mathematical description]{Alexander_2013}. We adopt the peak position as the measured lag and the 16$^{\rm th}$ and 84$^{\rm th}$ percentiles of the normalized likelihood function (or the fiducial distribution) as the uncertainties of the peak position. Due to the binning method, the search range of ZDCF is limited by the number of data pairs, especially with sparse light curves. 

\subsection{JAVELIN}\label{sec:javelin}

Another approach to measure lags is to assume {a statistical quasar variability model} and model the continuum light curves, line light curves and their lags simultaneously. {\tt JAVELIN} assumes that the quasar continuum light curve can be described by the DRW model and the line light curve is the shifted, scaled continuum light curve smoothed by a transfer function (a narrow top-hat function is usually assumed in {\tt JAVELIN}, though there are other options available in the code as well). This is a more empirically motivated method to interpolate the data than simple linear interpolation as in ICCF, especially when the observations are sparse or unevenly sampled. {Linear interpolations have minimum uncertainties halfway between data points, where there are no actual data points and the uncertainties are expected to be the largest. On the other hand, the DRW model (and other stochastic process models) is a model of data covariance and can interpolate unmeasured data points based on the statistical properties of the entire light curve.} For the timescales of interest here (e.g., days to months), the DRW model provides a reasonably good statistical description of stochastic quasar continuum variability \citep[e.g.,][]{Kelly_2009, Kozlowski_2010,Macleod_2010}.

The {\tt JAVELIN} code first fits a DRW model to the continuum light curve and then fits the lag, width and scale of the transfer function. Since our mock light curves typically do not have sufficient quality (in terms of cadence and baseline) to constrain the damping time scale or the width of the transfer function, we fix these parameters to 300 days and 2 days, respectively. {Since the damping timescale is fixed, we are merely using {\tt JAVELIN} to ``interpolate'' the light curves with a DRW model.} The damping timescale is chosen to be close to the median of the assigned values in the mock sample that mimic the observed distribution for SDSS quasars \citep[][]{Kelly_2009, Macleod_2010, Macleod_2012}, and the transfer function width is chosen to be smaller than the observing cadence. {Even though the transfer function width is different from our assigned value when generating light curves (1/10 of the assigned lag), it is sufficiently close to the widths for most of our detected lags (on a scale of few days) and the exact choice does not matter as the transfer function cannot be well constrained with our cadence.} We tested this assumption by fixing the damping timescale and the transfer function width at different values (damping timescales at 180, 300, 500 days and transfer function widths at 1, 2, 5, 10 days) in {\tt JAVELIN} and {found the lag measurements do not change with different damping timescale or transfer function widths}; we thus stress that our results are mostly insensitive to these assumptions. 

We ran {\tt JAVELIN} on the full length of light curves with a flat prior of lags, but only examine the posterior distribution within $\pm$100 days to match our ICCF analysis. {This practice is almost equivalent to limiting the lag search range to $\pm$100 days in {\tt JAVELIN}. We chose not to limit the search range in {\tt JAVELIN} so that we can examine the posterior to verify the lag limit is reasonable for the length of our light curves and examine the alias effects at the lag limit. In some cases, imposing the lag limit later can effectively remove the strong peaks near the edges in the posterior, which are caused by fits with only a small overlapping segment of the light curves.} The fitting uses MCMC to sample the probability distribution of all the fitted parameters. The posterior distribution function (PDF) is used in a similar fashion as the CCCD for ICCF to calculate the measured lag and its uncertainties, which will be further discussed in Section \ref{sec:aliases}. 

\begin{figure}
    \centering
    	\includegraphics[width=0.5\textwidth]{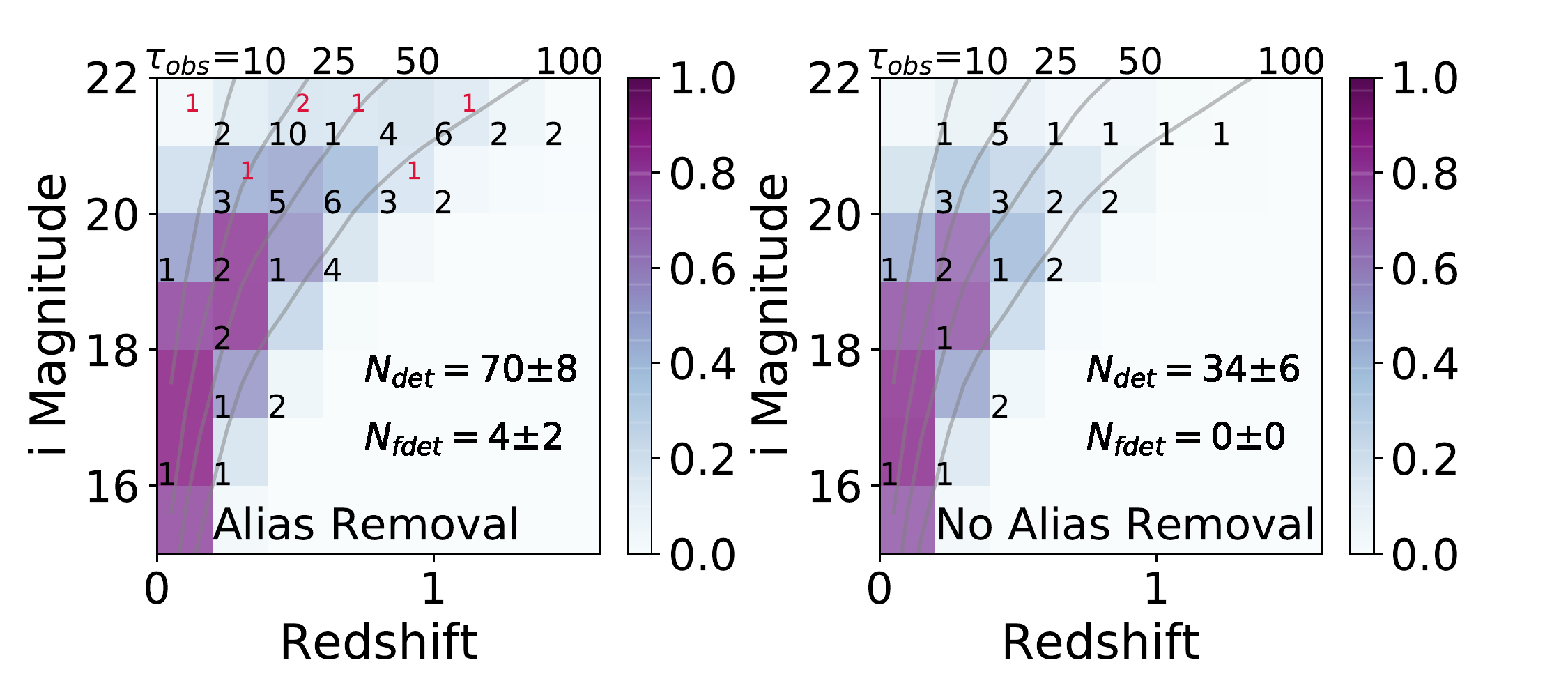}
    \caption{{Similar to Figure \ref{fig:peakdefinition}. Detection efficiency of the ICCF method for simulations with a cadence of 6 days and 30 epochs, with (left) and without (right) the alias removal procedure.}}
    \label{fig:aliasremoval}
    \end{figure}

\subsection{Alias Removal}\label{sec:aliases}

Upon examining the CCCD of the traditional ICCF and the PDF of {\tt JAVELIN}, we occasionally observe multiple peaks. These aliases may be caused by various reasons, including aliases from a segment of the light curves that may be coincidentally correlated, or from a local minimum in MCMC in the case of {\tt JAVELIN}. Here, we follow the quantitative alias removal procedure of \cite{Grier_2017}. First, we apply a weight $P$ to each point in the CCCD/PDF using the fraction of data points included in the calculation: $P=[N(\tau)/N(0)]^{2}$, where $N(x)$ is the number of overlapping points at time lag $x$. Next, we smooth the CCCD/PDF by convolving with a Gaussian filter with a dispersion of 5 days. The 5-day kernel is determined by visual inspection of the PDF. Finally, the primary peak of the weighted and smoothed CCCD/PDF is identified and all data points beyond the range of the peak, i.e. beyond the closest local minima on both sides of the peak, are excluded. Once the primary peak is identified, we adopt the median of the truncated (but not weighted or smoothed) CCCD/PDF as the final measured lag and the 16$^{\rm th}$ and 84$^{\rm th}$ percentiles as the lower and upper uncertainties. 

As discussed by \cite{Grier_2017}, the particular choice of the weights does not carry any physical significance. Instead, this empirical weighting form was found to perform well in recovering the true lags and reducing aliases. This step goes beyond the traditional ICCF and {\tt JAVELIN} approach in lag measurements, but is necessary for low-quality light curve data, specifically when the sample size is large and it is unknown whether or not a true lag will be detected. This additional alias removal step does not affect the results for good-quality light curve data where the CCCD/PDF has a well-defined primary peak. {Figure \ref{fig:aliasremoval} shows an example of measured detection efficiency with and without alias removal using ICCF. The alias removal procedure is effective in improving lag detection efficiency by doubling the number of detections in this case, despite introducing a small number of false detections.}

\begin{figure*}[t]
\centering
\includegraphics[width=\textwidth]{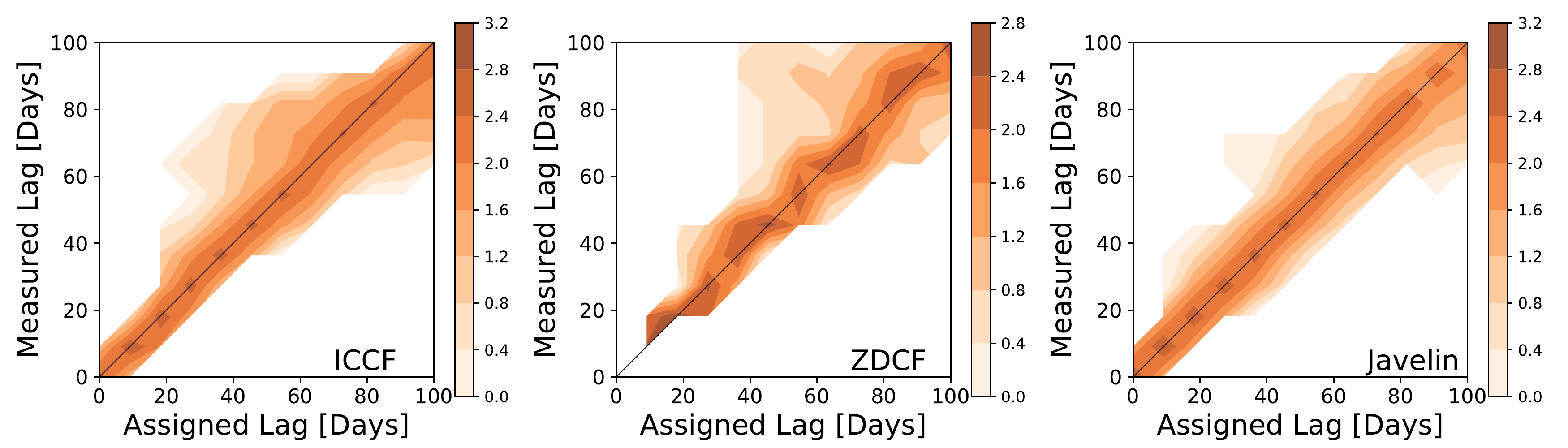}
\caption{Distribution of the measured lag and assigned lag {of the true detections in the uniform sample} for simulations with a cadence of 6 days and 30 epochs. The shaded area is the 2D histogram of the assigned and measured lags (in terms of number of detections in each bin; color bars are in logarithmic scale) and the solid vertical line segments are the uncertainties of the measured lags (randomly down-sampled from all detections for clarity). The black solid line is the 1:1 line for guidance.}
\label{fig:lag_density}
\end{figure*}

\begin{figure*}
\centering
\includegraphics[width=\textwidth]{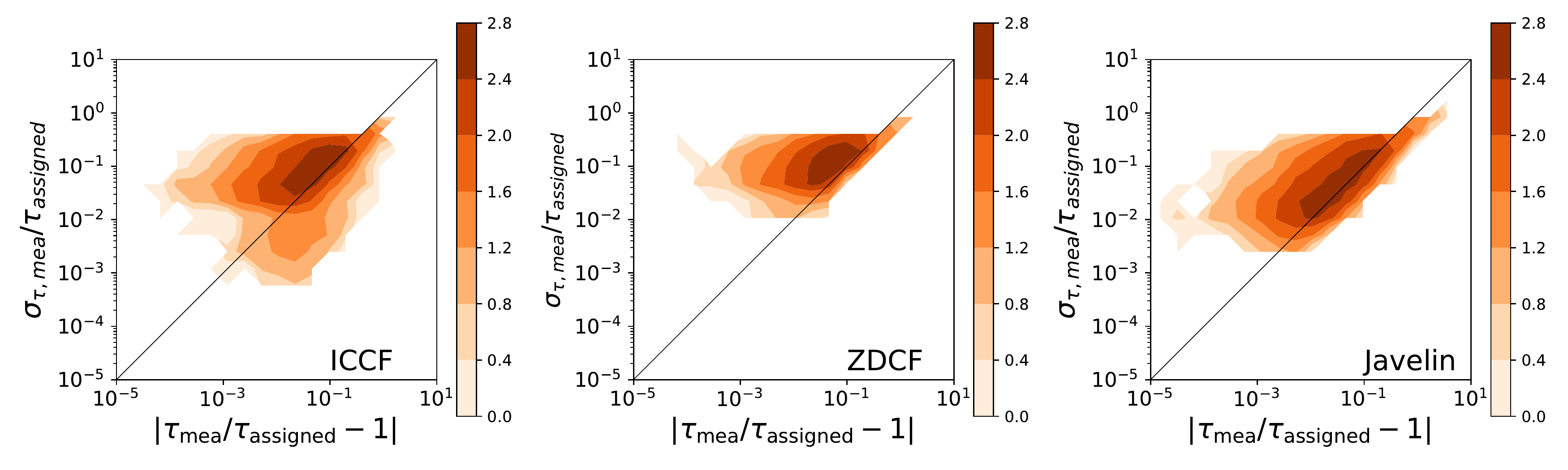}
\caption{Distribution of the normalized measurement uncertainty ($\sigma_{\tau, mea}/\tau_{true}$) and the fractional difference between measured and assigned lags {of the true detections in the uniform sample} for simulations with a cadence of 6 days and 30 epochs. The irregular edges in the upper right corner in the 2-d histograms are shaped by the detection criteria, that is, absolute difference $<$3 days (appears as the upper-right tip along the 1:1 line), $\delta$Lag$<$0.75 (cutoff in x-axis seen in ZDCF and {\tt JAVELIN}) or the normalized measured uncertainty $<$1/3 (cutoff in y-axis). {Lag measurement precision (accuracy) improves towards the lower (left) direction. }}
\label{fig:lag_relerr}
\end{figure*}

\begin{figure}
\centering
\includegraphics[width=0.5\textwidth]{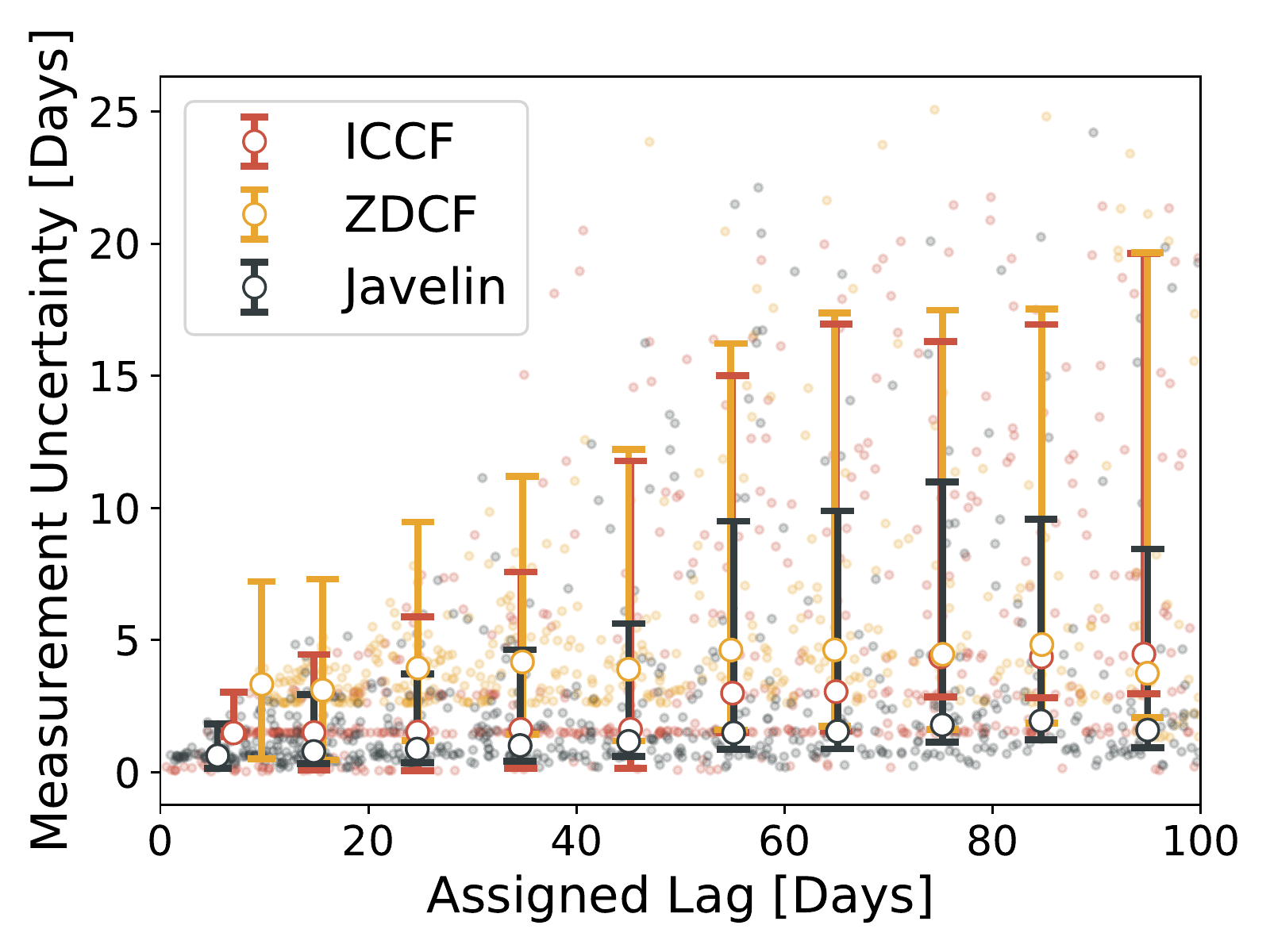}
\caption{Absolute uncertainties of lag measurements as a function of assigned lags, based on {the true detections in the  uniform sample}. The small dots represent the individual measurements in the mock sample. The open circles mark the median absolute uncertainty and the error bars show the 16$^{\rm th}$ and 84$^{\rm th}$ percentile in each 10-day bin of assigned lags.}
\label{fig:lag_abserr}
\end{figure}

\begin{figure}
\centering
\includegraphics[width=0.5\textwidth]{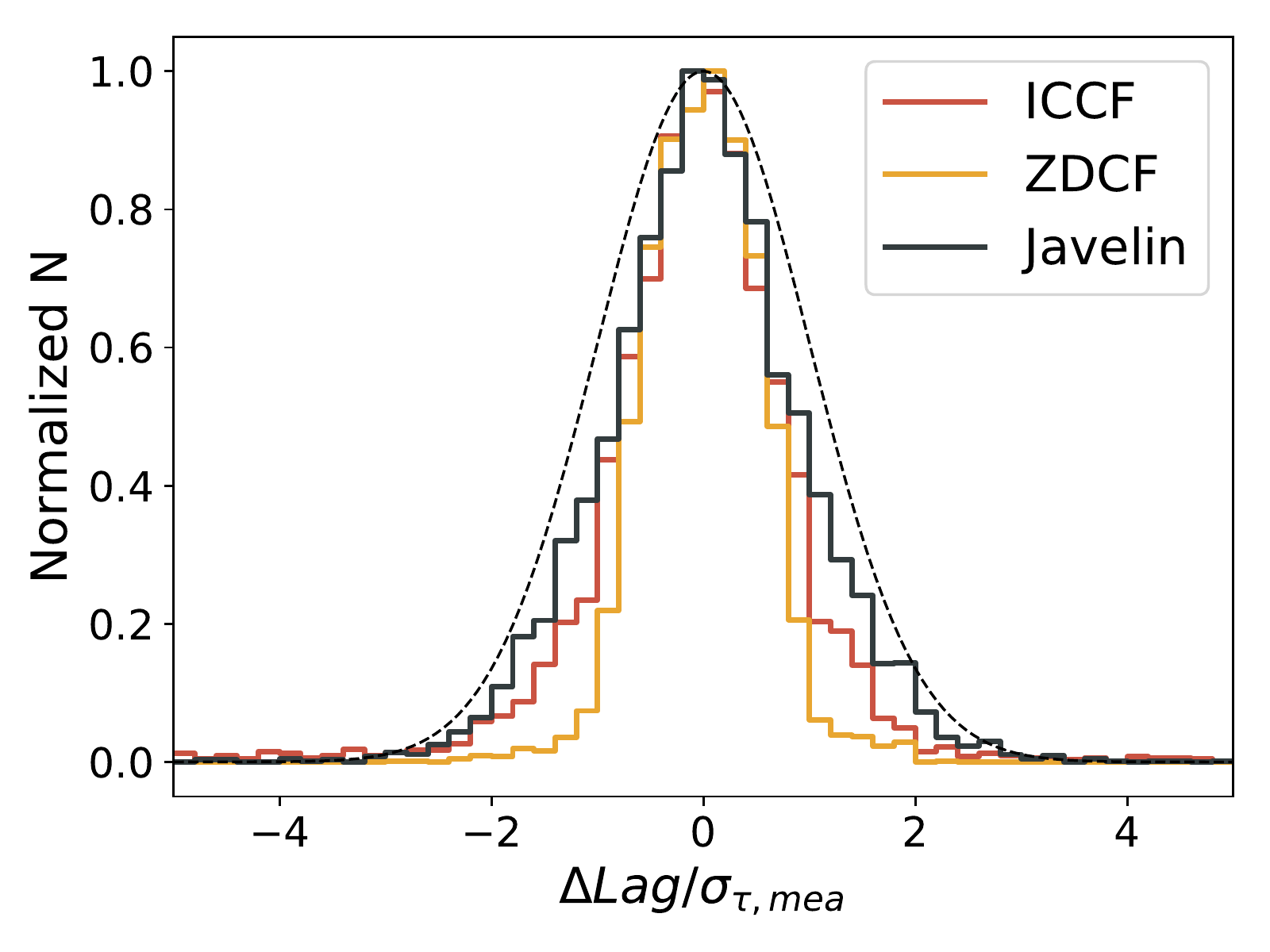}
\caption{Distribution of the difference between assigned and measured lags, normalized by the measurement uncertainty, of the uniform sample. The black dashed line is a Gaussian distribution with unity dispersion.}
\label{fig:norm_err}
\end{figure}

\subsection{Detection Criteria}\label{sec:criteria}

For a measured lag to be a detection, we require that it lie more than 3$\sigma$ away from zero, is positive (i.e., the line flux lags behind the continuum flux), and fewer than half of the CCCD points or MC realizations are rejected in the alias removal procedure (for ICCF and {\tt JAVELIN}). This approach assumes that there is no physical reason to produce a negative lag, and measured negative lags are likely to arise from aliases due to sampling properties of the light curves. 

We impose additional criteria for the measured lag to qualify as a ``true detection'' in our simulated data. For a true detection, the measured lag must fulfill at least one of the following criteria:
\begin{enumerate}
	\item[1.] Absolute difference from the true lag is $<$ 3 days. 
	\item[2.] Relative difference from the true lag is $<$ 25$\%$. 	
	\item[3.] Absolute difference from the true lag is $<3 \sigma$.
\end{enumerate}
{The first two criteria are introduced because the last criteria can be systematically biased against short lags: short lags are less likely to meet the 3$\sigma$ detection requirement given the same measurement error.} If all three criteria are not satisfied, the detection is classified as a false detection. False detections are inevitable even after imposing our alias removal procedure. 

In addition, we require detections (including both true and false detections) to have assigned lags $<$100 days, i.e., the search range for a 180-day observation baseline. A great majority of false detections are produced by light curve pairs of longer lags with variability on shorter timescales that leads to aliases. In our controlled experiment, where the true lags are known, we can simply choose to make a cut of the assigned true lags (which will not be detected with a search range of $\pm100$ days), and compare different lag measuring methods in Section \ref{sec:results}. Of our uniform sample, 9,942 ($\sim$10\%) of the mock quasars have assigned lags less than 100 days, thus only 10\% of the initial quasar sample can have lags detected from one season of observation. In reality, the true lags of quasars are unknown, therefore we develop a set of realistic selection criteria that can effectively remove false detections in \S\ref{sec:reallife}. By implementing these reasonable selection criteria, we can assess the reliability of observed lags in actual MOS-RM programs. 

Unless otherwise specified, the measured lags refer to the measured observe-frame lags in the following. 

\subsection{Flux-limited Down-sampling}\label{sec:downsample}
In order to mimic realistic MOS-RM surveys with flux-limited samples, we also compare the lag detection results by down-sampling from our uniform sample. The redshift and $i$-mag distribution is matched to that of SDSS-RM quasars using the quasar luminosity function from \cite{Richards_2006}. {For each simulation set, we generate 100 realizations of the down-sampling, which have a median of $956^{+47}_{-26}$ sources in total and $177^{+14}_{-11}$ ($\sim$19\%) sources with lag$<$100 days (uncertainties are derived from the 16$^{\rm th}$ and 84$^{\rm th}$ percentiles).} This sample will be referred to as the flux-limited sample. 

\section{Results}\label{sec:results}

\subsection{Measured Lags}

To evaluate the robustness of each technique, Figure \ref{fig:lag_density} shows the density distribution of the assigned lags versus the measured lags {of the true detections} for the uniform sample. {There are very few false detections and they can be ignored for now.} The results for {\tt JAVELIN} have the lowest scatter in the distribution: {the Pearson correlation coefficients are $r_{\rm ICCF}\sim$0.984, $r_{\rm ZDCF}\sim$0.980, $r_{\rm \tt{JAVELIN}}\sim$0.993}, indicating that {\tt JAVELIN} lags are more accurate. ICCF lags are consistent with their assigned lags in general, despite the larger scatter. ZDCF is the the least accurate at reproducing the assigned lags and is not capable for detecting lags shorter than the observation cadence by design. {For the flux-limited sample, the Pearson correlation coefficients are $r_{\rm ICCF}\sim$0.933, $r_{\rm ZDCF}\sim$0.925, $r_{\rm \tt{JAVELIN}}\sim$0.974.}

Figure \ref{fig:lag_relerr} evaluates the quality of lag measurements by comparing the normalized measurement uncertainties (normalized by the value of the assigned lag) and fractional difference between the assigned and measured lags ($\delta{\rm Lag\equiv |\tau_{\rm mea}/\tau_{\rm assigned}-1|}$) for the true detections {of the uniform sample}. At low measurement quality (high normalized uncertainties and $\delta$Lag), the irregular edges are caused by the detection criteria and are similar among all three methods. At high measurement quality, ICCF and {\tt JAVELIN} are able to make lag measurements with smaller uncertainties and $\delta$Lag than ZDCF. {The normalized measurement uncertainties and $\delta$Lag (median values and uncertainties derived from the 16$^{\rm th}$/84$^{\rm th}$ percentiles) are 7.5\%($^{+12}_{-0.50}$), 3.3\%($^{+8.0}_{-2.5}$) for ICCF, 11\%($^{+12}_{-6.1}$), 3.6\%($^{+6.3}_{-2.5}$) for ZDCF and 4.3\%($^{+12}_{-0.30}$), 2.2\%($^{+7.3}_{-1.7}$) for {\tt JAVELIN}.} In addition, more {\tt JAVELIN} lags lie in the higher quality regime (low normalized uncertainties and $\delta$Lag) than ICCF lags. {For the flux-limited sample, the normalized measurement uncertainties and $\delta$Lag are 15\%($^{+12}_{-8.7}$), 5.8\%($^{+12}_{-4.4}$) for ICCF, 17\%($^{+11}_{-9.2}$), 4.7\%($^{+11}_{-3.5}$) for ZDCF and 9.8\%($^{+13}_{-6.6}$), 4.2\%($^{+10}_{-3.3}$) for {\tt JAVELIN}.} 

Figure \ref{fig:lag_abserr} demonstrates that the absolute uncertainties of {\tt JAVELIN} lags are smaller than those from ICCF and ZDCF, a result confirmed in previous works \citep[e.g.,][]{Grier_2017,Edelson_2019}. With our controlled experiment with known lags, we are able to demonstrate that {\tt JAVELIN} can provide more accurate lag measurements, as already evident in Figure \ref{fig:lag_density}. In addition, Figure \ref{fig:lag_relerr} suggests that the {\tt JAVELIN} errors are reasonable and are not an underestimation of the actual uncertainties in general. To further illustrate this point, Figure \ref{fig:norm_err} shows the distribution of the difference between assigned and measured lags, normalized by the measurement errors, {for the uniform sample}. {The distribution for {\tt JAVELIN} ($\sigma_{gauss}\sim$0.85) is most consistent with a Gaussian with unity dispersion, while the ICCF ($\sigma_{gauss}\sim$0.69) and ZDCF ($\sigma_{gauss}\sim$0.54) lag errors are more overestimated, leading to narrower distributions. ICCF also produces more outliers with underestimated lag uncertainties ($\Delta {\rm Lag}/\sigma_{\tau, mea}>3$) ($\sim$5.1\%) compared to {\tt JAVELIN} ($\sim$0.57\%) and ZDCF does not have any outliers. For the flux-limited sample, $\sigma_{gauss}\sim$0.73 for ICCF, $\sigma_{gauss}\sim$0.49 for ZDCF and $\sigma_{gauss}\sim$0.76 for {\tt JAVELIN} and the fractions of outliers with underestimated lag uncertainties are 1.1\%, 0.0\% and 0.28\%, respectively.}

The reason why ICCF produces overestimated lag errors is not entirely clear \citep[e.g.,][]{Edelson_2019}. The flux resampling part of ICCF produces noisier light curves than the original light curve (i.e., the data points are perturbed twice by flux errors), and the random subset sampling procedure will remove epochs, which increases the uncertainty of lag detection due to the loss of temporal information and may become critical in the low-quality regime (i.e., sparse sampling and large light curve errors). {\tt JAVELIN} is a more statistically rigorous approach and does not suffer from these simplifications used in the ICCF.    

\begin{figure}[h!]
\centering
	\begin{tabular}{@{}cc@{}}
	\includegraphics[width=0.5\textwidth]{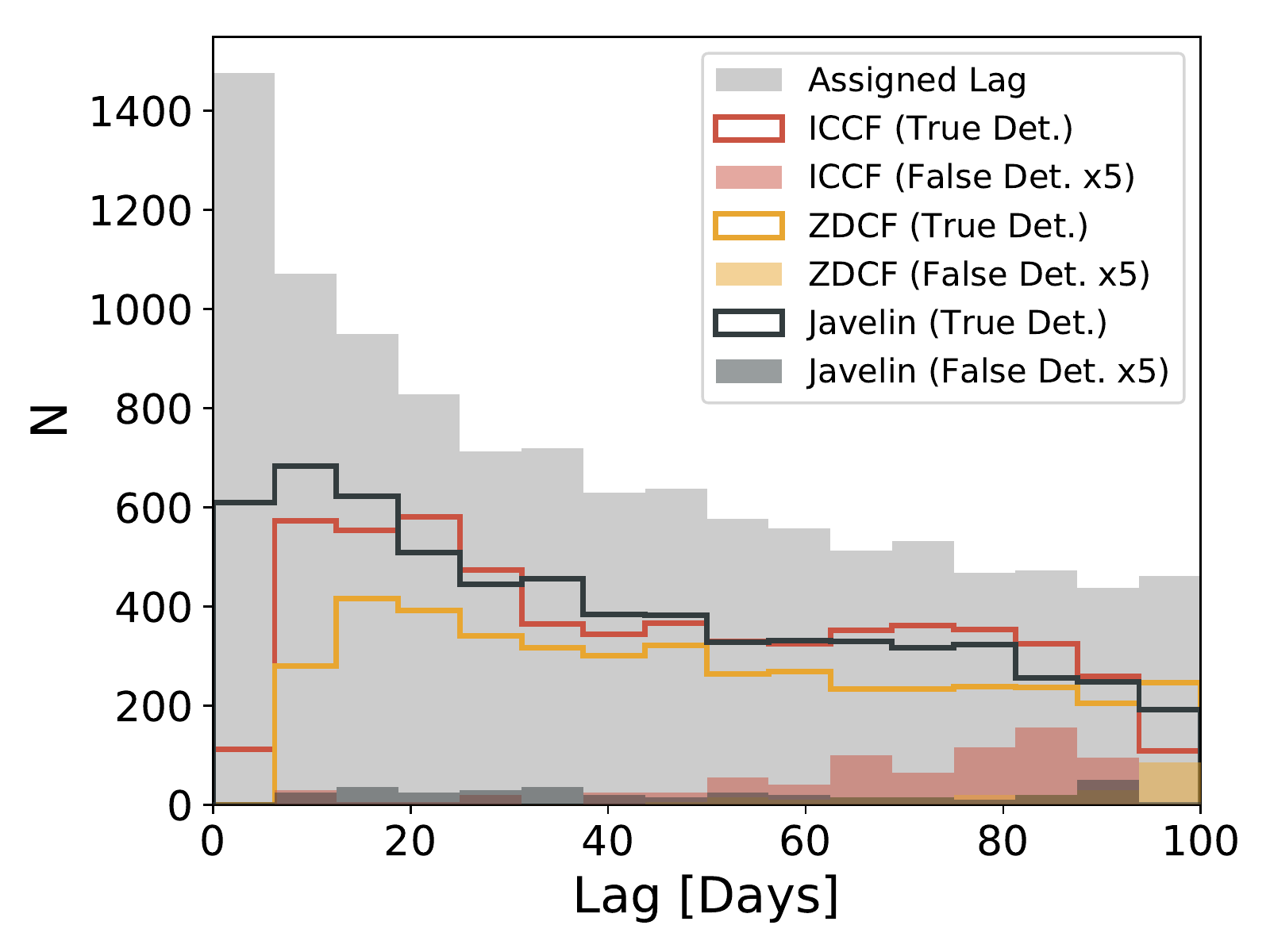}
	\end{tabular}
    \caption{Distribution of the measured lags of the uniform sample in observed frame. The grey solid histogram shows the number of detectable lags in each bin. The open histograms represent the number of true detections and the solid histograms are the number of false detections. {The number of false detections are inflated by a factor of five for clarity.}}
    \label{fig:hist_lim100}
\end{figure}

\begin{figure}[h!]
\centering
	\begin{tabular}{@{}cc@{}}
	\includegraphics[width=0.5\textwidth]{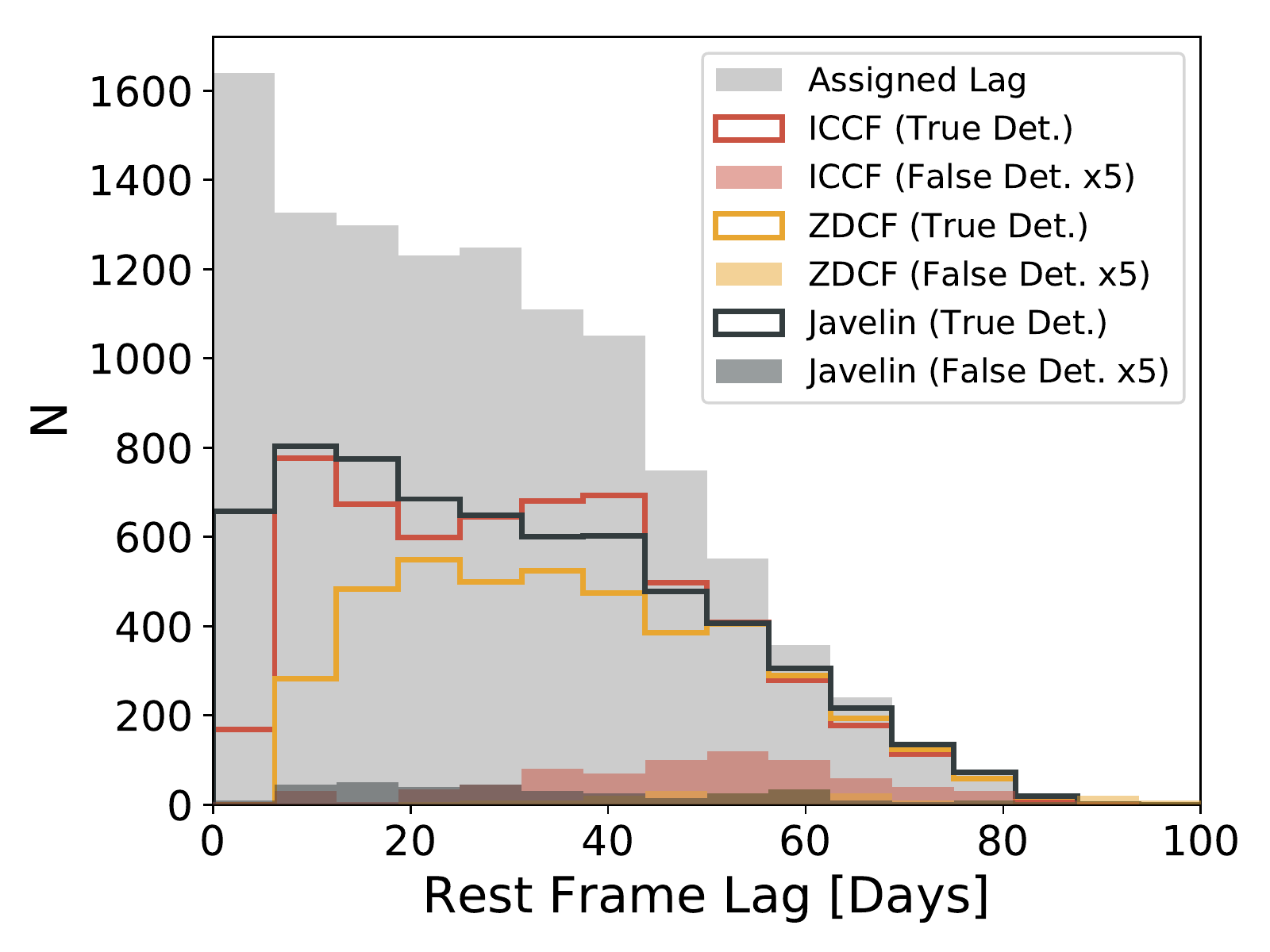}
	\end{tabular}
    \caption{Similar format to Figure \ref{fig:hist_lim100}. The distribution of measured lags of the uniform sample in rest frame.}
    \label{fig:hist_RF}
\end{figure}

\subsection{Distribution of Detected Lags}\label{sec:distribution_lags}

Figure \ref{fig:hist_lim100} compares the distribution of measured lags to that of assigned lags in the uniform sample. Lags in the range of $\sim$10--90 days are most likely to be detected with our fiducial cadence and baseline. In this range, {all methods have similar detection efficiency in each lag bins; the median detection efficiencies are $\sim$61$\%$ for ICCF, $\sim$52$\%$ for ZDCF and $\sim$64$\%$ for {\tt JAVELIN}}, which suggests the detections are not biased towards certain lag ranges. Interestingly, {\tt JAVELIN} detects many more short lags than ICCF and ZDCF. This behavior indicates that, by assuming the DRW model, {\tt JAVELIN} is capable of producing reasonable predictions of light curves on a grid finer than the cadence, and thus makes it possible to detect a lag below the formal cadence of the data under certain circumstances. {As shown in Figure \ref{fig:hist_lim100}, most of the false detections fall in the range of $>$60 days. ICCF is prone to producing false detections in the 60--100 days range, regardless of the input lag.}

Figure \ref{fig:hist_RF} shows the same distributions but in the rest frame. A uniform and wide distribution of rest-frame lags is critical to measuring an unbiased R-L relation. With our 180-day monitoring duration, we detect mostly rest-frame lags in the range of 20--40 days. The detection rate decreases at $\lesssim20$ days due to cadence limitations, and fewer detections are made at $>40$ days because the observed-frame lags are shifted beyond our search range. We further discuss the biases in measuring the R-L relation slope under different observing conditions and lag measurement methods in Section \ref{sec:rl}.

\begin{figure*}
    \centering
    	\includegraphics[width=1.0\textwidth]{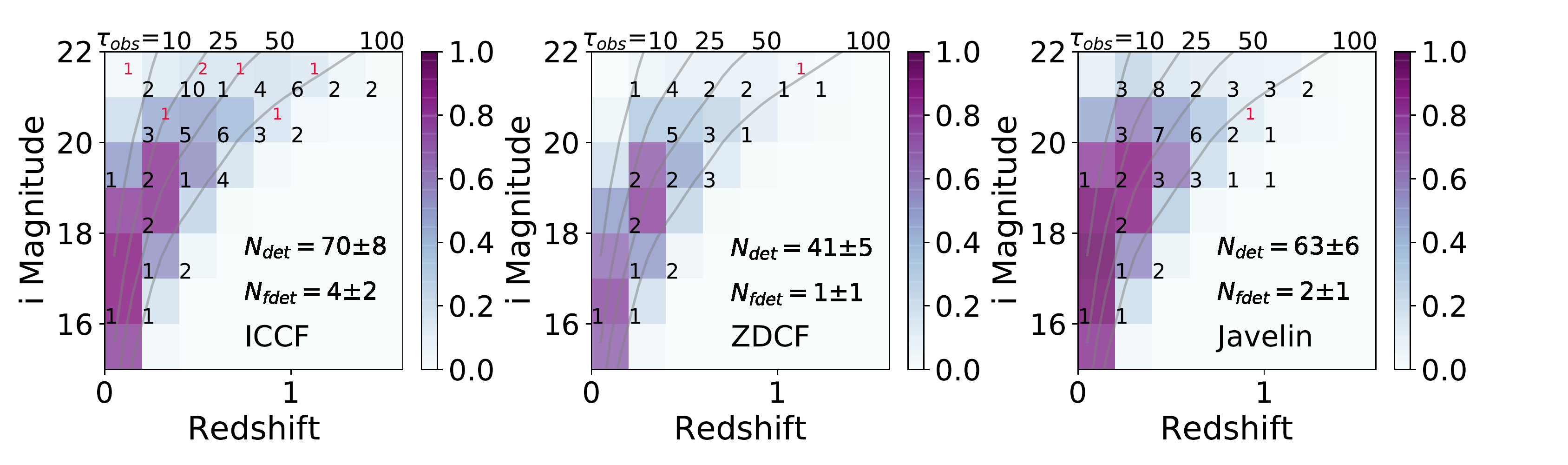}
    \caption{
    Detection efficiency in the simulated grid of quasars measured with each method in a simulated program of 6-day cadence and 30 epochs. From the top to bottom panels are the results from ICCF, ZDCF and {\tt JAVELIN}. The colormap represents the detection efficiency and the numbers are the detection counts (true detections in black and false detections in red) of a single down-sampling realization. The total numbers of true and false detections shown in the lower-right corner are the median and uncertainties derived from 100 down-sampling realizations. The grey contours show the approximate constant lags from the R-L relation from \cite{Bentz_2009b}.}
    \label{fig:detmap}
    \end{figure*}

\begin{figure*}
\centering
	\begin{tabular}{@{}cc@{}}
	\includegraphics[width=0.5\textwidth]{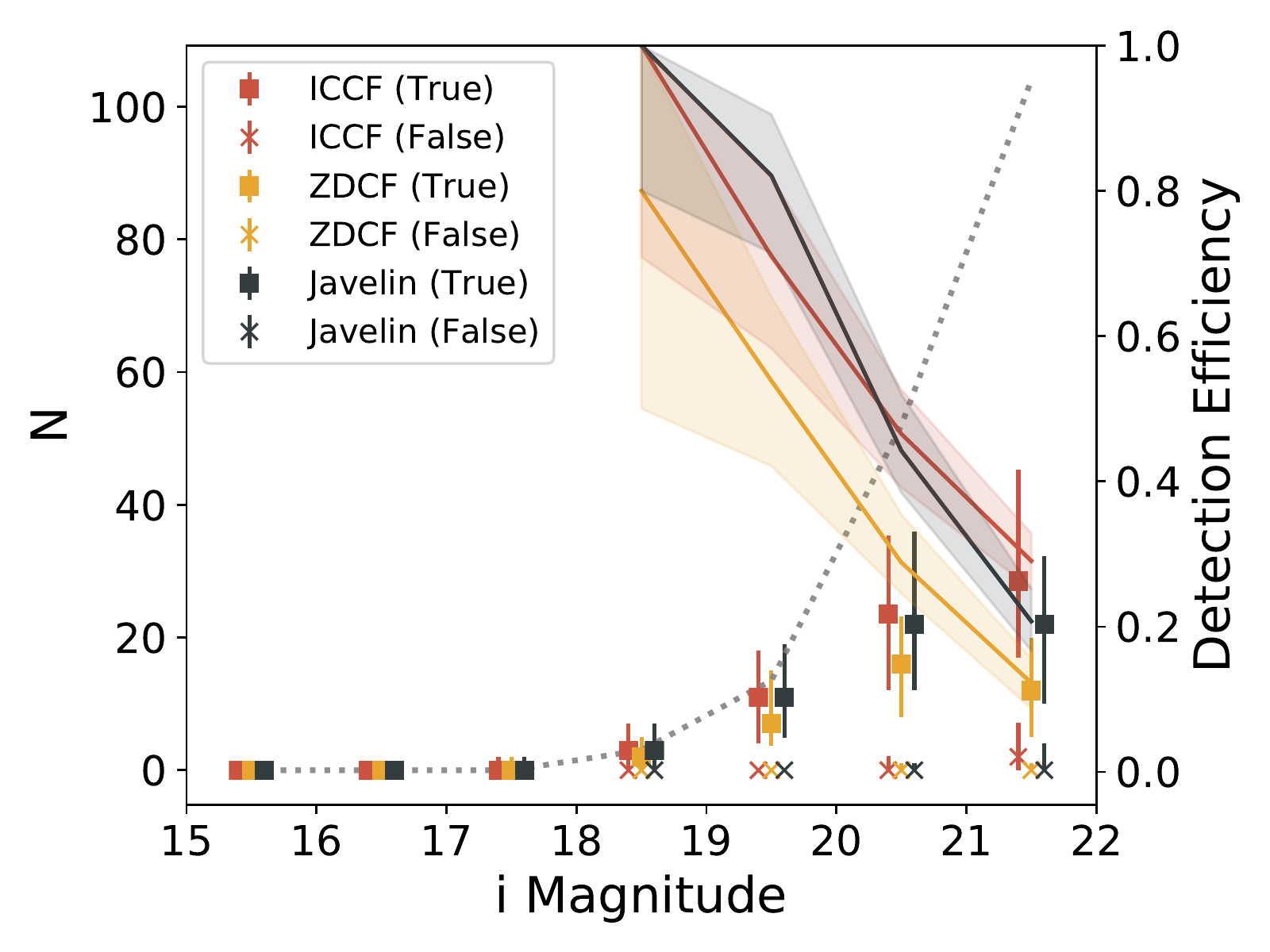}
	\includegraphics[width=0.5\textwidth]{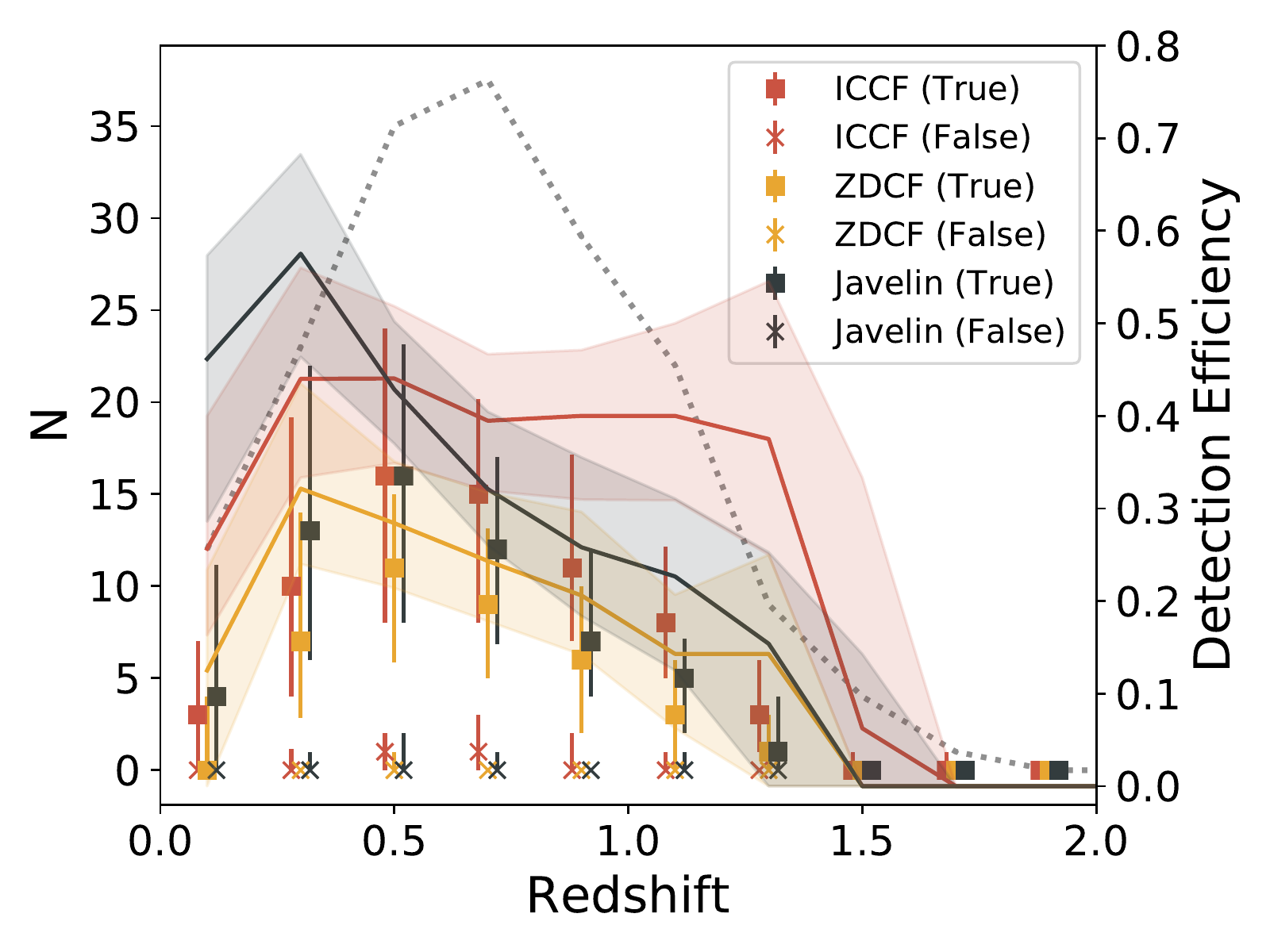}
	\end{tabular}
    \caption{Detection efficiency (solid line and shaded area) and true (square) and false (cross) detection counts of the three methods as functions of $i$-band magnitude (right panel) and redshift (left panel) in a simulated observation with 6 day cadence and 30 epochs. Detection counts are obtained using 100 down-sampling realizations, the median and the 16$^{\rm th}$ and 84$^{\rm th}$ percentiles are adopted as the final counts and their uncertainties. {The dotted lines show the number of sources with lags shorter than the search range (i.e. 100 days) in each magnitude or redshift bin.} For $i<$18, the detection efficiencies are not shown because there are no quasars selected in more than 80\% bootstrapping iterations.}
    \label{fig:corr}
\end{figure*}

\subsection{Detection Efficiency}\label{sec:det_eff}
Figure \ref{fig:detmap} displays the detection efficiency of true detections in each redshift and $i$-band magnitude bin with simulations of 6-day cadence and 30 epochs for the uniform sample. {The overall detection fractions are $\sim$40\% for ICCF and $\sim$36\% {\tt JAVELIN}, and $\sim$23\% for ZDCF out of all the detectable sources (i.e. assigned lag$<$100 days) in the flux-limited sample.} However, as previously shown in Figure \ref{fig:hist_lim100}, ICCF has a higher false detection rate {($\sim$5.4\%)} than the other two methods {({\tt JAVELIN}$\sim$3.1\% and ZDCF$\sim$2.4\%) for the flux-limited sample}. Most of these false detections lie in the fainter quasar population, where the quasar variability is buried in the flux measurement uncertainties. 

The detection efficiency, as the time lags, depends on redshift and $i$-band magnitude. Observed-frame lags are time dilated by $(1+z)$, so the lags will be shifted out of the search range at high redshifts. Our 100-day search range only allows detection of lags at redshifts $z<$1.5. Similarly, lags at low redshift are difficult to detect as the lags may fall below the observing cadence. Quasar variability is more likely to be diluted by noise for dimmer sources, so the detection efficiency naturally decreases as we approach the survey flux limit. In the faintest $i$-mag and lowest $z$ bin, the detection rate is low because luminous quasars with high Eddington ratios tend to vary more on longer timescales ($>100$ days) and their lags are not detectable within our observing baseline \citep{Macleod_2010}. 

The detection efficiency and detection counts as functions of $i$-band magnitude and redshift are shown in Figure \ref{fig:corr}, using the downsampled simulations that mimic the SDSS-RM sample. Detection efficiency decreases with $i$-band magnitude. However, since the number of quasars increases with $i$-band magnitude, the number of detections also increases. The detection efficiencies of ICCF and {\tt JAVELIN} are roughly the same and higher than that of ZDCF in all magnitude bins.

The detection efficiency is the highest for {\tt JAVELIN} and ZDCF at redshift $\sim$0.2 and decreases both towards lower and higher redshift. As the redshift increases, quasars with detectable lags tend to be fainter, thus decreasing the detection efficiency. The lags of the quasars in the lowest redshift bin are too short to detect. For ICCF, detection efficiency is relatively consistent in the range of $\sim$0.2--1.0, {because ICCF is more sensitive to lags around $\sim$100 days.}

Our down-sampled realizations demonstrate that most of the detected lags are from quasars around redshift $\sim$0.5 and with $i$>19. This behavior is a selection effect due to the sample characteristics and the range of lags where the fiducial survey design is sensitive.

\begin{figure}
\centering
	\begin{tabular}{@{}cc@{}}
	\includegraphics[width=0.5\textwidth]{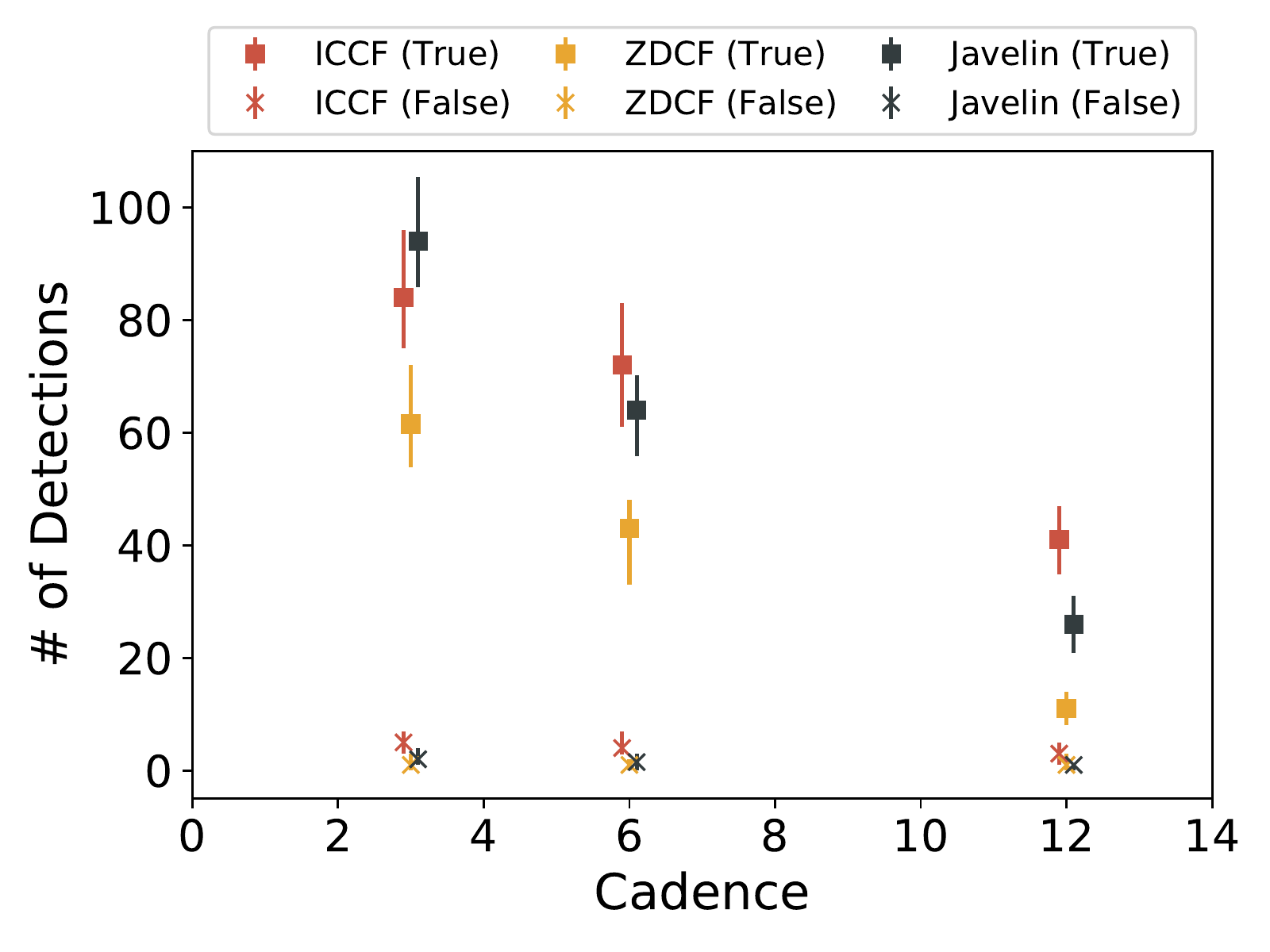}
	\end{tabular}
    \caption{Total counts of true (square) and false (cross) detections in the flux-limited sample of simulations with 3-, 6-, and 12-day cadence.}
    \label{fig:cad}
\end{figure}

\subsection{Effects of Cadence/$N_{epoch}$}\label{sec:diss_cad}

To investigate the effect of cadence on lag detection, we ran additional simulations with cadences of 3 and 12 days and the same 180-day observation baseline to compare to our fiducial cadence of 6 days. We start from the daily-sampled light curves described in Section \ref{sec:data} and resample the cadence and flux measurements based on the same mock quasars and light curves, and follow the same procedures in measuring lags as described in Section \ref{sec:measurelags}.

As shown in Figure \ref{fig:cad}, using the same method, the overall detection efficiency decreases as the monitoring cadence increases. Again {\tt JAVELIN} and ICCF have higher detection efficiencies than ZDCF. The increase of detection efficiency as cadence improves is mainly a result of more data points in the light curves, since most of the expected lags will be resolved even with a 12-day cadence, but a higher cadence can lead to more lag detections on shorter timescales. These results are already confirmed in earlier simulations with ICCF \citep{Shen_2015a}.

\begin{figure*}
\centering
	\begin{tabular}{@{}cc@{}}
	\includegraphics[width=0.33\textwidth]{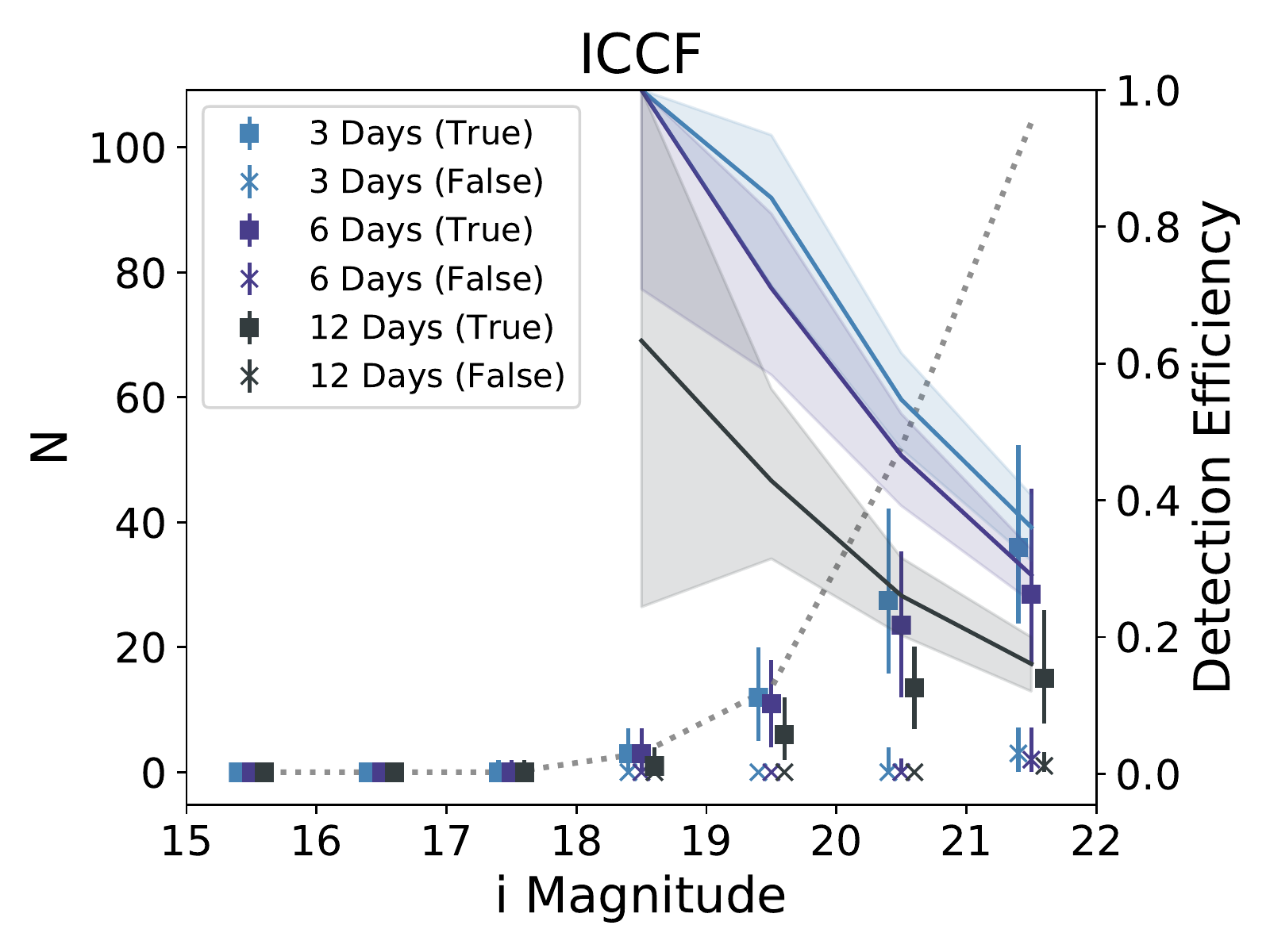}
	\includegraphics[width=0.33\textwidth]{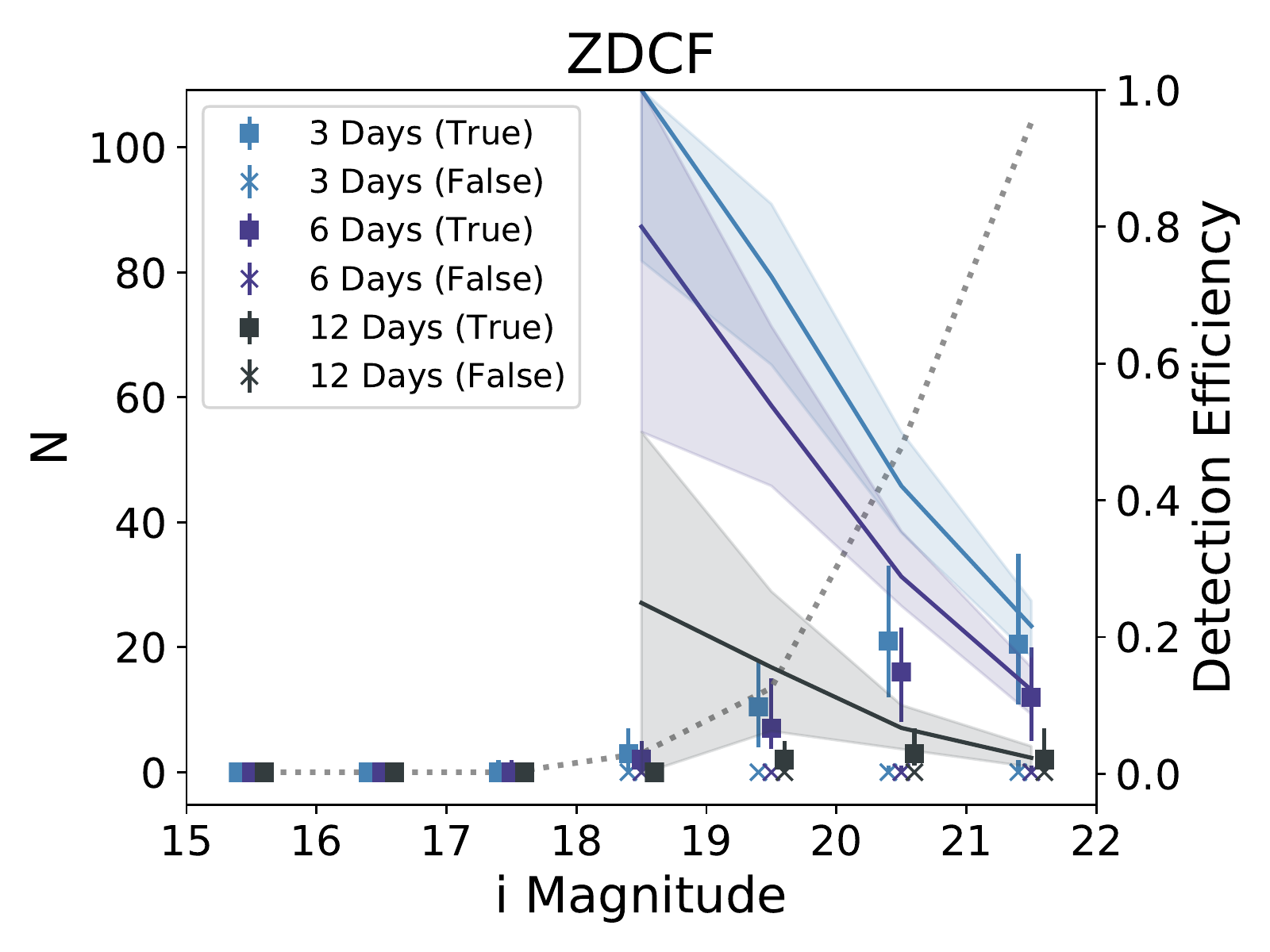}
	\includegraphics[width=0.33\textwidth]{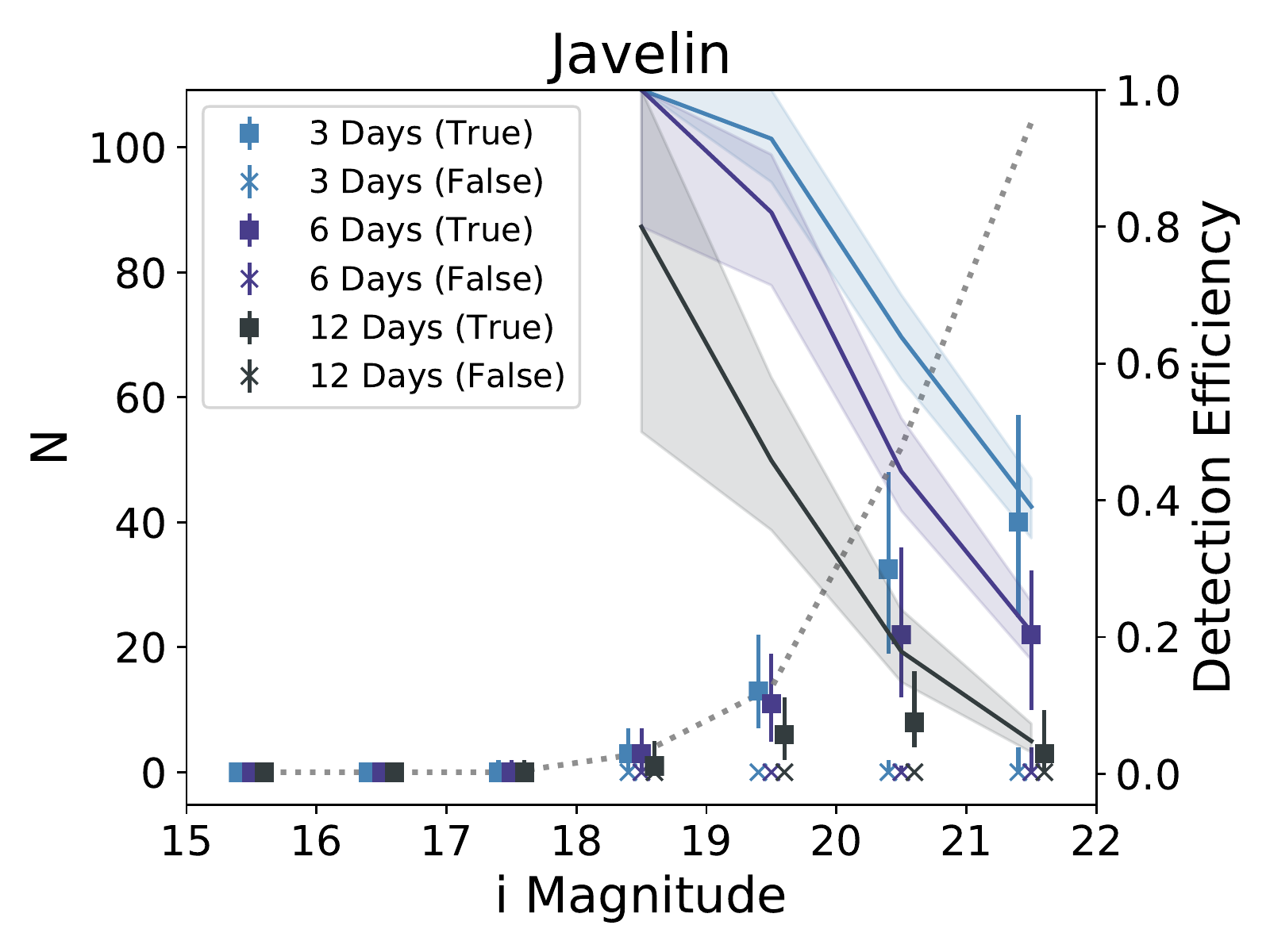}
	\end{tabular}
    \caption{
    Detection efficiency (solid line and shaded area) and true (square) and false (cross) detection counts as functions of $i$-band magnitude in simulated observations with 3-, 6- and 12-day cadence. The dotted lines show the number of sources with lags shorter than the search range (i.e. 100 days) in each magnitude or redshift bin. For $i<$18, the detection efficiencies are not shown because there are no quasars selected in more than 80\% bootstrapping iterations.}
    \label{fig:corr_cad_imag}
\end{figure*}

\begin{figure*}
\centering
	\begin{tabular}{@{}cc@{}}
	\includegraphics[width=0.33\textwidth]{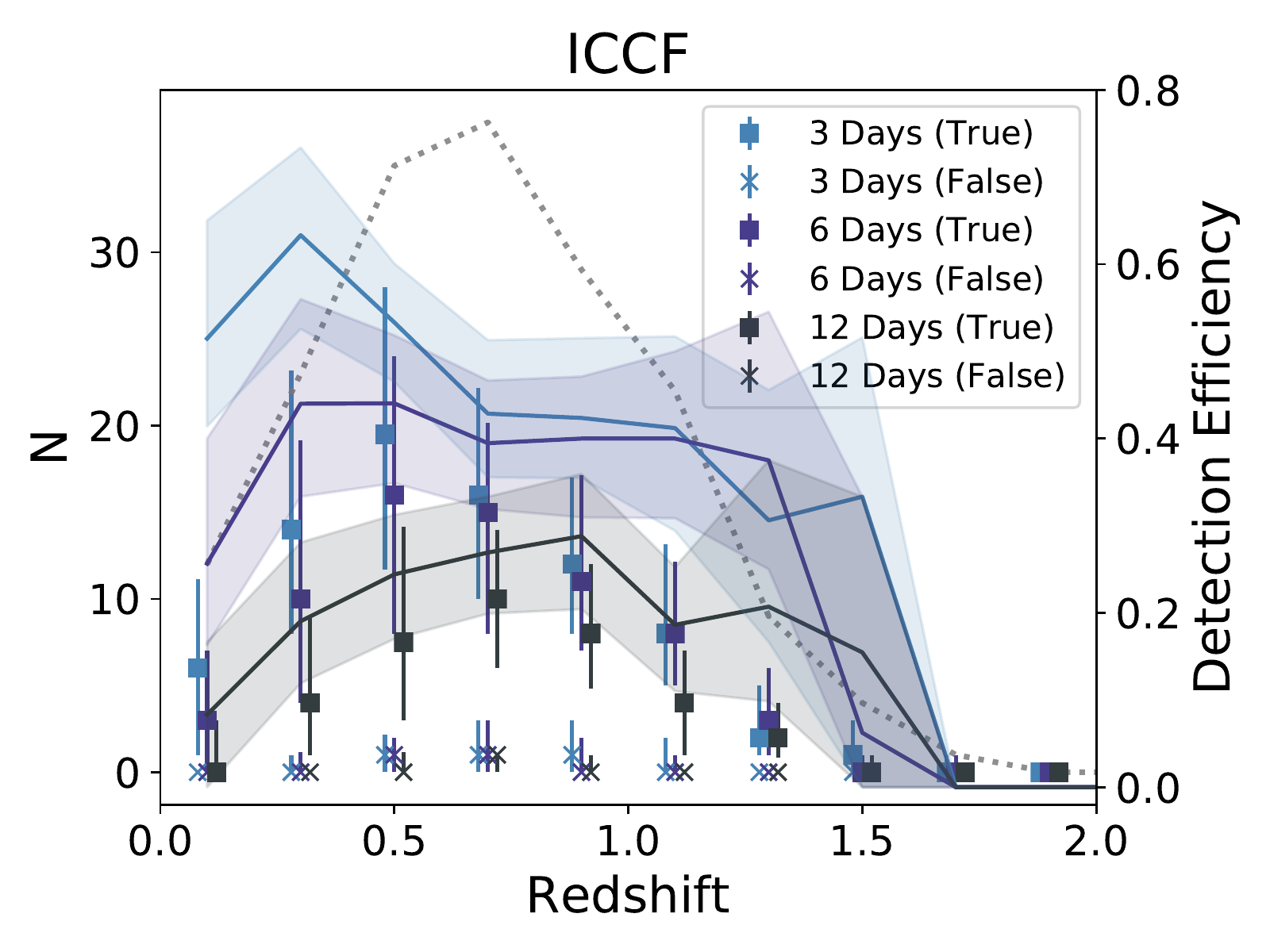}
	\includegraphics[width=0.33\textwidth]{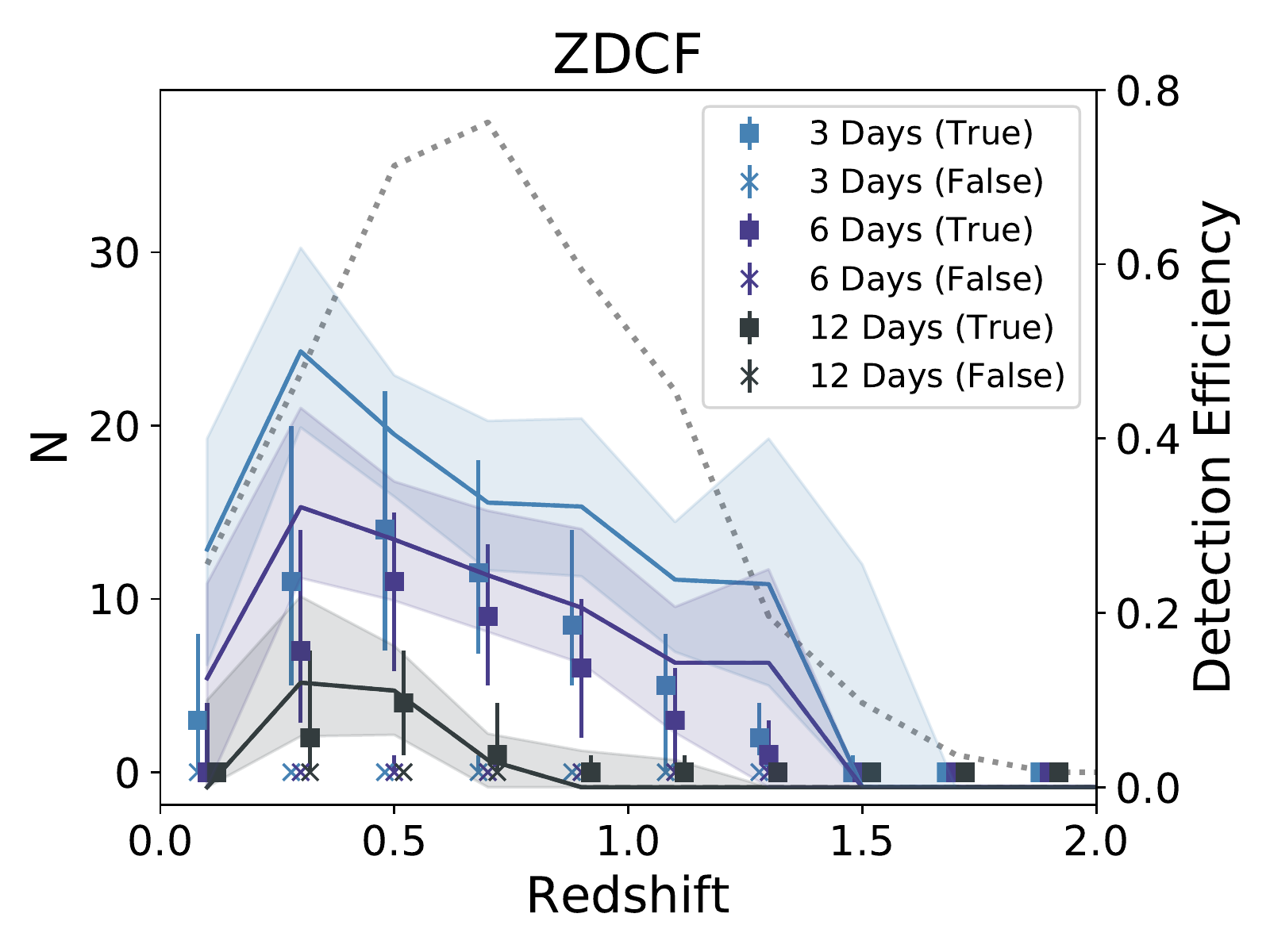}
	\includegraphics[width=0.33\textwidth]{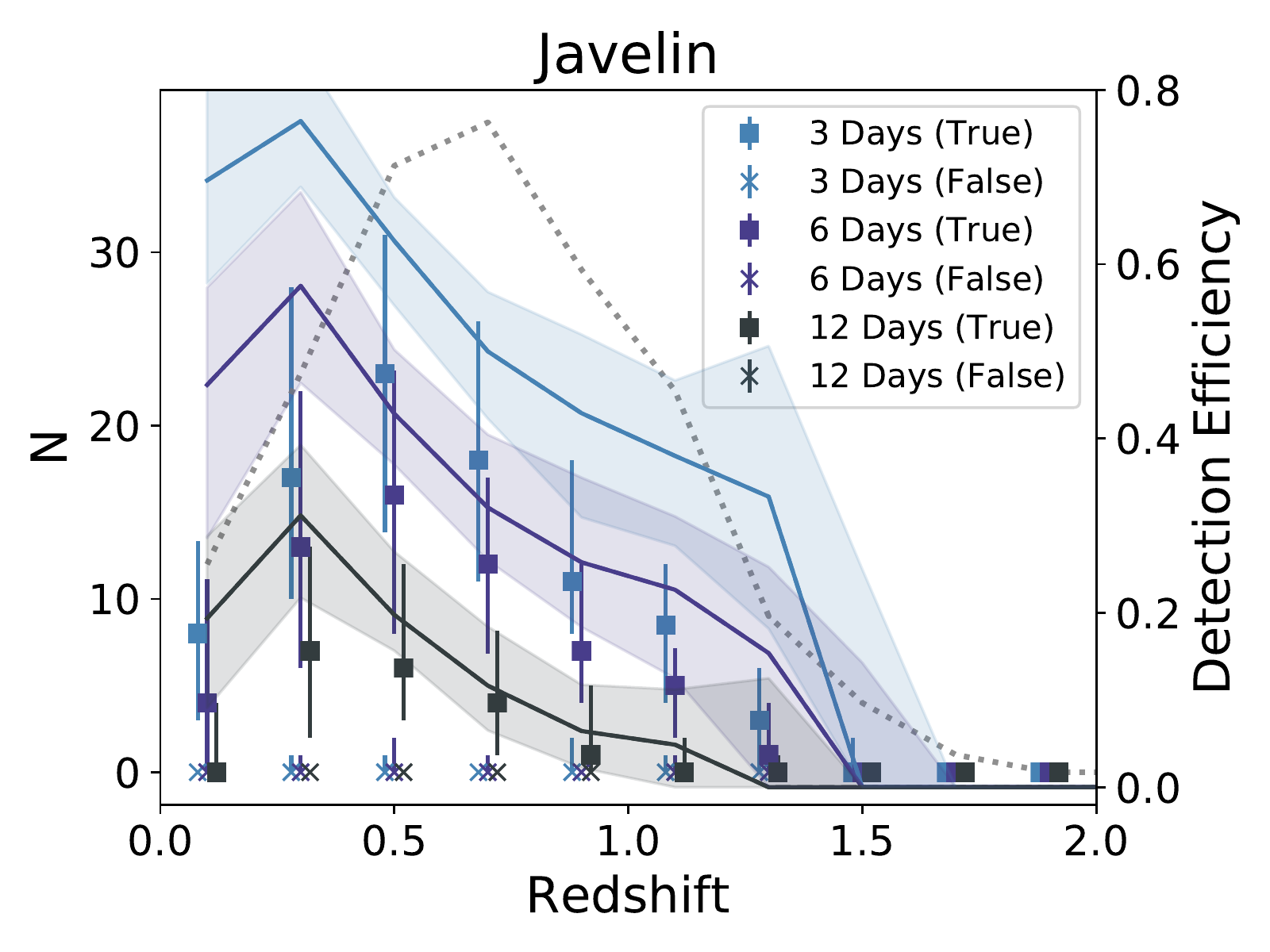}
	\end{tabular}
    \caption{
    Detection efficiency (solid line and shaded area) and true (square) and false (cross) detection counts as functions of redshift in simulated observations with 3-, 6- and 12-day cadence. The dotted lines show the number of sources with lags shorter than the search range (i.e. 100 days) in each magnitude or redshift bin.}
    \label{fig:corr_cad_z}
\end{figure*}

Figure \ref{fig:corr_cad_imag} and Figure \ref{fig:corr_cad_z} show the breakdown of the detection efficiency in each redshift and $i$-band magnitude bin for different cadences/$N_{epoch}$ (but with fixed baseline). The number of detections and detection efficiency decrease in all bins with increasing cadence as expected. The only exception is the high-redshift bins in the ICCF case, which remains similar in the 3-day and 6-day cadence simulation; this result may simply be due to small number statistics. 

We also ran our simulated observations with non-uniform cadence using the first-year SDSS-RM spectroscopic observations that have an average cadence of 5.7 days (median cadence of 4 days) and 32 epochs. For ICCF and {\tt JAVELIN}, the overall detection efficiency and number of detections after downsampling are consistent with our uniform-cadence simulations. While correlated variations in the poorly-sampled sections of the light curves may be missed by the correlation analysis for some sources, lags of other sources might be identified in more densely-sampled parts of the light curves, so the non-uniform cadence does not significantly change the results for the overall sample. However, this is not the case for ZDCF, where the detection efficiency for the non-uniform cadence case is only about half of that for the uniform-cadence case. ZDCF detects fewer lags in all bins with non-uniform cadence, but especially so at lag $\lesssim 20$ days and $\sim40$ days. This lack of detections arises because the ZDCF binning algorithm is less sensitive to lag in this range with the non-uniform cadence. Having a reasonable interpolation scheme, such as with ICCF or {\tt JAVELIN}, helps detect lags when the cadence is not uniform.

\begin{figure}
\centering
	\begin{tabular}{@{}cc@{}}
	\includegraphics[width=0.5\textwidth]{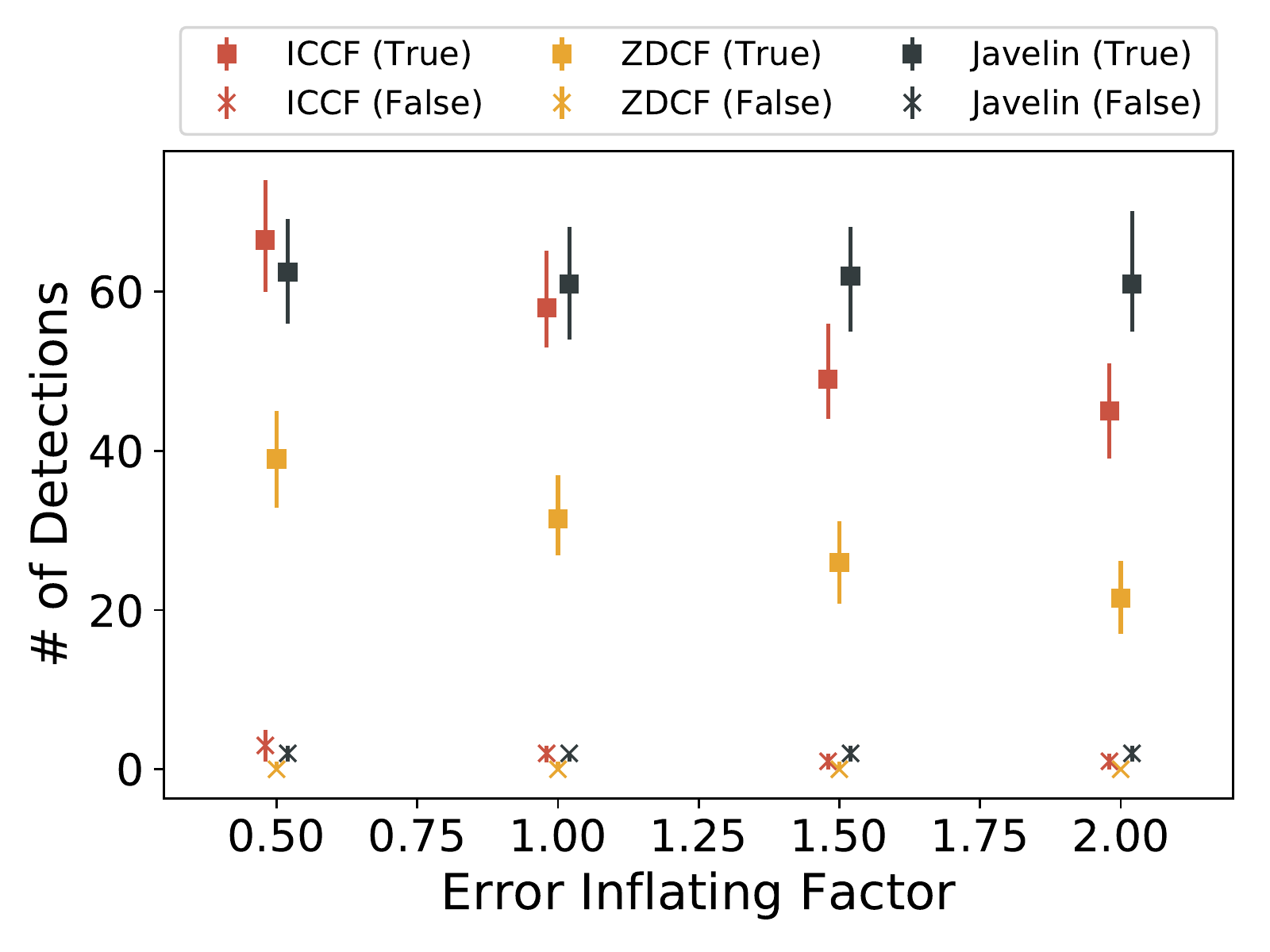}
	\end{tabular}
    \caption{Total counts of true (square) and false (cross) detections in the flux-limited sample at different error inflating factor for the line light curve. Continuum light curve errors are all 3.5 times higher compared to previous figures (i.e., Figures \ref{fig:lag_density} to \ref{fig:corr_cad_z}).}
    \label{fig:snr}
\end{figure}
   
\begin{figure*}
\centering
	\begin{tabular}{@{}cc@{}}
	\includegraphics[width=0.33\textwidth]{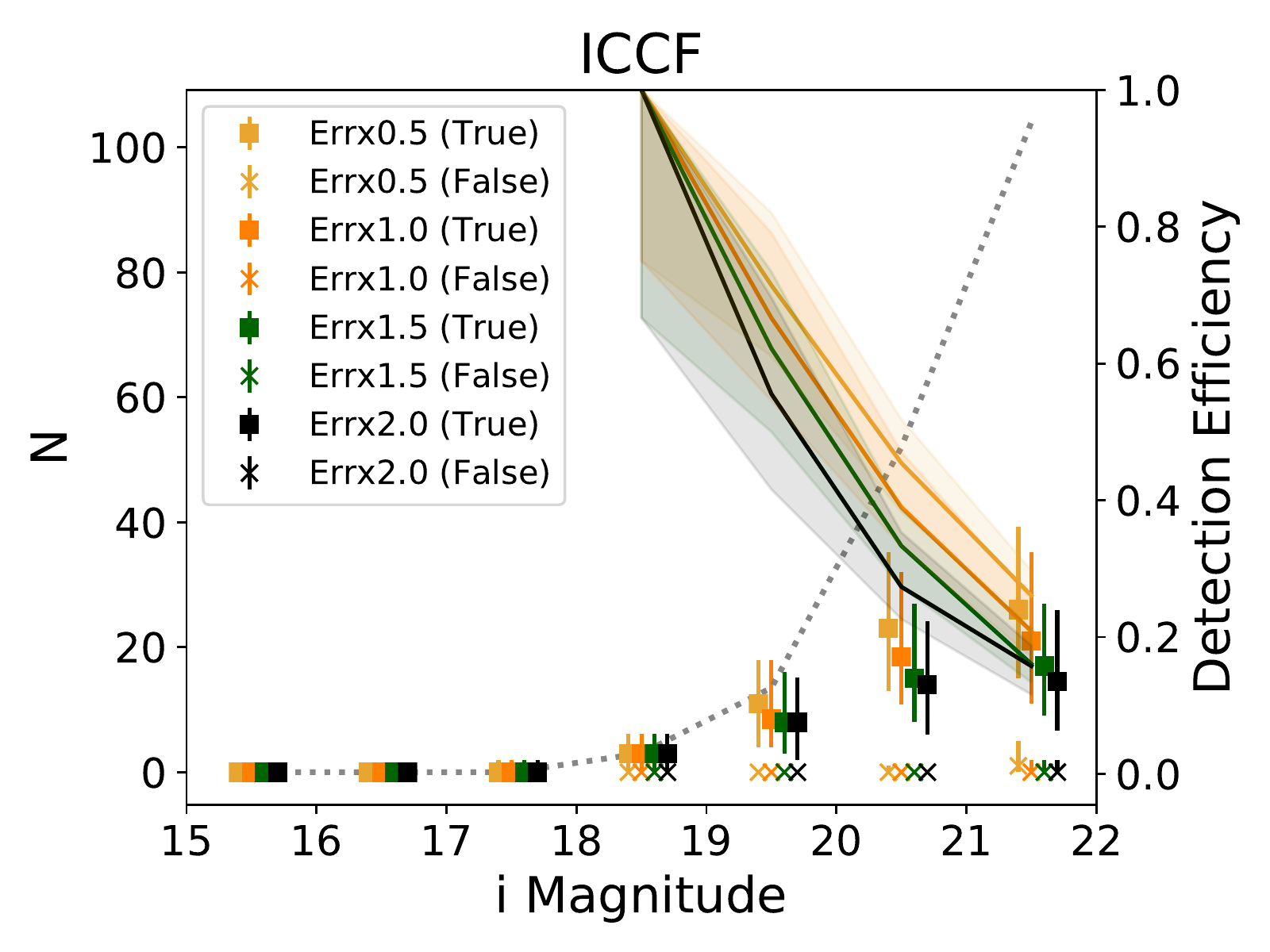}
	\includegraphics[width=0.33\textwidth]{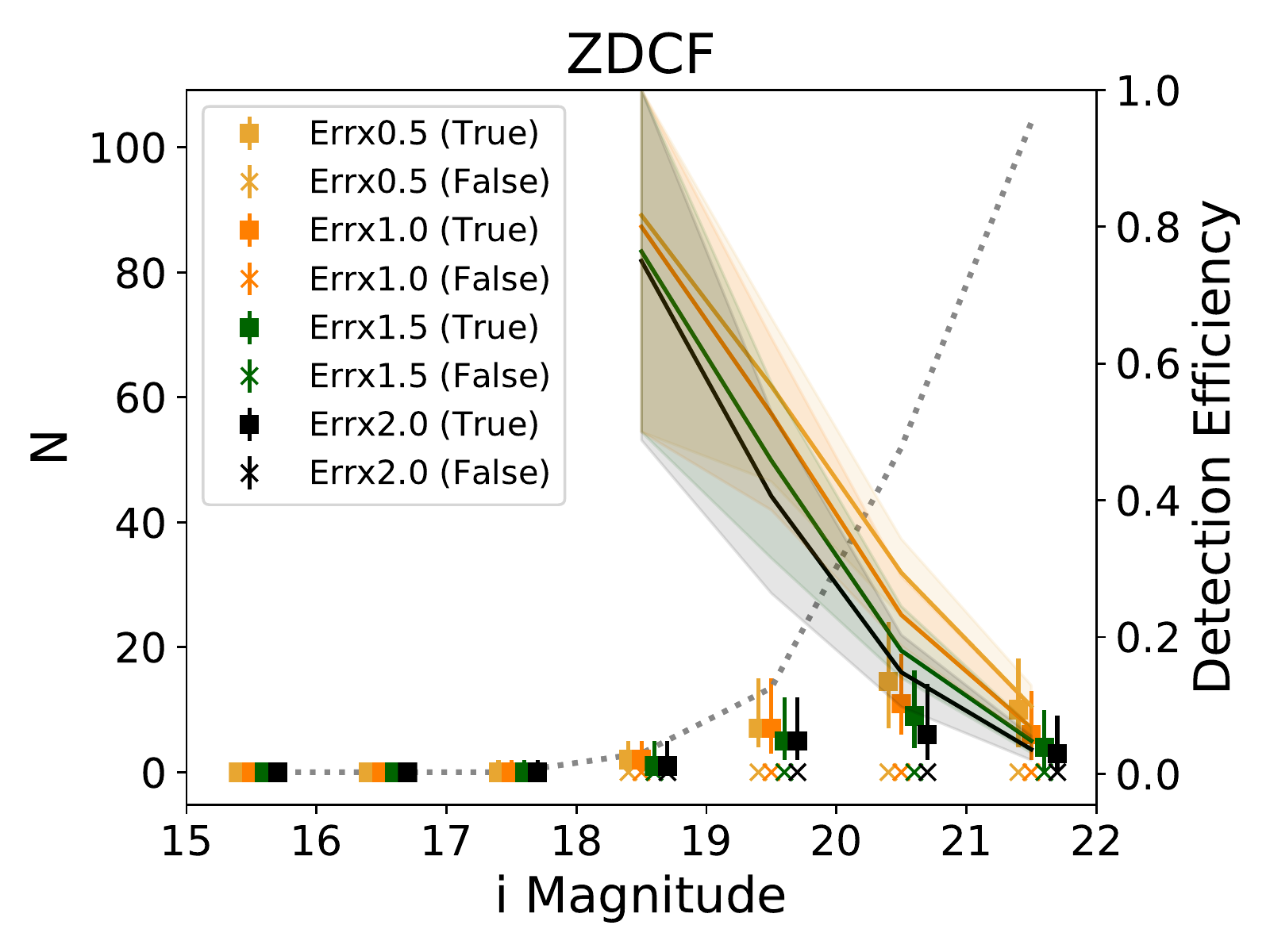}
	\includegraphics[width=0.33\textwidth]{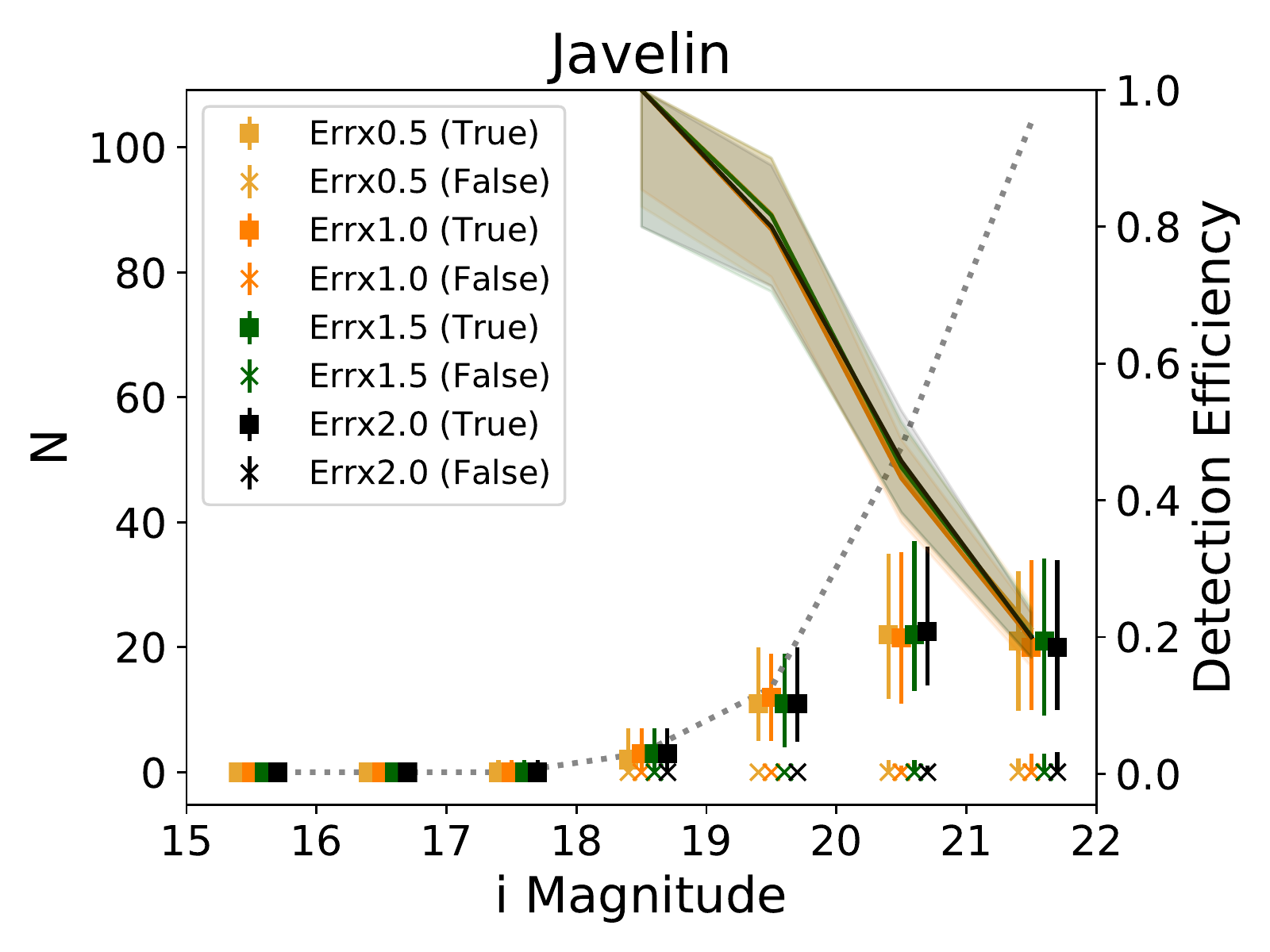}
	\end{tabular}
    \caption{
    Detection efficiency (solid line and shaded area) and true (square) and false (cross) detection counts as functions of $i$-band magnitude in simulated observations with inflated error bars. The dotted lines show the number of sources with lags shorter than the search range (i.e. 100 days) in each magnitude or redshift bin. For $i<$18, the detection efficiencies are not shown because there are no quasars selected in more than 80\% bootstrapping iterations.}
    \label{fig:corr_snr_imag}
\end{figure*}   
    
\begin{figure*}
\centering
	\begin{tabular}{@{}cc@{}}
	\includegraphics[width=0.33\textwidth]{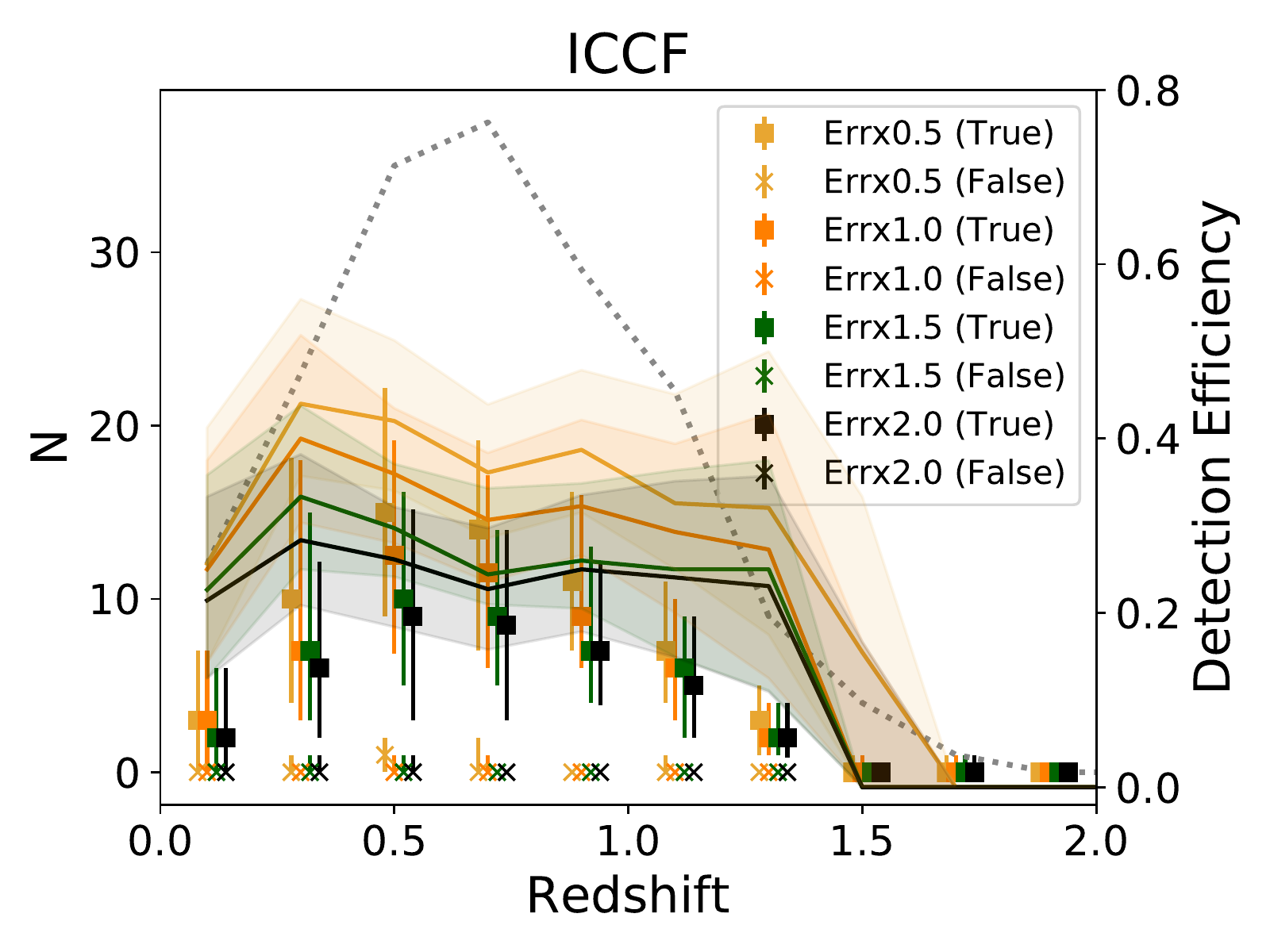}
	\includegraphics[width=0.33\textwidth]{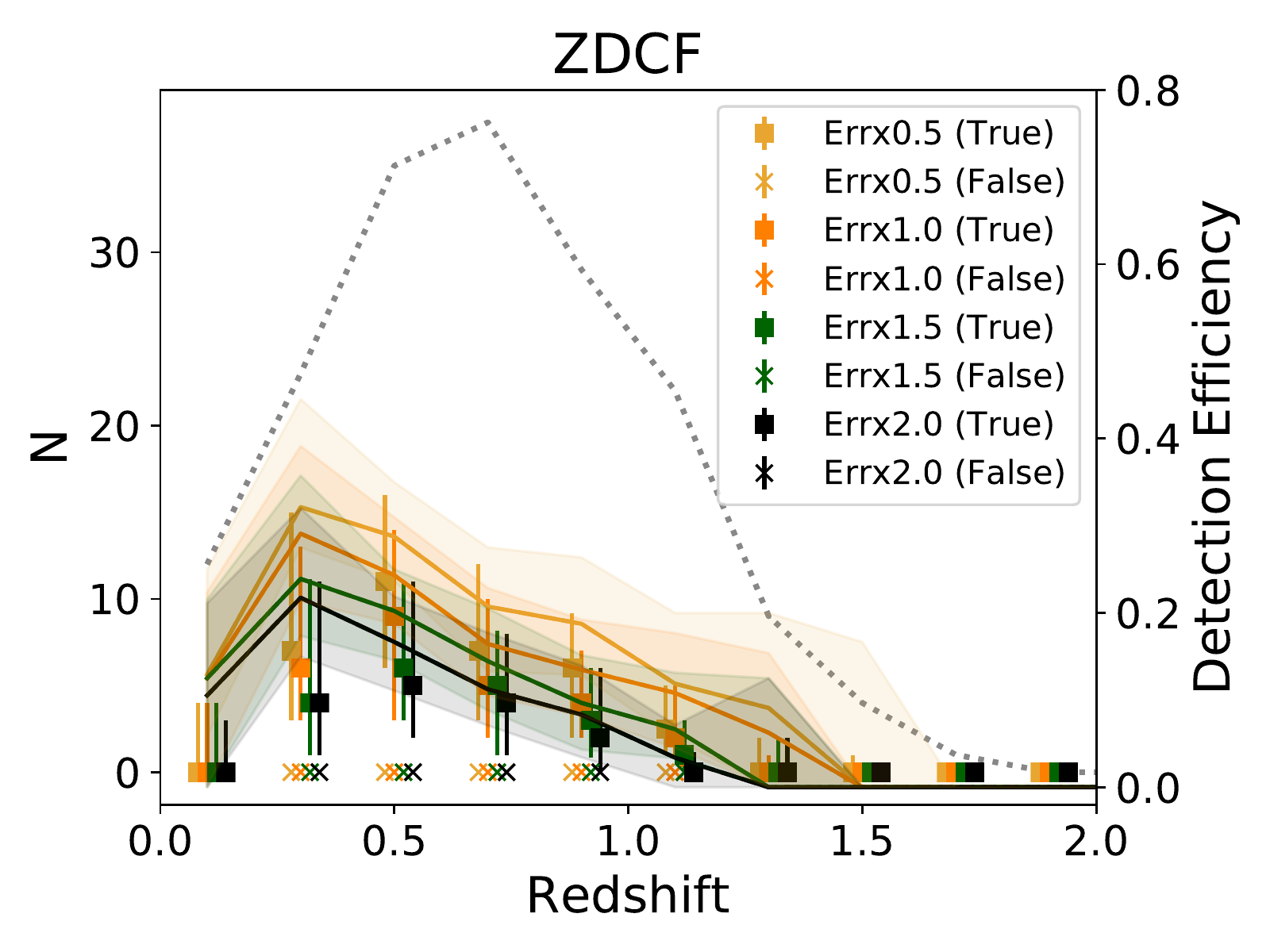}
	\includegraphics[width=0.33\textwidth]{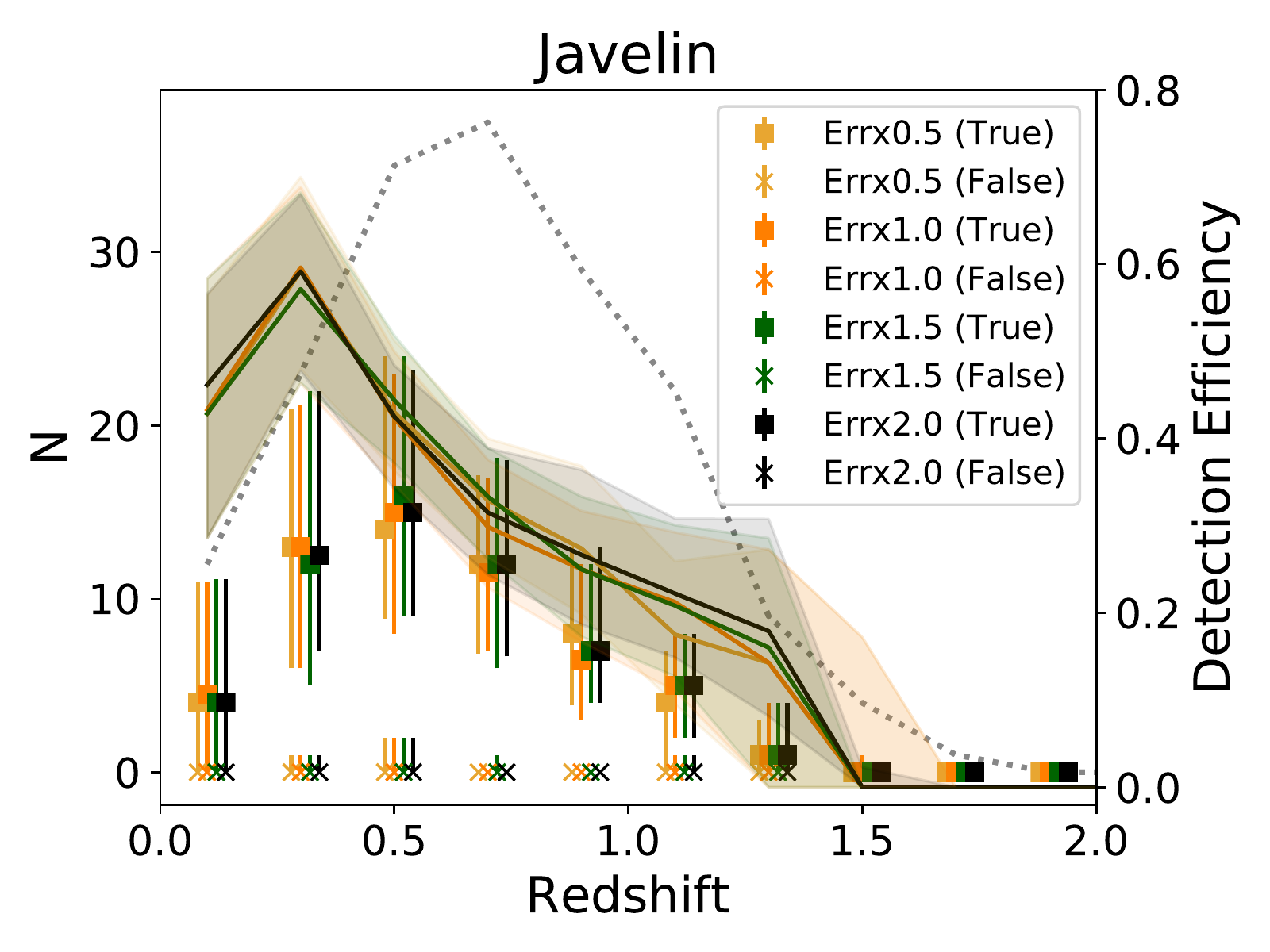}
	\end{tabular}
    \caption{
    Detection efficiency (solid line and shaded area) and true (square) and false (cross) detection counts as functions of redshift in simulated observations with inflated error bars. The dotted lines show the number of sources with lags shorter than the search range (i.e. 100 days) in each magnitude or redshift bin.}
    \label{fig:corr_snr_z}
\end{figure*}

\subsection{Effects of Light Curve S/N}\label{sec:diss_err}

Sufficient light curve S/N is required for any lag measuring method to identify correlated variability in the presence of flux errors. In this section, we decrease our continuum light curve S/N by a factor of 3.5, to match the S/N of the continuum light curves of \cite{Grier_2017}, and line light curve S/N by a factor of 0.5, 1.0, 1.5 (closest to those in the \cite{Grier_2017} sample), and 2.0 to investigate the performance of each method under various flux S/N. 

In Figure \ref{fig:snr}, the total detection count decreases as the light curve S/N decreases for ICCF and ZDCF as expected. However, for {\tt JAVELIN}, the total number of detections remains approximately constant as light curve S/N decreases. {The individual bootstrapping realizations indicate that most of the {\tt JAVELIN} lags are still detected when the light curve quality is degraded to these levels, but with slightly larger lag uncertainties.} For ICCF and ZDCF, however, the dimmer and higher-$z$ quasars are no longer detected when the light curve S/N decreases. {Measurement uncertainties for {\tt JAVELIN} are always the most reliable (i.e. $\sigma_{gauss}\sim$0.75 for all simulations) for different S/N levels, but ICCF and ZDCF measurement uncertainties become more overestimated when light curve S/N decreases.} Similar trends are observed in the detection efficiency when broken down into $i$-band magnitude and redshift bins in Figure \ref{fig:corr_snr_imag} and Figure \ref{fig:corr_snr_z}. 

\subsection{Effects of the Power Spectral Density (PSD) of the Driving Light Curve}\label{sec:psd}
{
Our mock light curves are simulated using DRW models, which is also the assumption used in {\tt JAVELIN} for lag measurements. If the actual quasar light curves are approximately described by DRW models, as observed for large samples of quasars for the timescales of interest here \citep[e.g.,][]{Macleod_2010}, then using {\tt JAVELIN} is the correct approach to interpolate the light curves between the epochs in the lag calculation. However, one concern is that if the actual quasar light curve significantly deviates from a DRW model, then the basic assumption in {\tt JAVELIN} is violated and the lag measurement may be problematic. } 

{To test this possibility, we generate long, {daily-sampled} continuum light curves with a power-law PSD$\propto f^{\alpha}$ with slope $\alpha$ of $-1$, $-2$ and $-3$ {using the astroML \citep{astroML} package in python, which follows the approach described in \citet{TK95}}. The light curve variances are scaled to match the same rms variability as for the {uniform sample} in our fiducial simulations {before adding gaussian measurement uncertainties}. We then follow the same procedures of generating the line light curves, {assigning light curve uncertainties} and measure the time lags with ICCF, ZDCF and {\tt JAVELIN} as described in Section \ref{sec:measurelags}. {After down-sampling to sparse, shorter light curves, there will not be sufficient data points or baseline to properly sample the frequency space and the measured PSD slope might change, which is similar to the situation that our light curves can not constrain DRW parameters.} Specifically, the DRW model has a broken power-law PSD with a slope of $-2$ at high frequencies and a slope of $0$ at low frequencies \citep[the characteristic timescale is about a few hundred days, e.g.,][]{Macleod_2010}. Recent PSD measurements for several AGN observed by the Kepler satellite {\citep{Mushotzky_2011, Kasliwal_2015, Kasliwal_2017, Smith_2018}} suggested a PSD slope steeper than $-2$ for timescales below a few days, indicating less variability on the shortest timescales than the DRW model. Using a single power-law slope for the PSD over all relevant timescales is probably a bad assumption, as the quasar variability PSD is usually a broken power-law in the optical \citep[e.g.,][]{Simm_2016, Smith_2018}, but nevertheless this allows us to test the impact of any deviations from the DRW models. }

{
The measured lags are most correlated with the assigned lags ($r_{\rm ICCF}\sim$0.95, $r_{\rm ZDCF}\sim$0.95, $r_{\rm \tt{JAVELIN}}\sim$0.98 for the flux-limited sample) when $\alpha=-1$ and least correlated with the assigned lags when $\alpha=-3$ ($r_{\rm ICCF}\sim$0.81, $r_{\rm ZDCF}\sim$0.75, $r_{\rm \tt{JAVELIN}}\sim$0.92 for the flux-limited sample). {\tt JAVELIN} remains the best among the three in reproducing the assigned lags in terms of being close to the true lags, even when the input PSD is significantly different from the one assumed (i.e., DRW) in {\tt JAVELIN}.
}

\begin{figure}
\centering
	\begin{tabular}{@{}cc@{}}
	\includegraphics[width=0.5\textwidth]{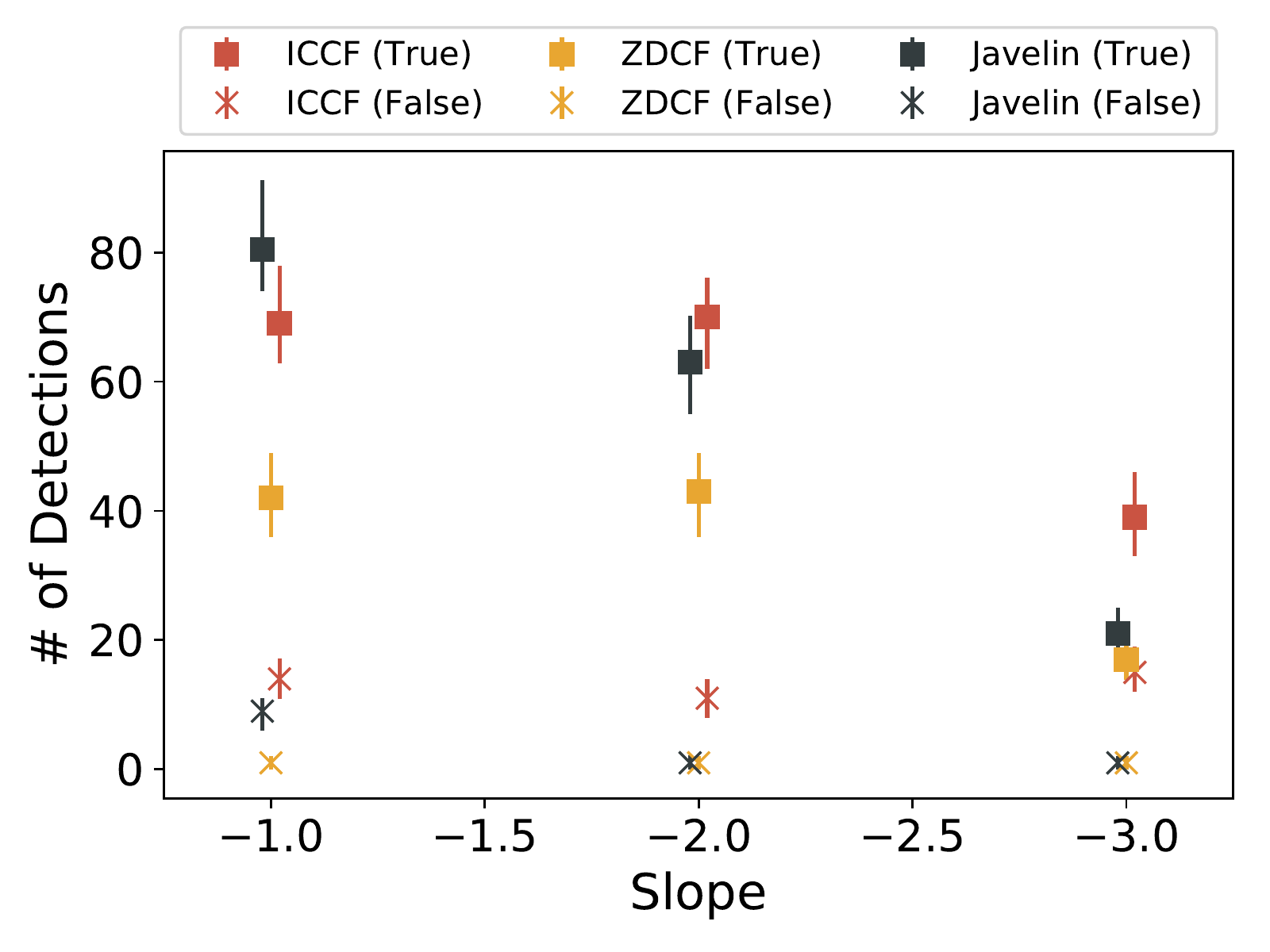}
	\end{tabular}
    \caption{{Total counts of true (square) and false (cross) detections for the flux-limited sample with mock light curves generated from single power-law PSDs (as opposed to the DRW model) with different slopes. The overall detection fraction decreases for steeper PSDs, where the light curves are more and more dominated by slow varying (or even monotonic) trends. }}
    \label{fig:psd}
\end{figure}

\begin{figure*}
\centering
	\begin{tabular}{@{}cc@{}}
	\includegraphics[width=0.33\textwidth]{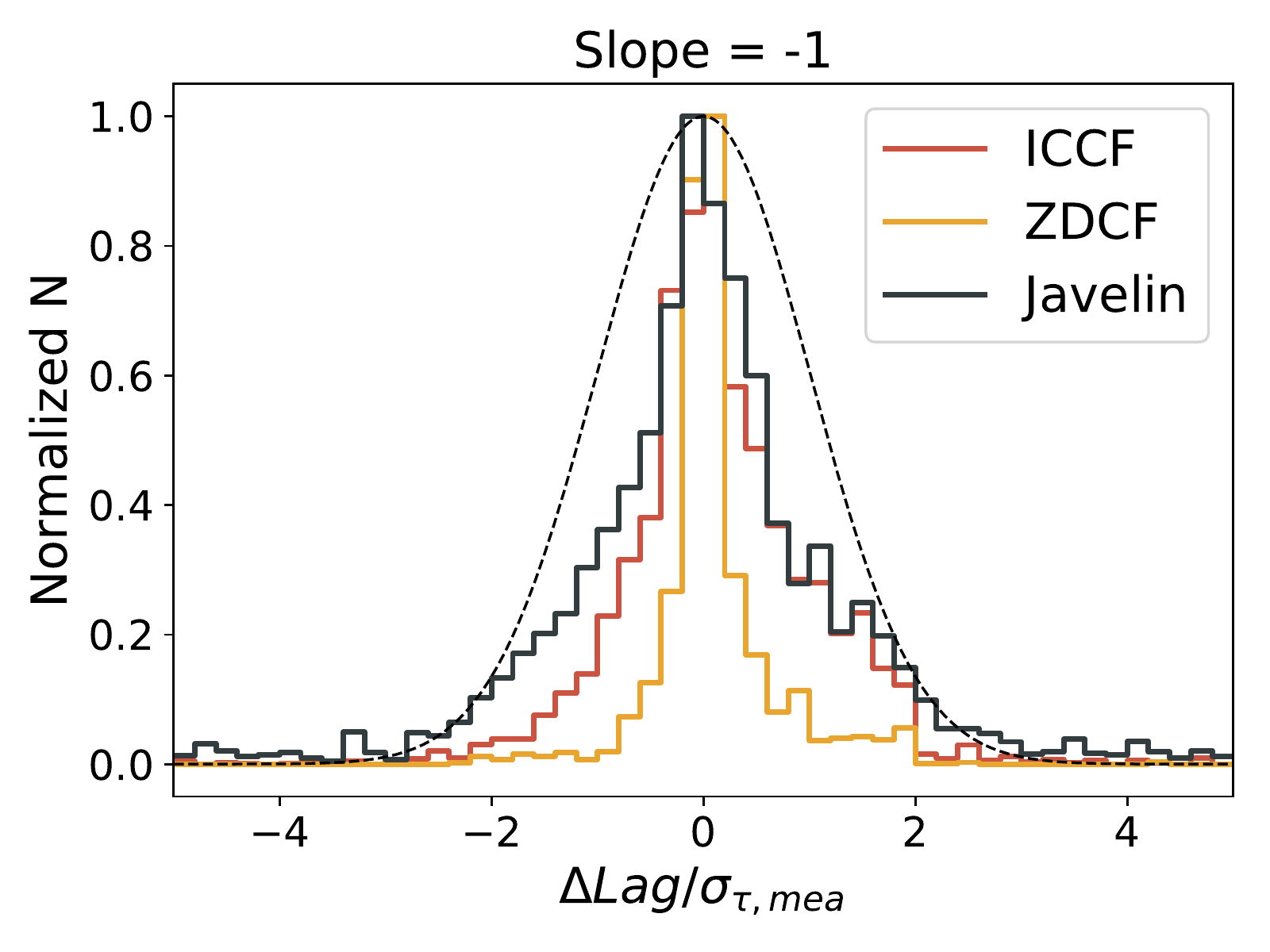}
	\includegraphics[width=0.33\textwidth]{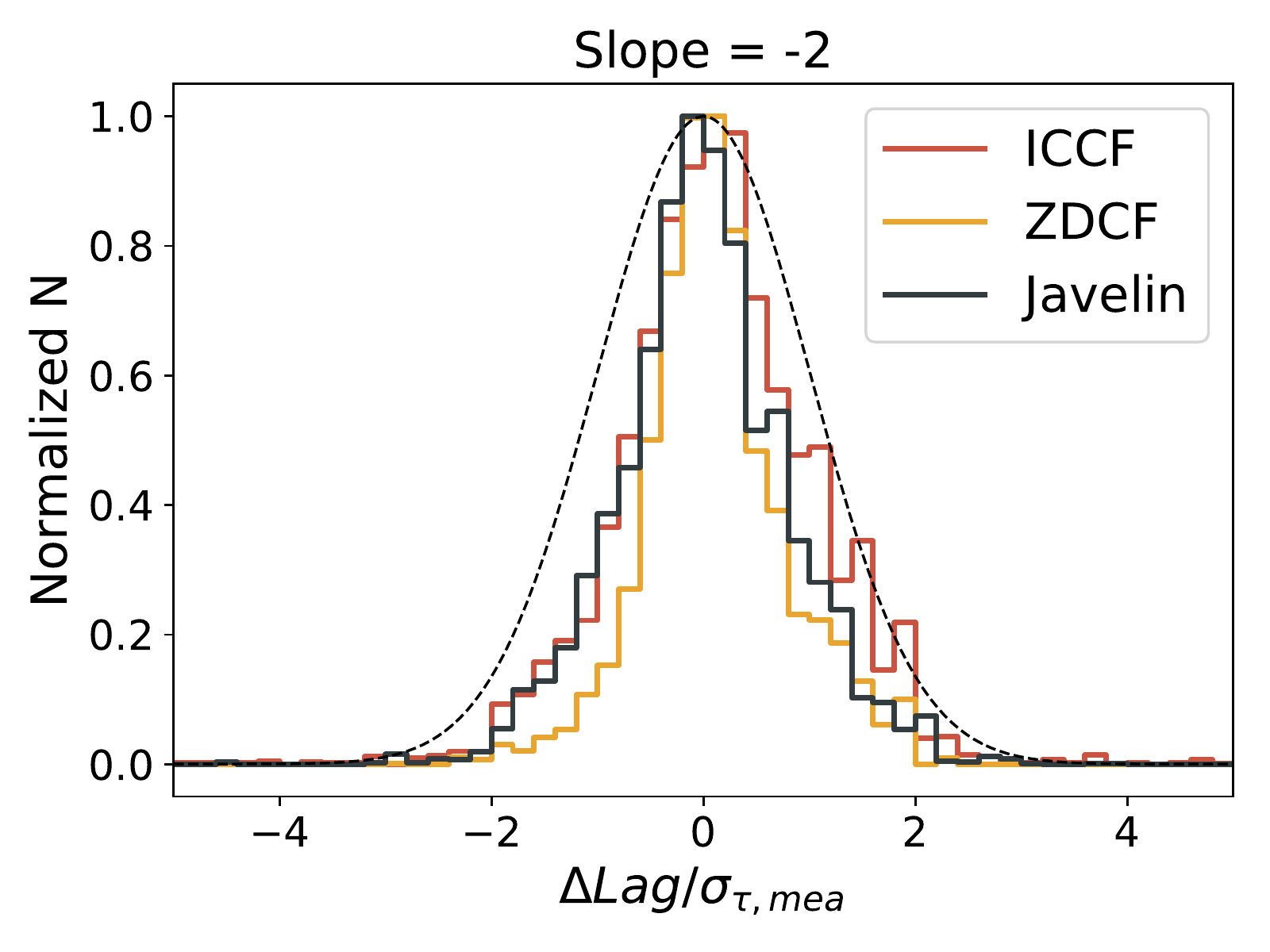}
	\includegraphics[width=0.33\textwidth]{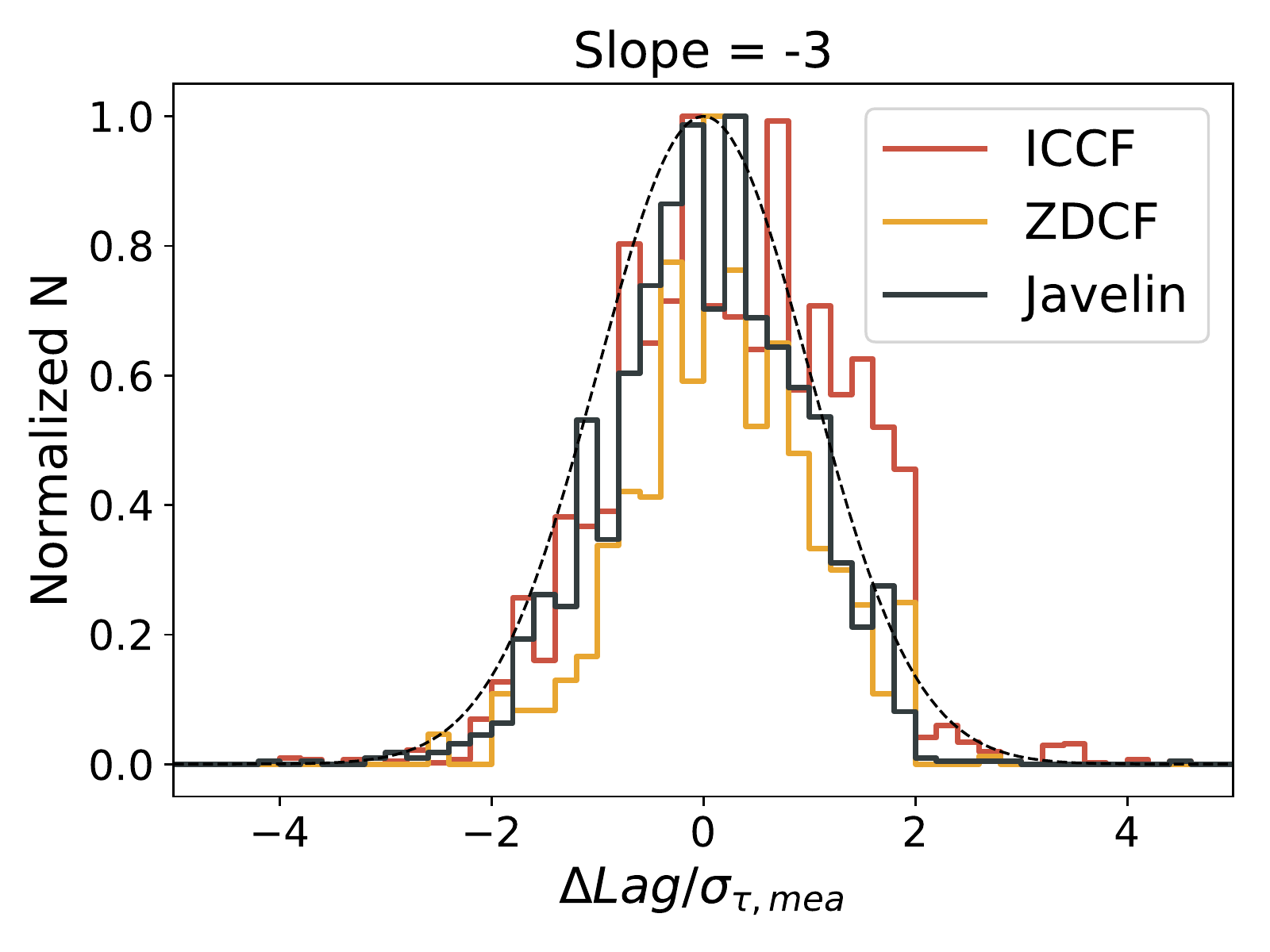}
	\end{tabular}
\caption{
{Distribution of the difference between assigned and measured lags, normalized by the measurement uncertainty, for the flux-limited sample using mock light curves generated with single power-law PSDs with different slopes. The black dashed line is a Gaussian distribution with unity dispersion.}}
\label{fig:psd_gauss}
\end{figure*}

{
As shown in Figure \ref{fig:psd}, all methods detect the most lags when $\alpha=-1$ (detection efficiencies for the flux-limited sample: $\sim$38\% for ICCF, $\sim$24\% for ZDCF and $\sim$45\% for {\tt JAVELIN}) and the fewest lags when $\alpha=-3$ (detection efficiencies for the flux-limited sample: $\sim$22\% for ICCF, $\sim$9.6\% for ZDCF and $\sim$12\% for {\tt JAVELIN}). When there is more variability on short timescales ($\alpha=-1$), there are more features in the light curves for the methods to model and correlate. When $\alpha=-3$, light curves tend to be slowly varying or even monotonic for almost the entire 180-day baseline, which makes detecting lags more difficult for all methods. In all cases, ICCF has the highest false detection rate, and the false detections tend to cluster around the search limit ($\sim 60\%$ of the monitoring period), which will lead to a biased lag distribution as discussed in Section \ref{sec:distribution_lags}. {\tt JAVELIN} also has a higher false detection rate when $\alpha=-1$ compared to the two other $\alpha$ cases (but still fewer false detections than ICCF). {\tt JAVELIN} is unable to reproduce the high-frequency variations in these light curves for this shallowest PSD slope, leading to more false detections (8\% of all sources, compared to $<$1\% when $\alpha=-2$ or $-3$), and more overestimated uncertainties as seen in the left panel in Figure \ref{fig:psd_gauss}. Overall, ICCF uncertainties are overestimated ($\sigma_{gauss}\sim$0.6) when $\alpha=-1$, but underestimated ($\sigma_{gauss}\sim$1.1) when $\alpha=-3$, ZDCF uncertainties are overestimated in all simulations, and {\tt JAVELIN} uncertainties are slightly overestimated but more consistent among all simulations. 
}

{These additional tests demonstrate that when the actual light curve PSD is different from the DRW model, the relative performance in terms of lag detection efficiency, rate of false detections, and the reliability of reported lag uncertainties remains more or less the same among the three methods. {These tests also show that the DRW model is extremely flexible and is capable of fitting non-DRW light curves (see \cite{Kozlowski_2016} for a similar conclusion). Even though the DRW parameters cannot be constrained by our light curves, the DRW model can produce reasonable interpolation (better than linear interpolation) and thus outperforms ICCF and ZDCF in most test cases.} Of course we have not exhausted the variety of PSD shapes and it is possible that certain peculiar PSD shapes will change the relative performance among the three methods.}

{Finally, we point out that we have not tested the effect of deviations in the line transfer function from the assumed top-hat function in {\tt JAVELIN}. \cite{Yu_submitted} performed a more detailed study on the impact of transfer function forms on the performance of {\tt JAVELIN}, in the regime of high quality light curves typically achieved for local RM programs. It is always a possibility that {\tt JAVELIN} will fail badly for specific cases with unusual transfer functions or variability PSD. However, for the bulk of typical quasar light curves, and especially for the regime of light-curve quality (e.g., S/N and sampling) of interest to most MOS-RM programs, {\tt JAVELIN} is favored over the other two methods.}


\section{Lag Detection in Real Surveys}\label{sec:reallife}

\subsection {{Selection with light curve quality cuts}}

In reality, the true lags of quasars in the MOS-RM sample are unknown. Instead of comparing with the true lag, we can use the quality of light curve fits and the properties of the light curves to evaluate the quality of the lag measurements. Traditionally, visual inspection is often invoked to assess the quality of the lag measurements.

\begin{figure*}
    \centering
    	\includegraphics[width=1.0\textwidth]{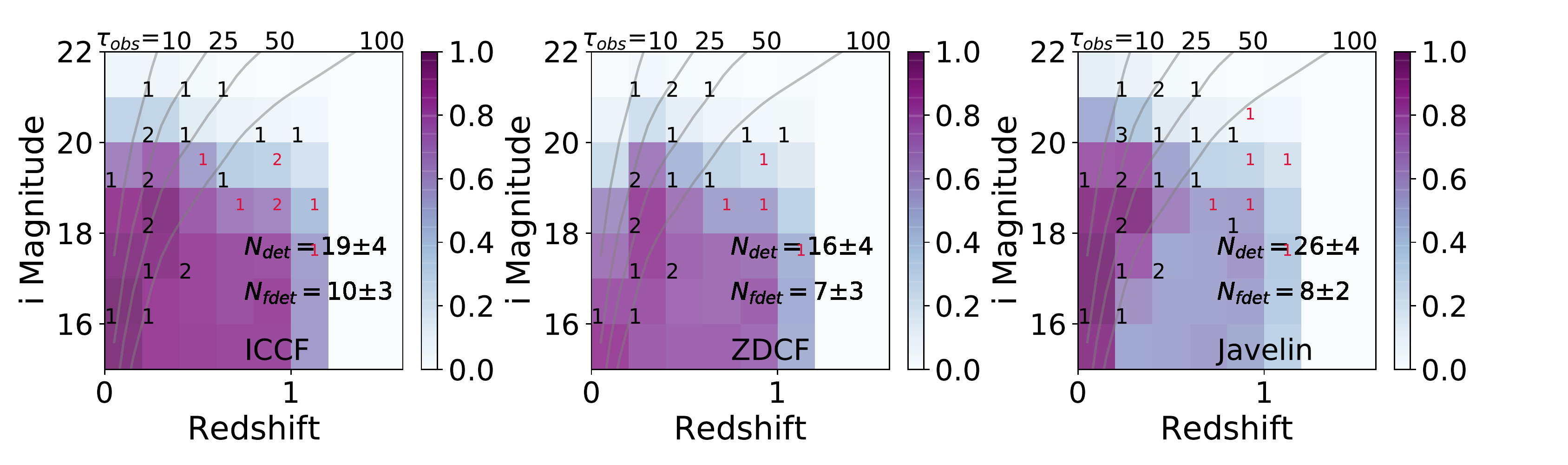}
    \caption{
    {Detection efficiency of simulations with a cadence of 6 days and 30 epochs, based on the lag-significance criteria of \cite{Grier_2017}. The colormap represents the detection efficiency and the numbers are the detection counts of a single down-sampling realization. The total numbers of true and false detections shown in the lower-right corner are the median and uncertainties derived from 100 down-sampling realizations, defined by the detection criteria described in Section \ref{sec:criteria} (true detections in black and false detections in red). The grey contours show the approximate constant lags from the R-L relation from \cite{Bentz_2009b}.}}
    \label{fig:detmap_g17}
    \end{figure*}

{\cite{Grier_2017} applied cuts on the minimum ICCF correlation coefficient ($r_{max}$) and the continuum and line light curve RMS variability S/N (defined as the intrinsic variability of the light curve about a fitted linear trend, divided by the uncertainty of the estimated intrinsic variability). $r_{max}$ can be used to evaluate if the light curves are well-correlated. The continuum and line RMS variability can be used to identify short-time variability and exclude spurious correlations for noisy light curves or light curves with long, monotonic trends. The selected cutoff values strongly depend on the desired balance between completeness and purity of lag detections, for example, to achieve an acceptable false-detection rate. Figure \ref{fig:detmap_g17} shows the simulated detections by imposing the additional lag-significance criteria of \cite{Grier_2017}, with simulated light curve S/N matched to the \cite{Grier_2017} sample (continuum light curve uncertainties inflated by 3.5 times and line light curve uncertainties inflated by 1.5 times). We evaluate the robustness of the detections based on Section \ref{sec:criteria}, which is more stringent than \cite{Grier_2017} as they only require a 2$\sigma$ deviation from zero-lag for a significant lag. Roughly half of the detections from the original test in Section \ref{sec:diss_err} are removed due to low correlation or low variability amplitude. The number of detections is slightly lower than for \cite{Grier_2017} due to the more stringent detection criteria, and the false-detection rate is around 20\% to 30\%. Since these additional quality cuts remove the same objects from the detected sample in each method, they do not affect our conclusions about the relative performance of different lag measuring methods. }

\subsection {{Selection with statistical test}}

Here we introduce a statistical approach to remove false detections in MOS-RM surveys without knowing the true lags or expected lags from an assumed R-L relation. Since detectable lags in a specific survey design depend on the quasar magnitude and redshift, we filter out false detections by removing all sources in a redshift-magnitude bin that are unlikely to host detectable lags. When analyzing light curve pairs with undetectable lags, statistically, all lag detection methods should have an equal chance to produce positive and negative lags, all which are false detections. We compute the ratio of positive lags to negative lags in each of the magnitude-redshift bins and set a cutoff to exclude bins with a low positive-to-negative measurement ratio. If a grid has a ratio below this cutoff, we assume the time lags of all sources in that grid are not reliable and all the detections (both true and false detections) are removed in the grid. In this work, we start with the uniform sample and impose redshift bins of 0.2 and magnitude bins of 0.5 as an example. Each grid element has $\sim$600 quasars. {The cutoff ratio of positive to negative lags of 1.5 is selected, which is optimized by searching for a ratio that eliminates the most false detections while keeping the most true detections.} {After the statistical selection, we apply the same down-sampling procedure as in \S \ref{sec:downsample} to mimic the SDSS-RM program.}

\begin{figure*}
    \centering
    	\includegraphics[width=1.0\textwidth]{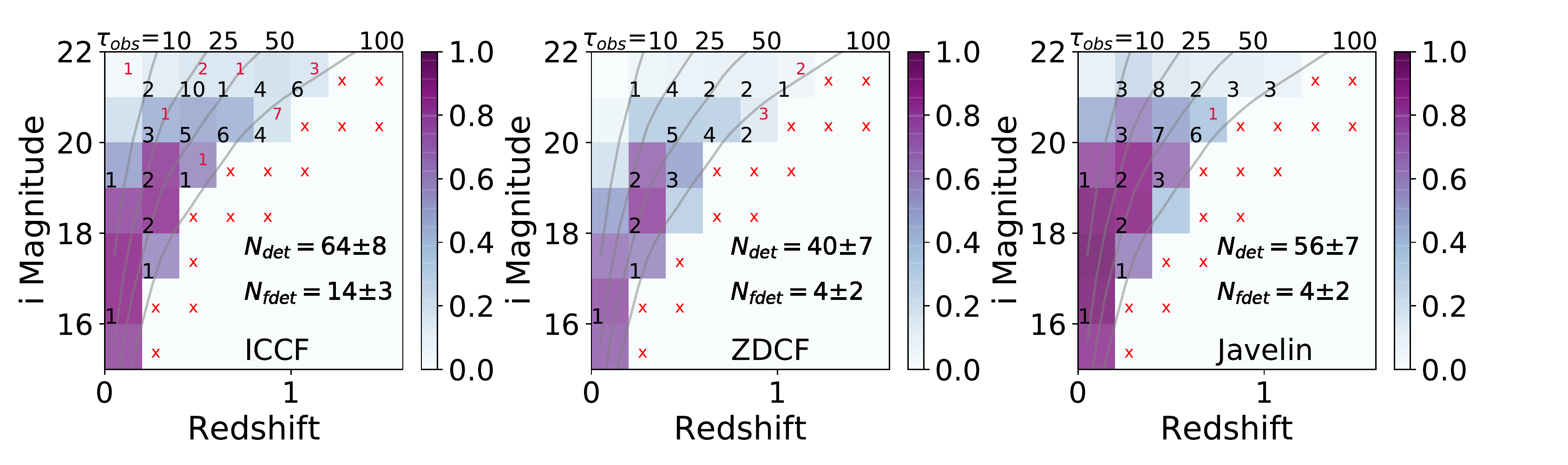}
    \caption{
    Detection efficiency of simulations with a cadence of 6 days and 30 epochs, selected by the statistical test described in Section \ref{sec:reallife}. The colormap represents the detection efficiency and the numbers are the detection counts (true detections in black and false detections in red) of a single down-sampling realization. The total numbers of true and false detections shown in the lower-right corner are the median and uncertainties derived from 100 down-sampling realizations. The grey contours show the approximate constant lags from the R-L relation from \cite{Bentz_2009b}. {The bins removed by the statistical test are labeled with a red cross.} }
    \label{fig:detmap_pntest}
    \end{figure*}

Using this ratio criterion, we regenerate the true and false detection map in Figure \ref{fig:detmap_pntest}. We recover most ($>$90\%) of the lags found previously with the knowledge of the true lags. {In the following analysis, we still label the lags according to the true/false detection criteria, but they are not selected by the assigned lags and should be indistinguishable in real surveys.}  Because we are selecting the redshift and $i$-band magnitude bins without knowing the true lags, false detections increase in the bins where $\tau_{obs}$ is $\sim$100 days, where some longer lags could be falsely detected with a smaller measured lag. {These falsely-detected long lags make up roughly a third (ICCF) to half (ZDCF and {\tt JAVELIN}) of all the false detections.} The median false detection rate is roughly 18\%, 9.0\%, and 6.7\% for ICCF, ZDCF and {\tt JAVELIN} for the flux-limited sample, again with {\tt JAVELIN} having the lowest false detection rate. These results are similar to the estimated false detection rate of \cite{Grier_2017}, roughly 10\%. Most of the sources in eliminated i-mag and redshift bins have lags of $\gtrsim$100~days, above the limit used in the lag search. 

With this statistical approach to mimic the reality of MOS-RM programs, the average lag distribution from the 100 down-sampling realizations is shown in Figure \ref{fig:hist_lag_ds}. {\tt JAVELIN} and ZDCF measure a relatively uniform lag distribution. ICCF favors lags around $\sim$60--90 days and measures more true and false detections in this range. {This result suggests that the R-L relation derived with ICCF lags are more biased, especially for samples with a narrow redshift distribution, where the limited observed-frame lag distribution would correspond to a limited rest-frame lag distribution).} In Figure \ref{fig:corr_pntest}, ICCF has higher detection rate in the low-luminosity and high-redshift bins compared to {\tt JAVELIN} and ZDCF, but the false detection rate is also high in those grids. The total number of detections deceases significantly beyond $z\sim$1 as the light curves have lower S/N and the lags are closer to the $\sim$100 day search limit. Overall these results are similar to those in Section \ref{sec:diss_cad} and Section \ref{sec:diss_err}.

\begin{figure}
\centering
	\begin{tabular}{@{}cc@{}}
	\includegraphics[width=0.5\textwidth]{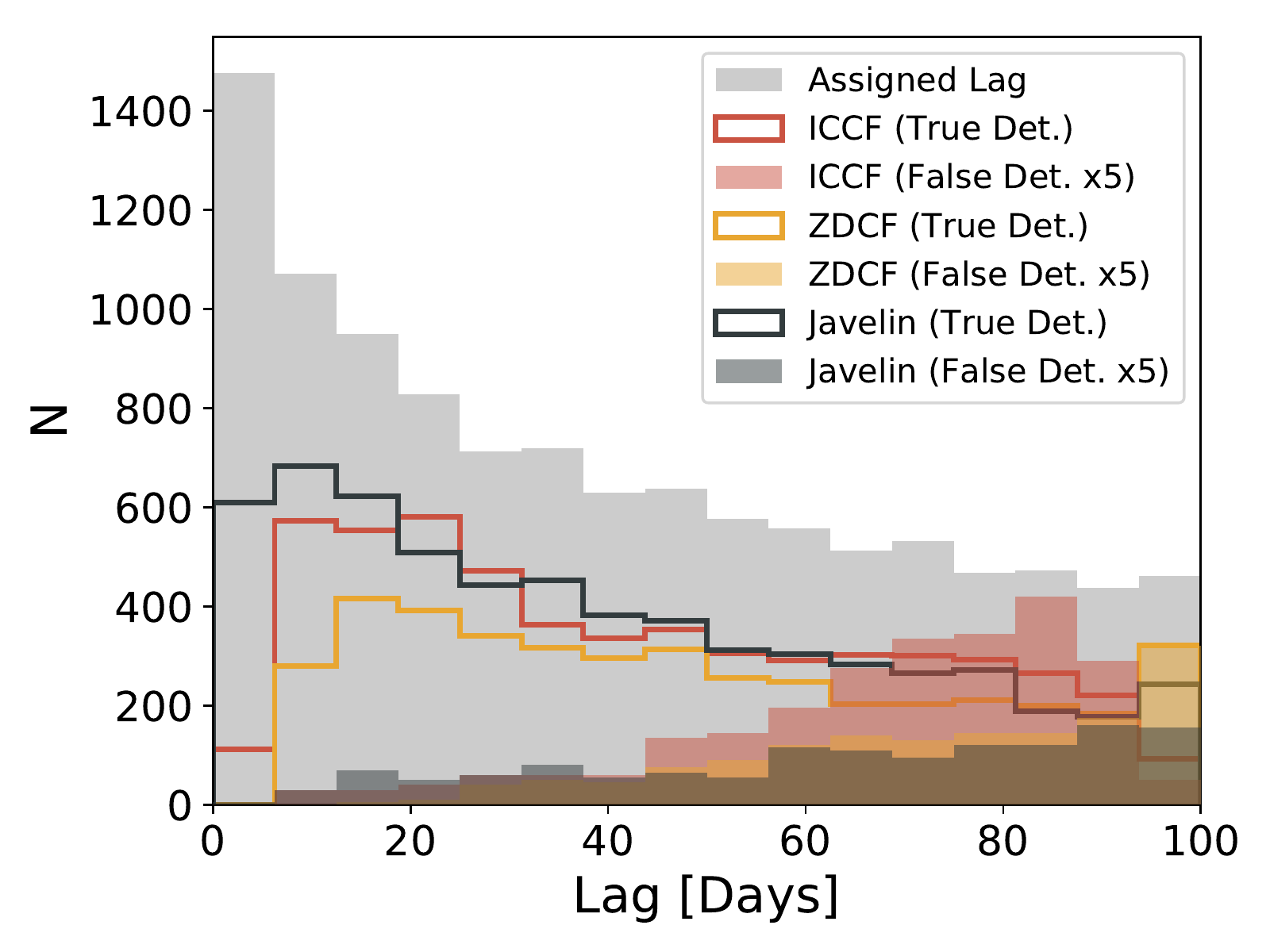}
	\end{tabular}
    \caption{
    Distribution of the measured lags of the uniform sample in observed frame, selected by the statistical test described in Section \ref{sec:reallife}. The grey solid histogram shows the number of detectable lags in each bin. The open histograms represent the number of true detections and the solid histograms are the number of false detections. The number of false detections are inflated by a factor of five for clarity.}
    \label{fig:hist_pntest}
\end{figure}

\begin{figure}
\centering
	\begin{tabular}{@{}cc@{}}
	\includegraphics[width=0.5\textwidth]{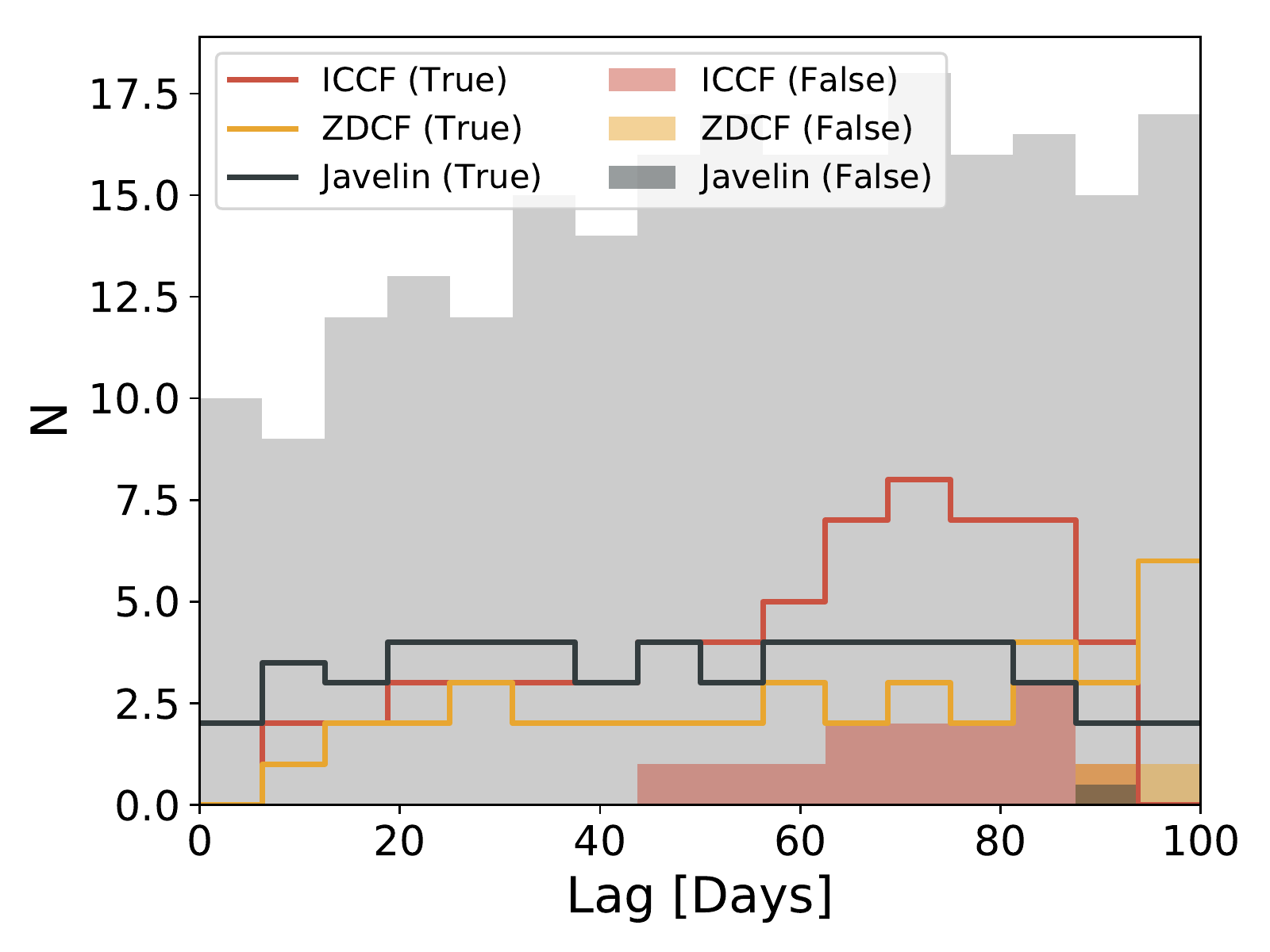}
	\end{tabular}
    \caption{Median distribution of the detected lags in the 100 down-sampling realization. The open histograms show the number of true detections and the solid histograms indicate the number of false detections. The grey shaded area represents the median assigned lag distribution.}
    \label{fig:hist_lag_ds}
\end{figure}

\begin{figure*}
\centering
	\begin{tabular}{@{}cc@{}}
	\includegraphics[width=0.5\textwidth]{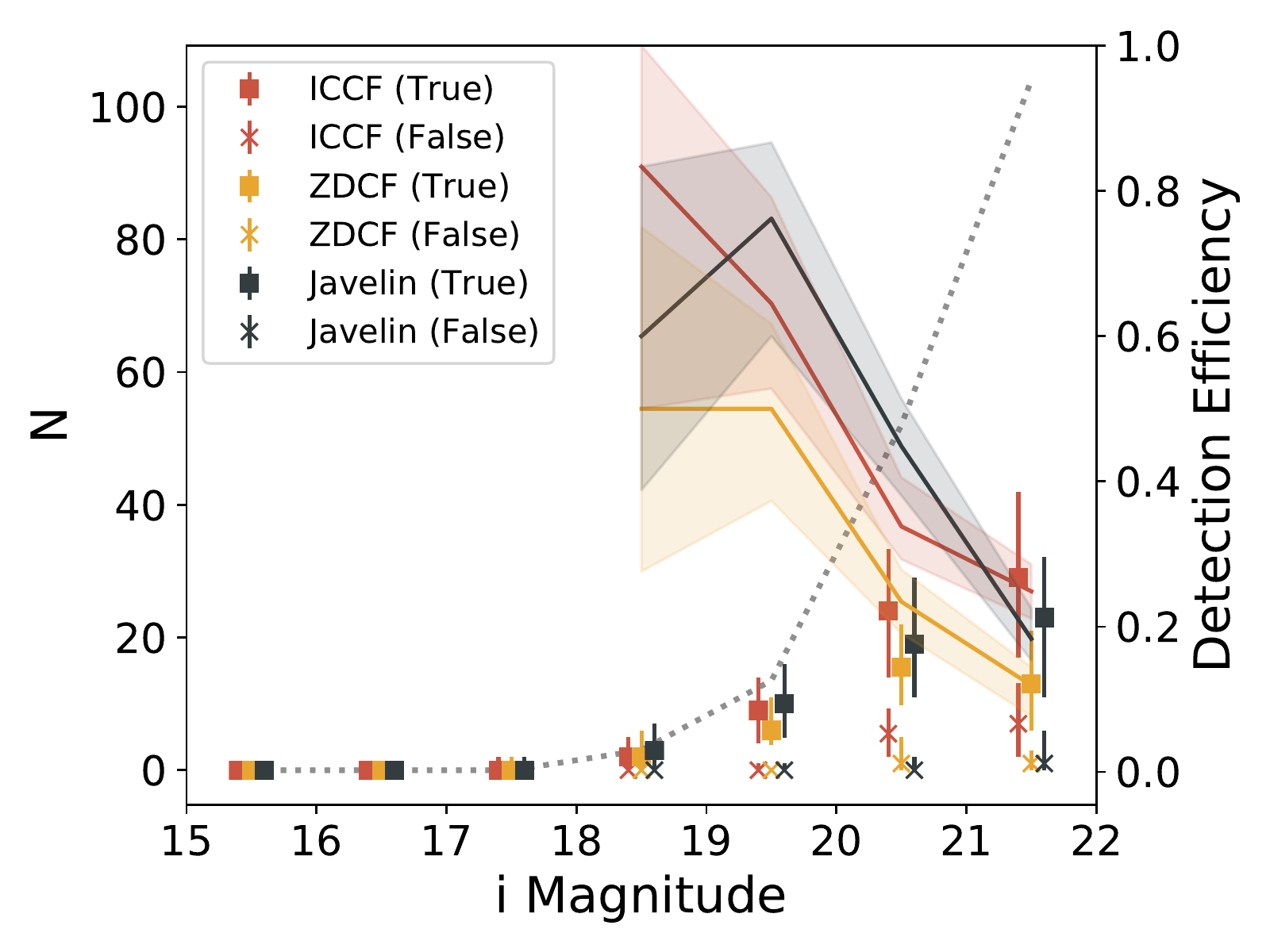}
	\includegraphics[width=0.5\textwidth]{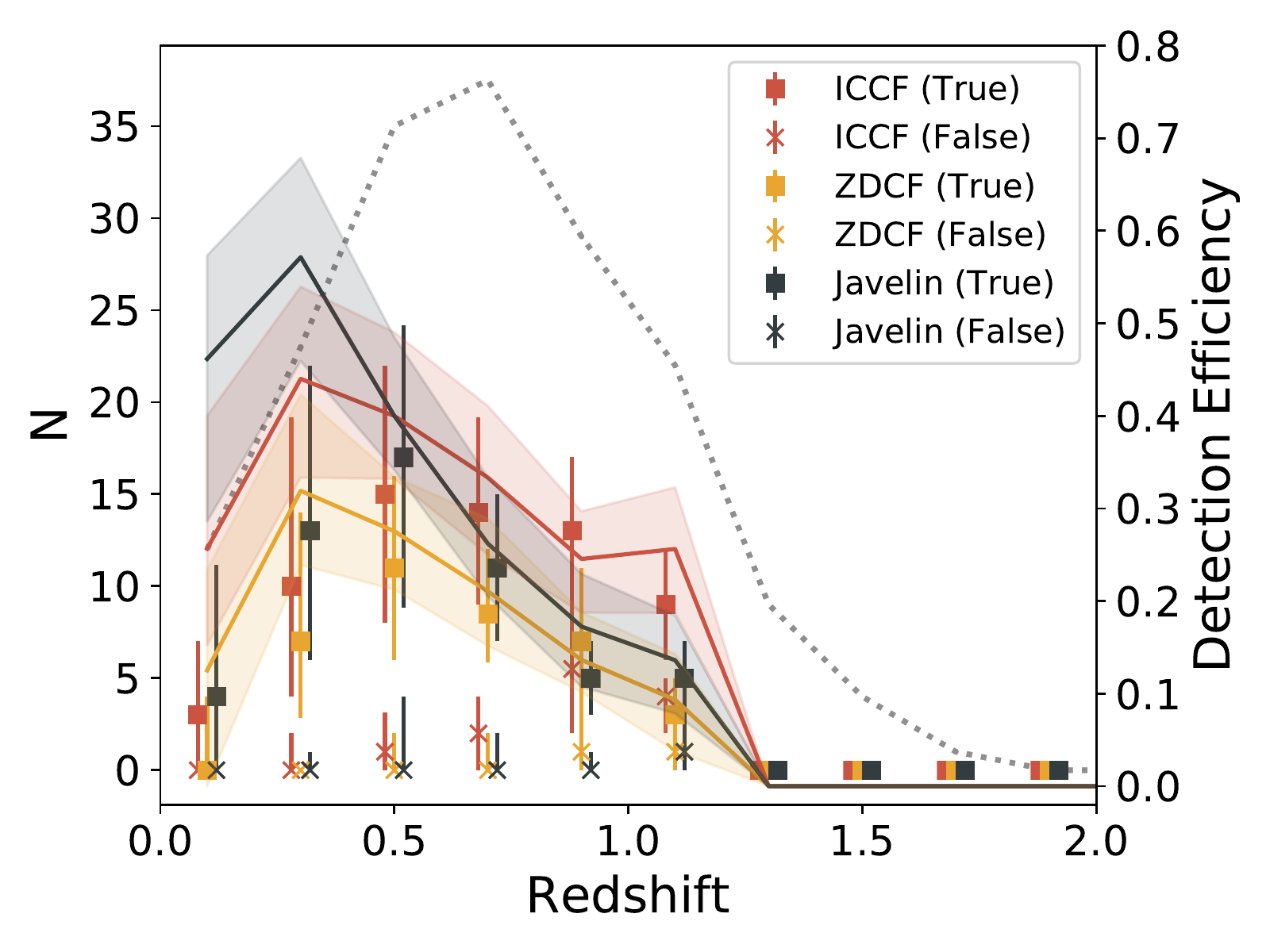}
	\end{tabular}
    \caption{
    Detection efficiency (solid line and shaded area) and true (square) and false (cross) detection counts of the three methods as functions of $i$-band magnitude (right panel) and redshift (left panel) in a simulated observation with 6 day cadence and 30 epochs. The detections are selected with the statistical approach described in Section \ref{sec:reallife}, i.e., assuming no knowledge of the true lags. The dotted lines show the number of sources with lags shorter than the search range (i.e. 100 days) in each magnitude or redshift bin. For $i<$18, the detection efficiencies are not shown because there are no quasars selected in more than 80\% bootstrapping iterations.}
    \label{fig:corr_pntest}
\end{figure*}


\section{Discussion}\label{sec:discussion}

\subsection{The R-L relation}\label{sec:rl}

\begin{figure*}
    \centering
    	\begin{tabular}{@{}cc@{}}
	\includegraphics[width=0.33\textwidth]{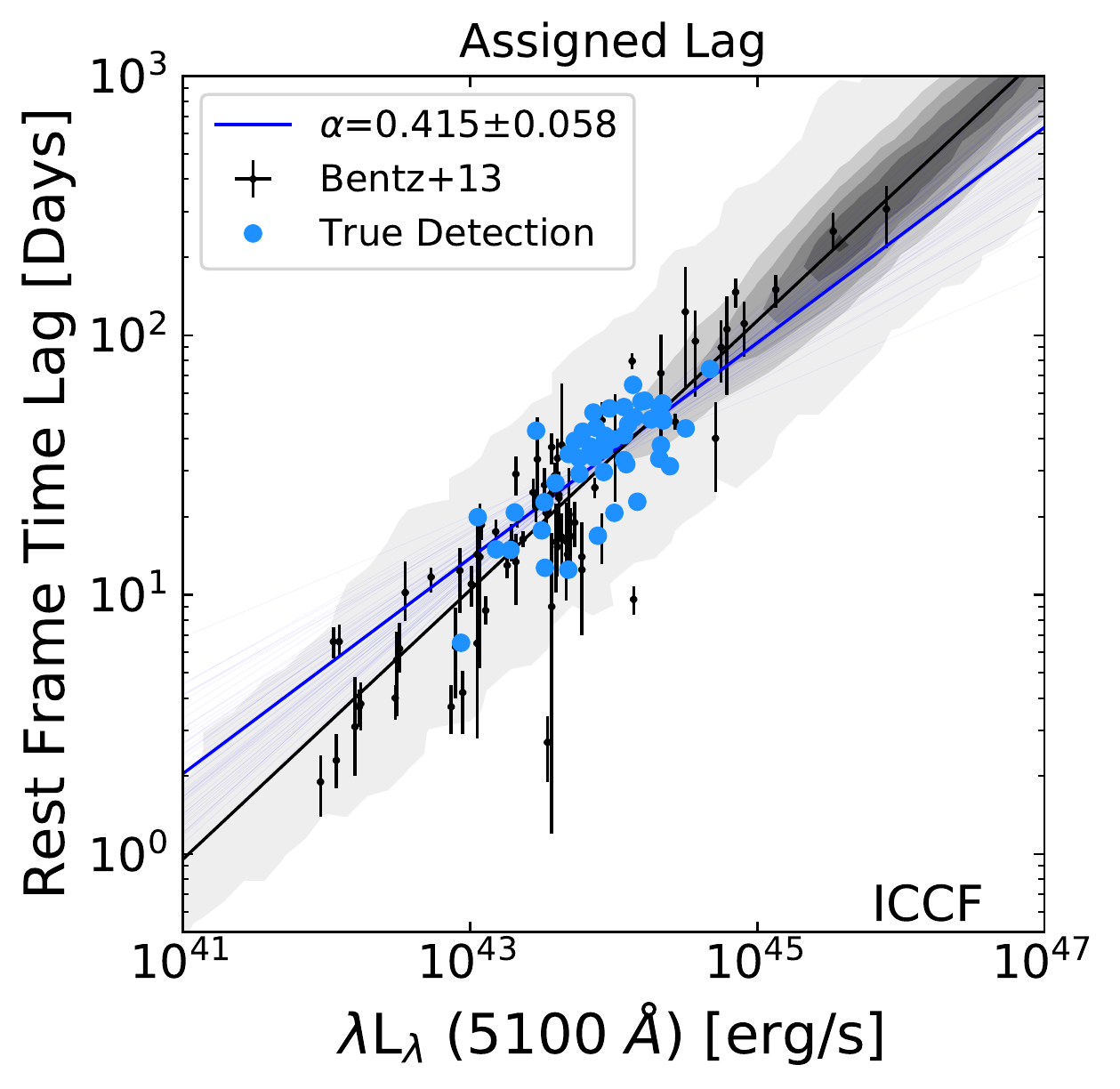}      
	\includegraphics[width=0.33\textwidth]{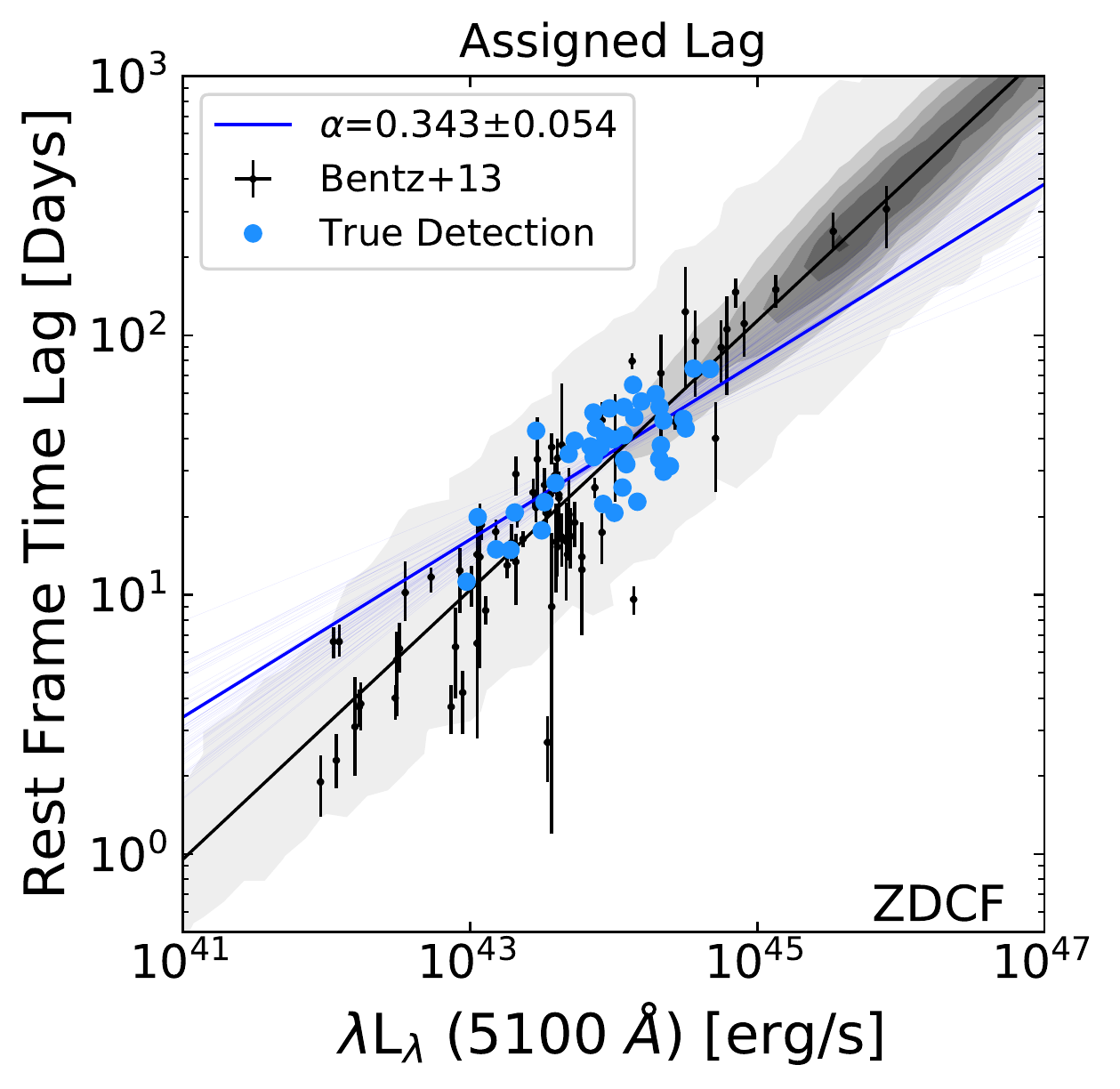}      
        \includegraphics[width=0.33\textwidth]{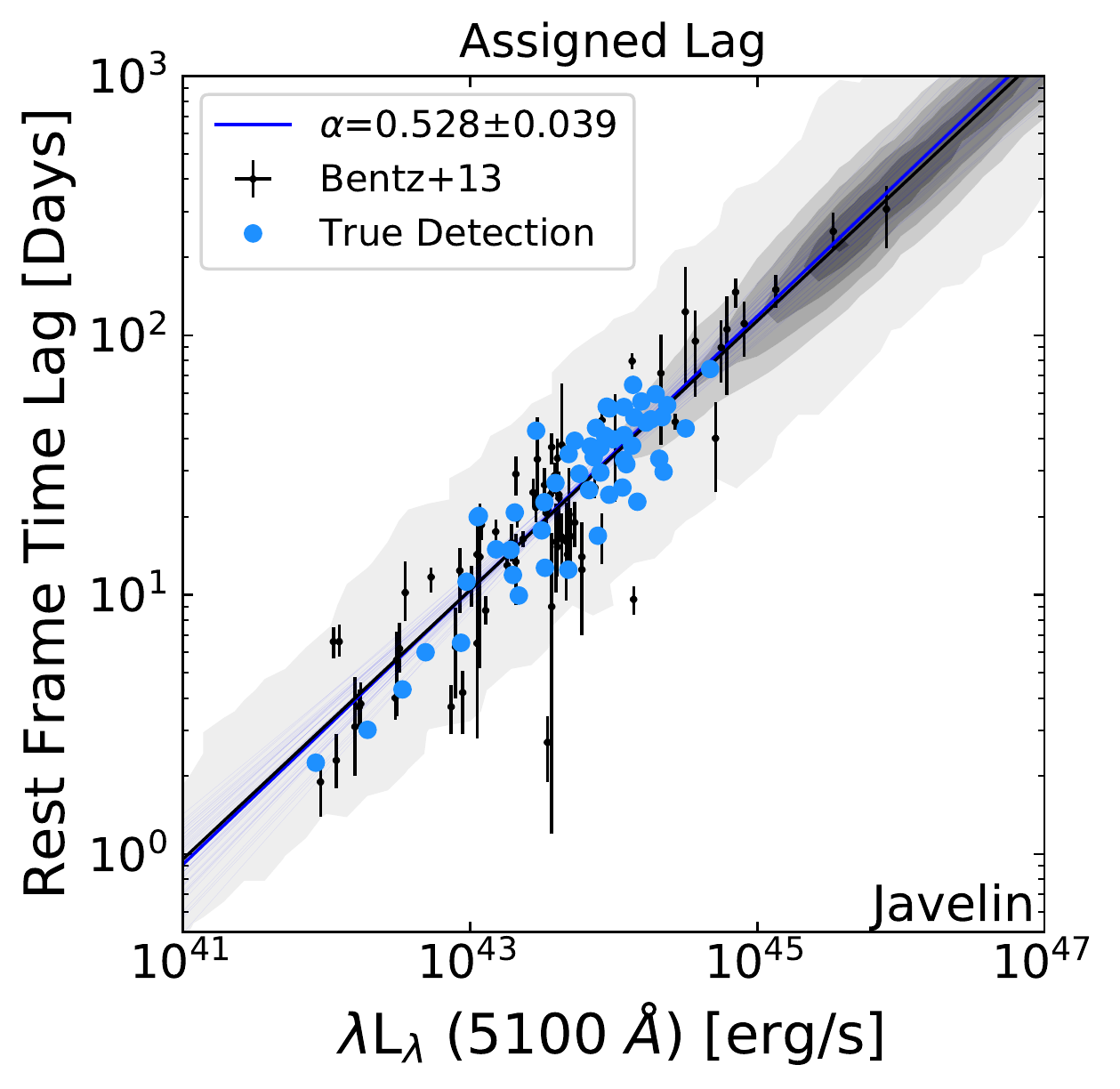}      \\
        \includegraphics[width=0.33\textwidth]{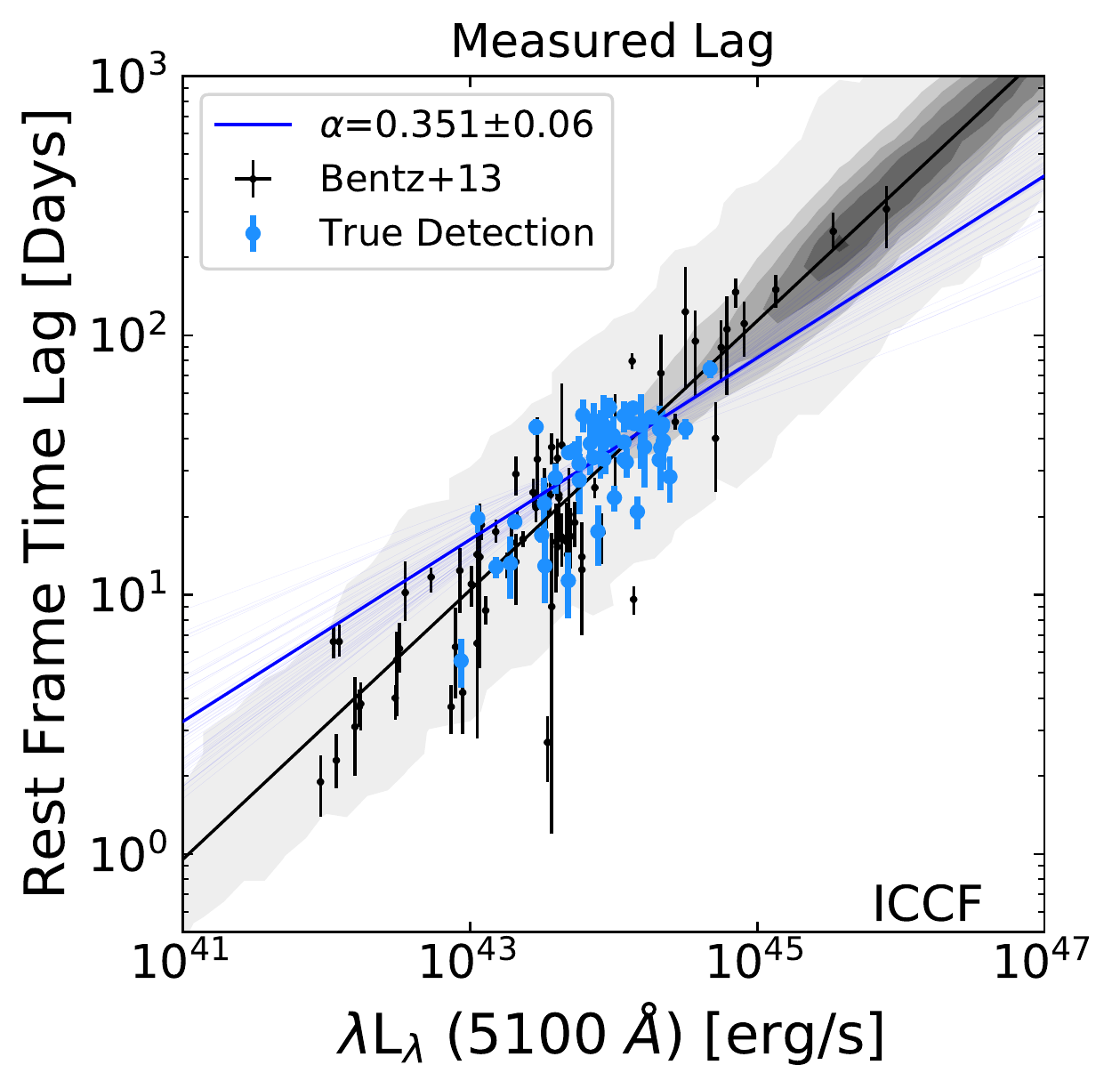}
        \includegraphics[width=0.33\textwidth]{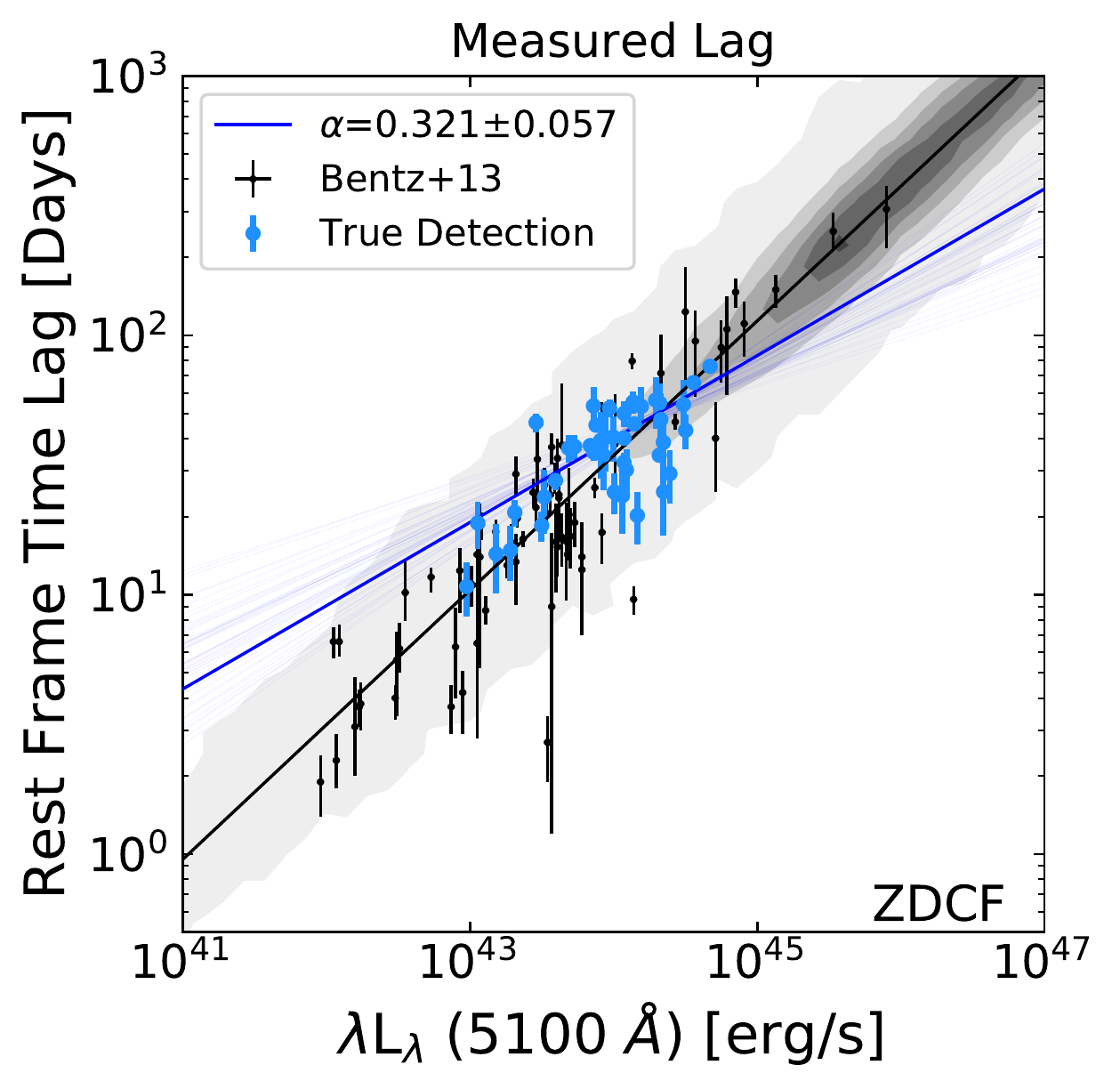}
        \includegraphics[width=0.33\textwidth]{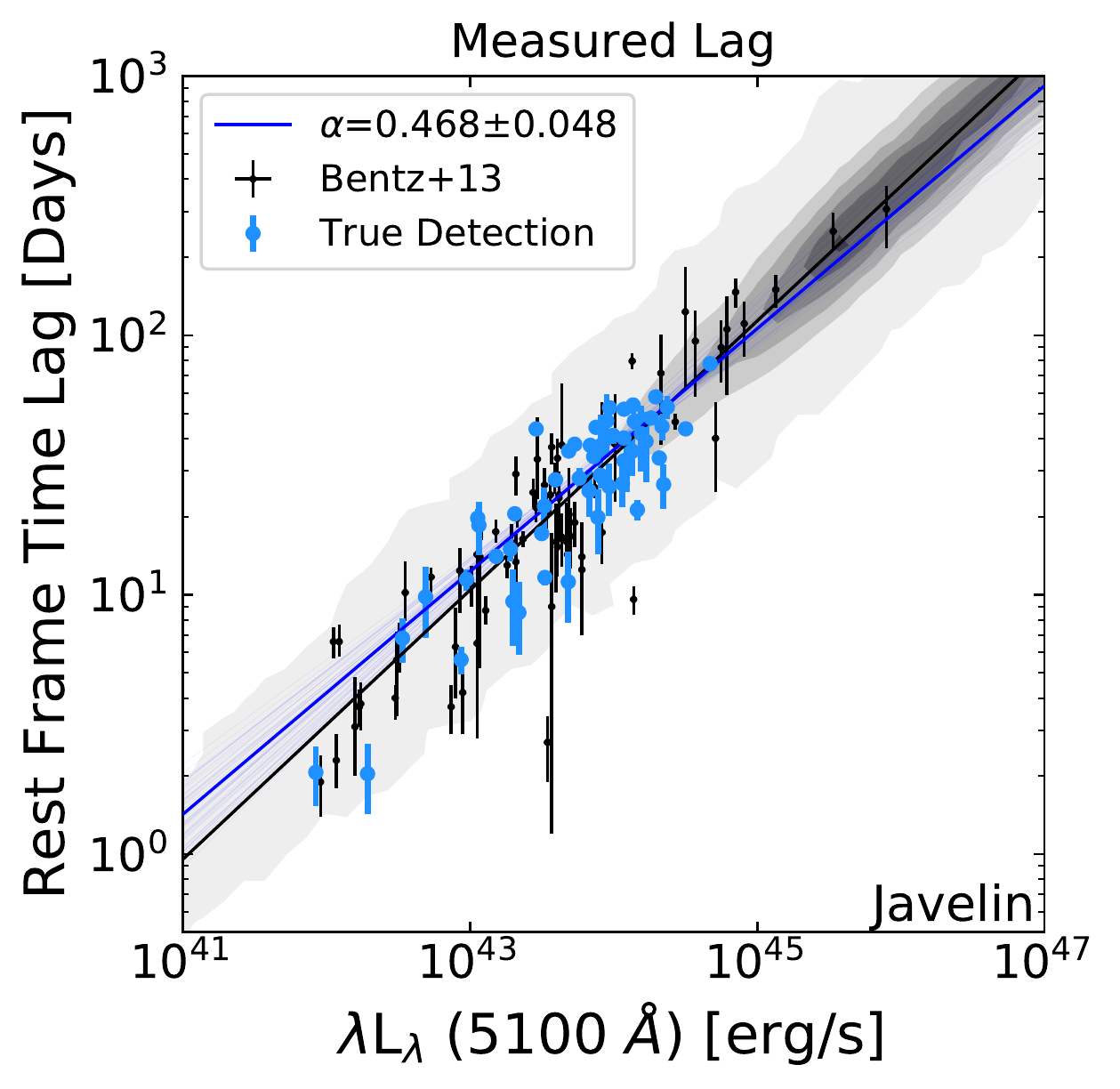}\\
        \includegraphics[width=0.33\textwidth]{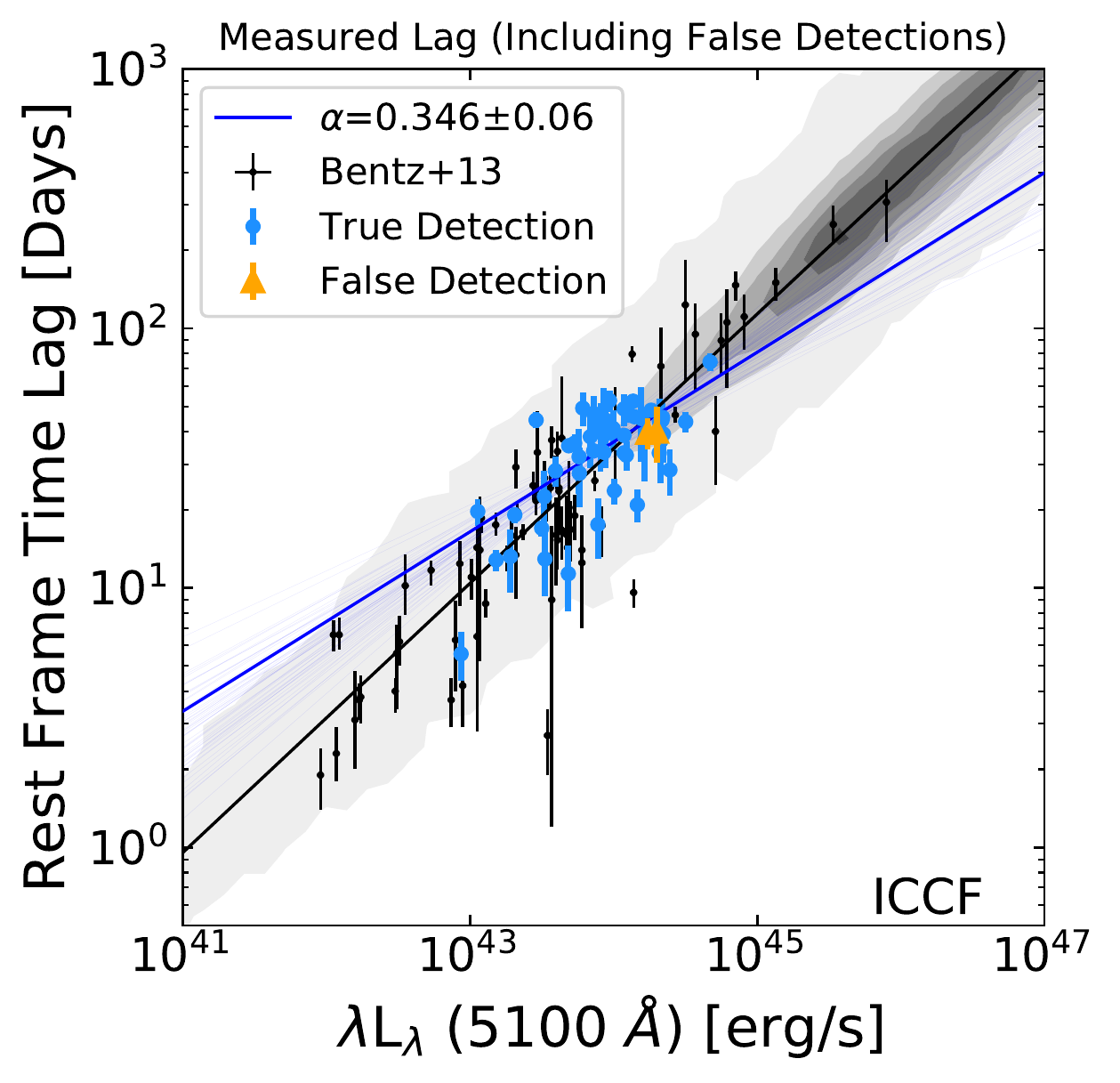}
        \includegraphics[width=0.33\textwidth]{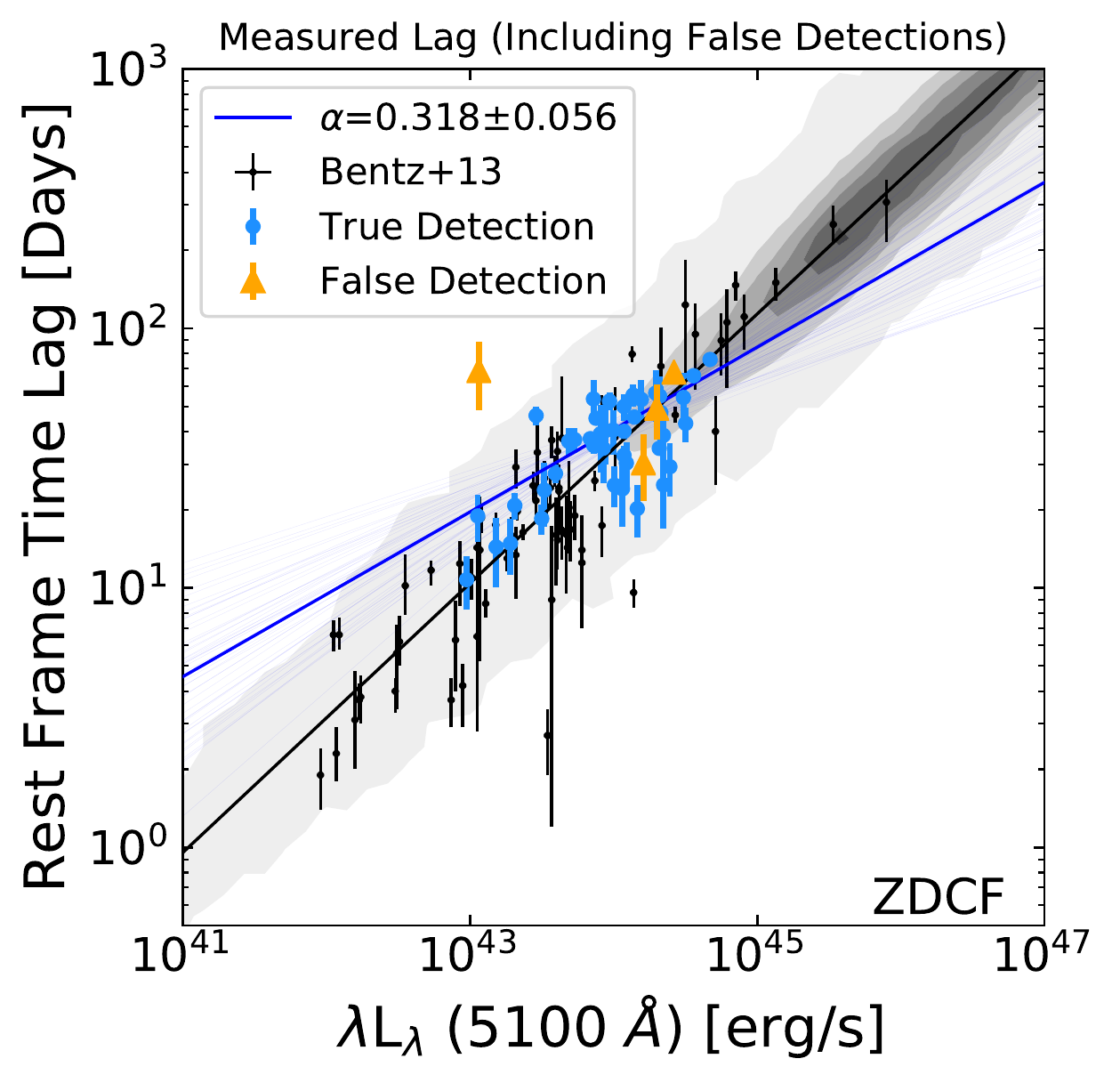} 
        \includegraphics[width=0.33\textwidth]{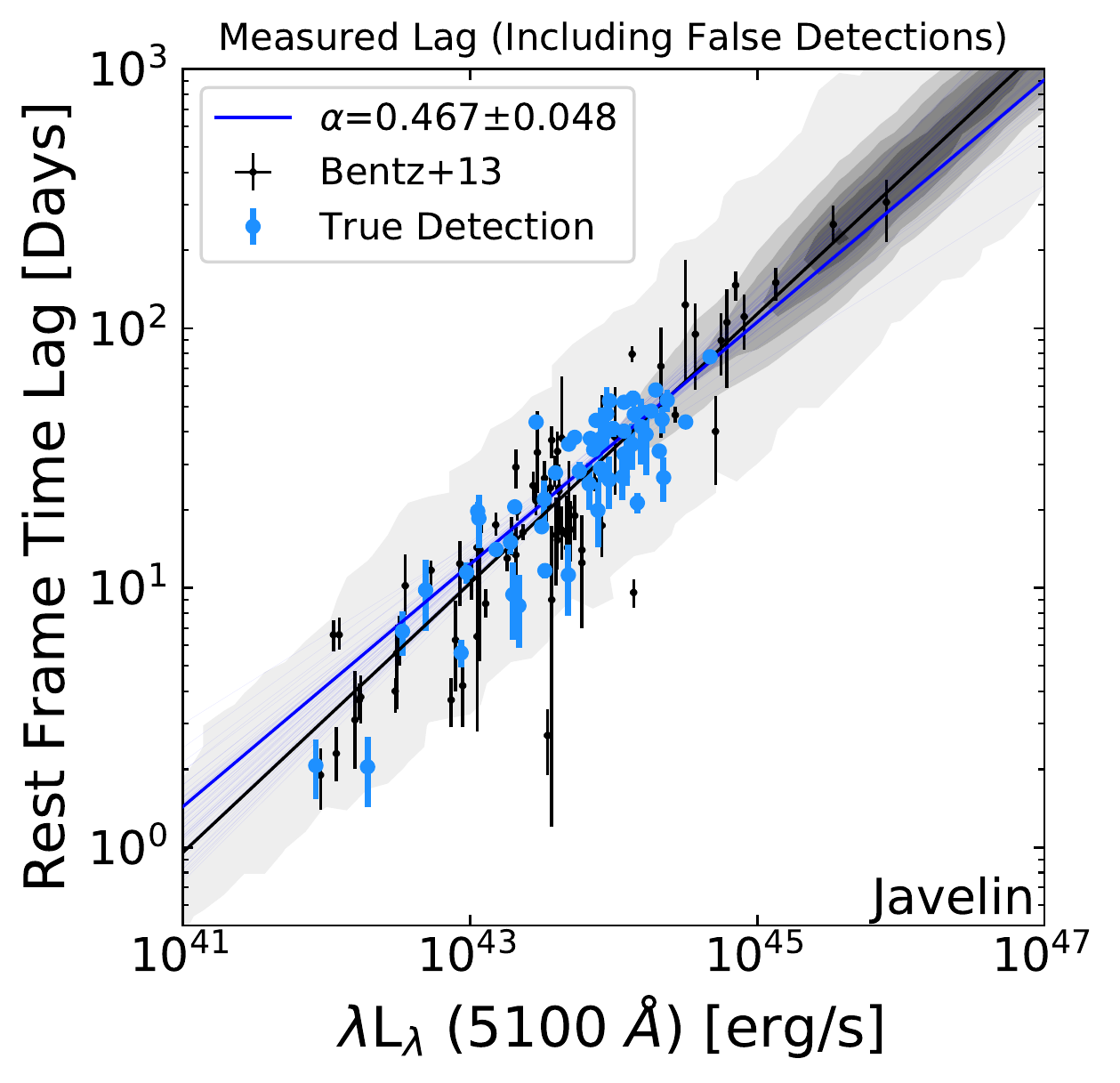}\\
    	\end{tabular}
    \caption{H$\beta$ R-L relation derived from one down-sampling realization. In each panel, the grey contours represent the uniform quasar sample, and the blue and orange points are the true and false detections. The top row shows the R-L relation derived using the assigned lags of the true detections, the middle row presents the result using measured lags with error bars of the true detections, and the bottom row displays the result using measured lags of both true and false detections. {The black solid line is the input R-L relation \citep{Bentz_2009b}} used to generate the uniform sample and the blue lines are 50 random realizations drawn from the posterior of the Bayesian regression fit to the R-L relation. {The black points are the \cite{Bentz_2013} local RM AGN sample for reference.}}
    \label{fig:rl}
    \end{figure*}

Now we examine how the selection effects from the sample and survey design, as well as the uncertainties in the measured lags, can affect the slope of the R-L relation, as compared to the R-L relation and scatter used to assign lags to our simulated quasars as described in \S \ref{sec:data}. We use the 100 down-sampled realizations of the flux-limited sample with the statistical approach described in Section \ref{sec:reallife}. The observed-frame lags are shifted to rest-frame by dividing by a factor of $(1+z)$, and then fit the slope in the $\tau-L$ relation with the linear regression code \code{LINMIX} \citep{Kelly_2007}. \code{LINMIX} uses a Bayesian approach to perform linear regression with measurement errors in both coordinates and produces more consistent fitting results than traditional regression methods when the data have large intrinsic scatter or are poorly measured. Since the R-L relation is derived with H$\beta$ lags which are only measured at $z<0.9$, we exclude all measured lags with $z>0.9$ during the fitting. The fitting results are shown in Figure \ref{fig:rl}. 

We first examine the effects of selection bias due to sample/survey design by fitting the R-L relation with the assigned lags of the true detections (top row in Figure \ref{fig:rl}). Due to our limited observation period, we cannot detect observed-frame lags longer than 100 days with any method and most short lags (less than the cadence) with ICCF and ZDCF. This constraint limits the dynamical range of luminosity and time lag in the R-L relation fitting, resulting in the fitted R-L relation slope being shallower than the nominal slope. 

\begin{figure*}
\centering
\begin{tabular}{@{}cc@{}}
\includegraphics[width=0.33\textwidth]{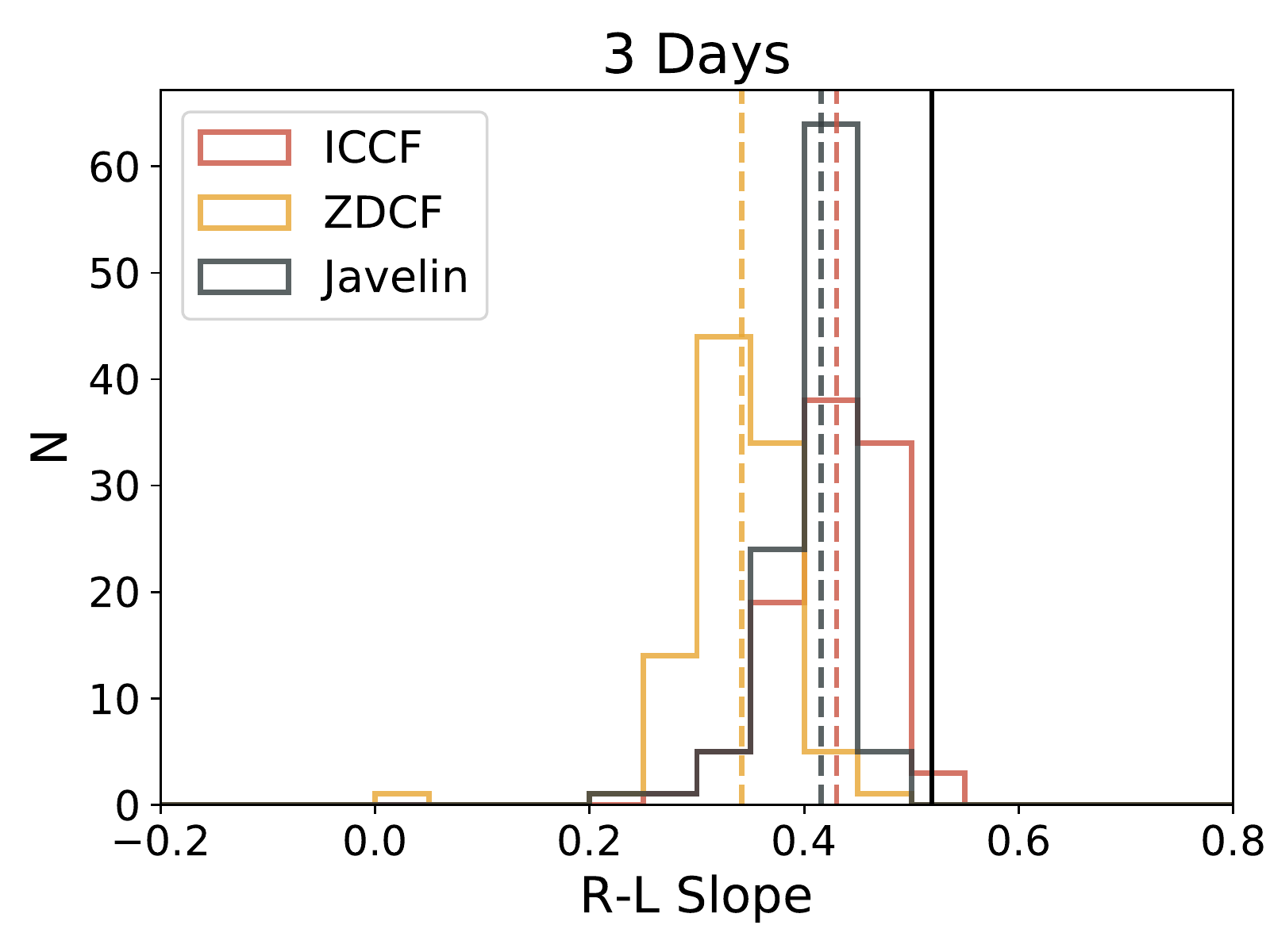}
\includegraphics[width=0.33\textwidth]{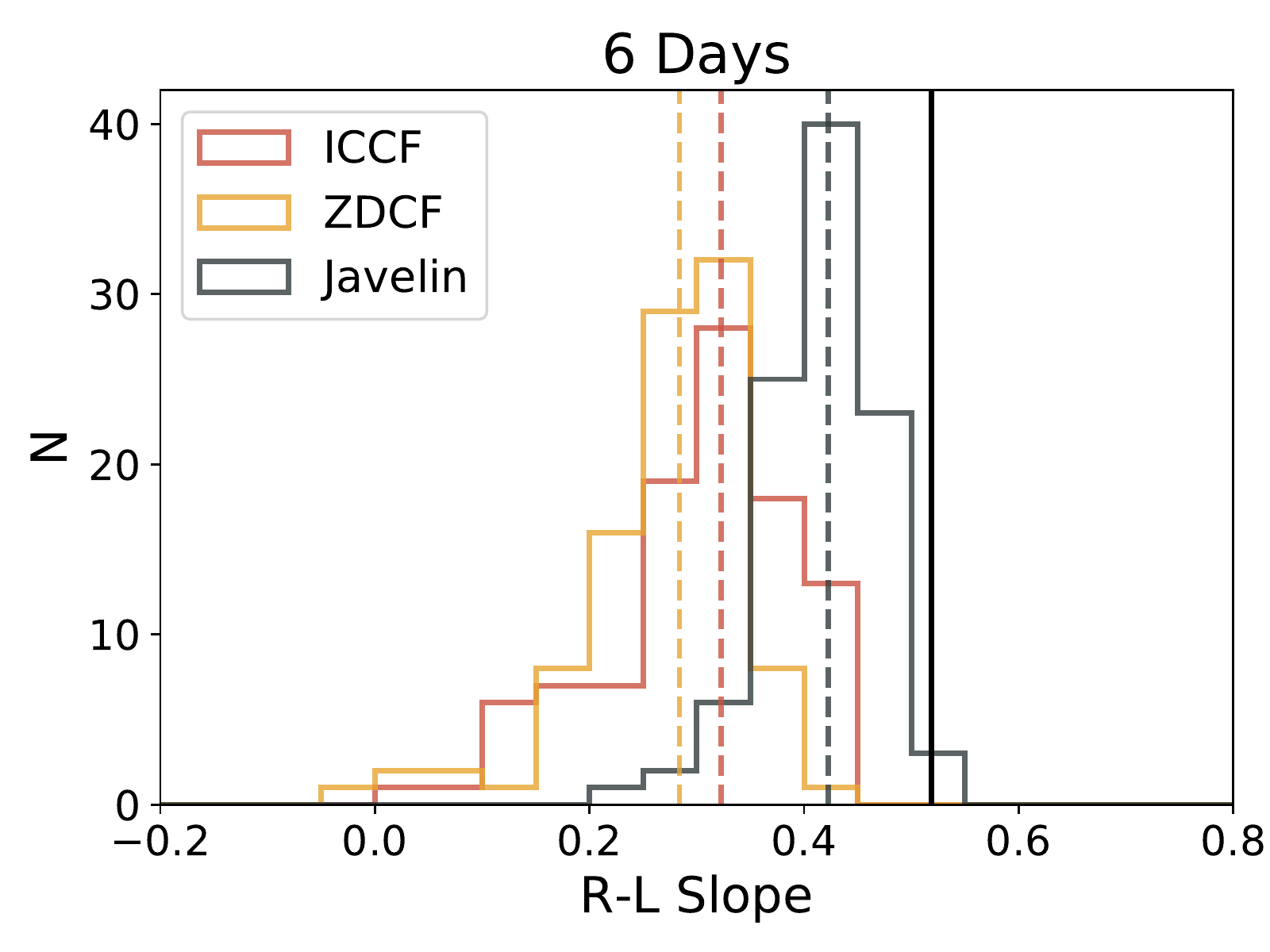}
\includegraphics[width=0.33\textwidth]{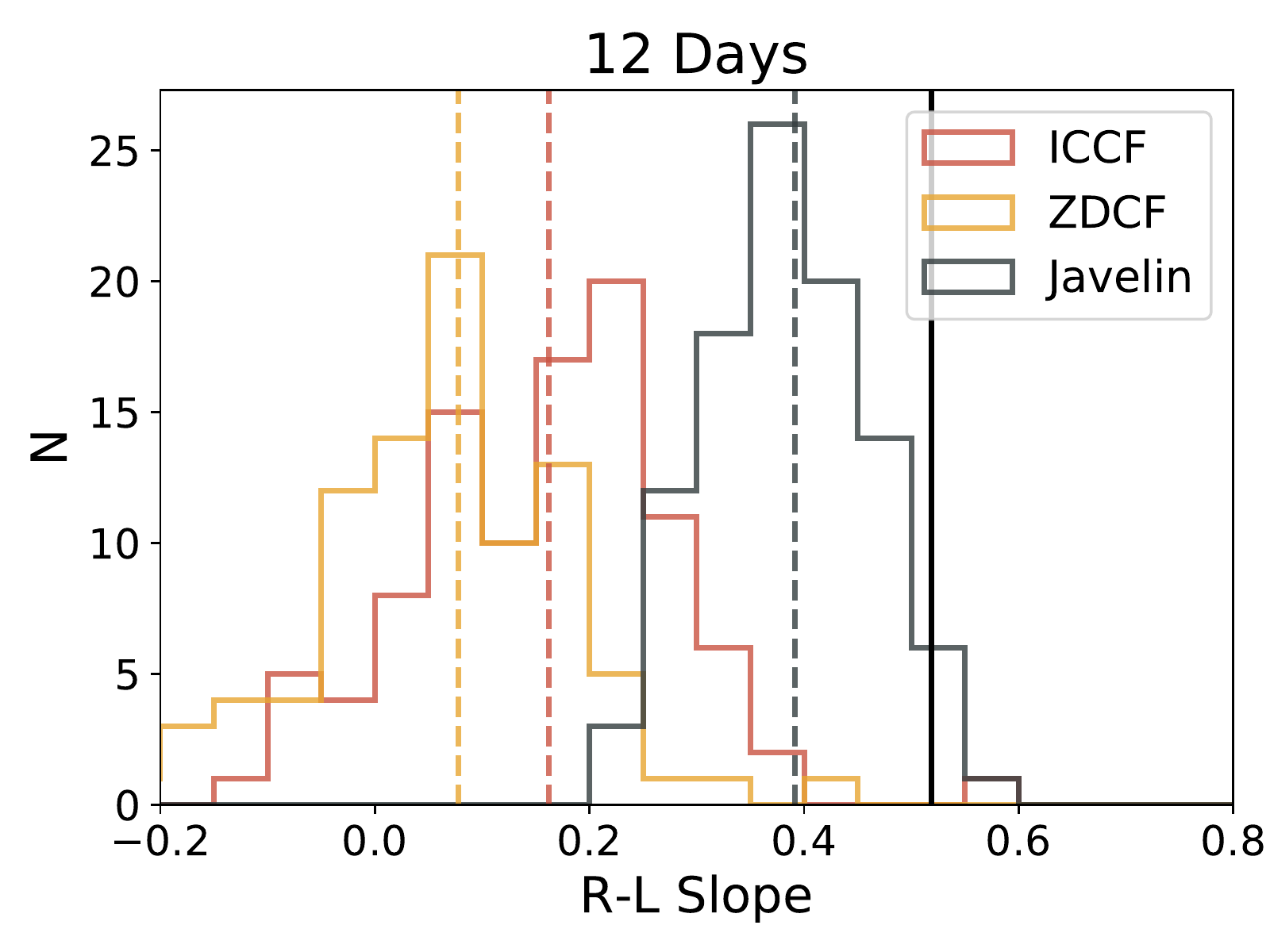}
\end{tabular}
\caption{Distribution of the best-fit R-L slopes from 100 down-sampling realizations. From the left to right are the simulations of 3-day, 6-day and 12-day cadences. The vertical dashed lines indicate the median of each distribution and the solid vertical lines mark the slope of the input R-L relation ($\beta$=0.519).}
\label{fig:slope_hist}
\end{figure*}

Next, we examine the effects of lag measurement uncertainties by fitting the R-L relation with the measured lags (incorporating the lag uncertainties) for the true detections only (middle row in Figure \ref{fig:rl}) and for all detections (bottom row in Figure \ref{fig:rl}). When only considering the true detections (which cannot be identified in real surveys), the fitted R-L slopes are slightly shallower but still consistent with the previous values. When including both true and false detections, the fitted R-L slopes are the same within the uncertainties for {\tt JAVELIN} compared to the fitting with only true detections. However, the fitted R-L slopes for ICCF and ZDCF are biased by the false detections at $\sim$100 days (observed-frame), as indicated by the larger scatter in Figure \ref{fig:slope_hist}. In some realizations, the measured lags of false detections deviate significantly from the nominal R-L relation for ICCF and lead to a highly biased R-L slope ($\sim$0). Therefore, in practice, it is important to examine questionable lag measurements and establish criteria to discard them from the sample. In general, R-L slopes measured from {\tt JAVELIN} are more robust and accurate than those from ICCF and ZDCF, which is due to the combined benefits of having more detected lags (especially the short lags) and higher lag measurement quality. 

Figure \ref{fig:slope_hist} presents the histograms of the fitted R-L slope {from the measured lags (including false detections)} in the 100 down-sampled realizations. In the 6-day cadence simulations, fitted slopes are $\sim$0.4 for {\tt JAVELIN} and $\sim$0.3 for ICCF and ZDCF, {and the normalized median absolute deviations (NMAD) are $\sim$0.08 for {\tt JAVELIN} and $\sim$0.14 for ICCF and ZDCF.} When the cadence decreases (number of epochs increases), there are fewer false detections in ICCF, so the fitted slope approaches $\sim$0.4, where most of the remaining bias is due to the limited lag range. With the 12-day cadence, the detections from ICCF and ZDCF decrease and false detections increase, causing the R-L relation fitting to become unreliable, as indicated by the broader range of the slope distribution {(NMAD $\sim$0.51 for ICCF and $\sim$1.35 for ZDCF). For {\tt JAVELIN}, the fitted slope converges around 0.4 for all three cadences and the NMAD only increase slightly to 0.14 at cadence of 12 days, which is comparable to NMAD for ICCF and ZDCF simulations at cadence of 6 days.} 

For the light curve S/N dependence, the median of fitted R-L slopes is consistent at different S/N for all three methods, because the number of detections and their distribution in the R-L plane do not change drastically as light curve S/N varies. When the S/N is degraded, the R-L slope uncertainties increase for ICCF and ZDCF, but not for {\tt JAVELIN} --- this is because the scatter in the R-L plane primarily originates from the increased lag uncertainties as light curve S/N decreases, which is not the case for {\tt JAVELIN} (see \S \ref{sec:diss_err}).
    
\subsection{Scatter of the R-L relation}

The slope of the R-L relation derived from our simulation is consistently shallower than the assigned value. {However, the \cite{Grier_2017} R-L relation shows more scatter than the \cite{Bentz_2009b} and \cite{Bentz_2013} R-L relations (the \cite{Bentz_2013} R-L relation is an updated version of the \cite{Bentz_2009b} R-L relation that includes more low-luminosity sources).} There are many possible reasons for this discrepancy. For example, \cite{Grier_2017} used spectral decomposition to correct for host galaxy light in the estimation of quasar-only luminosity instead of high resolution imaging decomposition as with \cite{Bentz_2009b, Bentz_2013}. There may also be intrinsic differences in the R-L relations due to the difference in samples (e.g., the SDSS-RM \hbeta\ lag sample is at substantially higher redshift and spans a broader range of quasar parameter space than the Bentz et al. sample). {\cite{Du_2016} suggested that the R-L relation might depend on quasar luminosity and accretion rate \citep[also see ][]{Loli_2019}.} This discrepancy motivates us to investigate how the observed R-L relation changes with different assumptions of the intrinsic scatter.  We produced another set of simulations while applying increased scatter in the input R-L relation, following the same procedures described in Section \ref{sec:data} but doubling the scatter in the initial R-L relation to generate a new set of mock quasars and light curves. 

\begin{figure*}
    \centering
    	\begin{tabular}{@{}cc@{}}
        \includegraphics[width=0.33\textwidth]{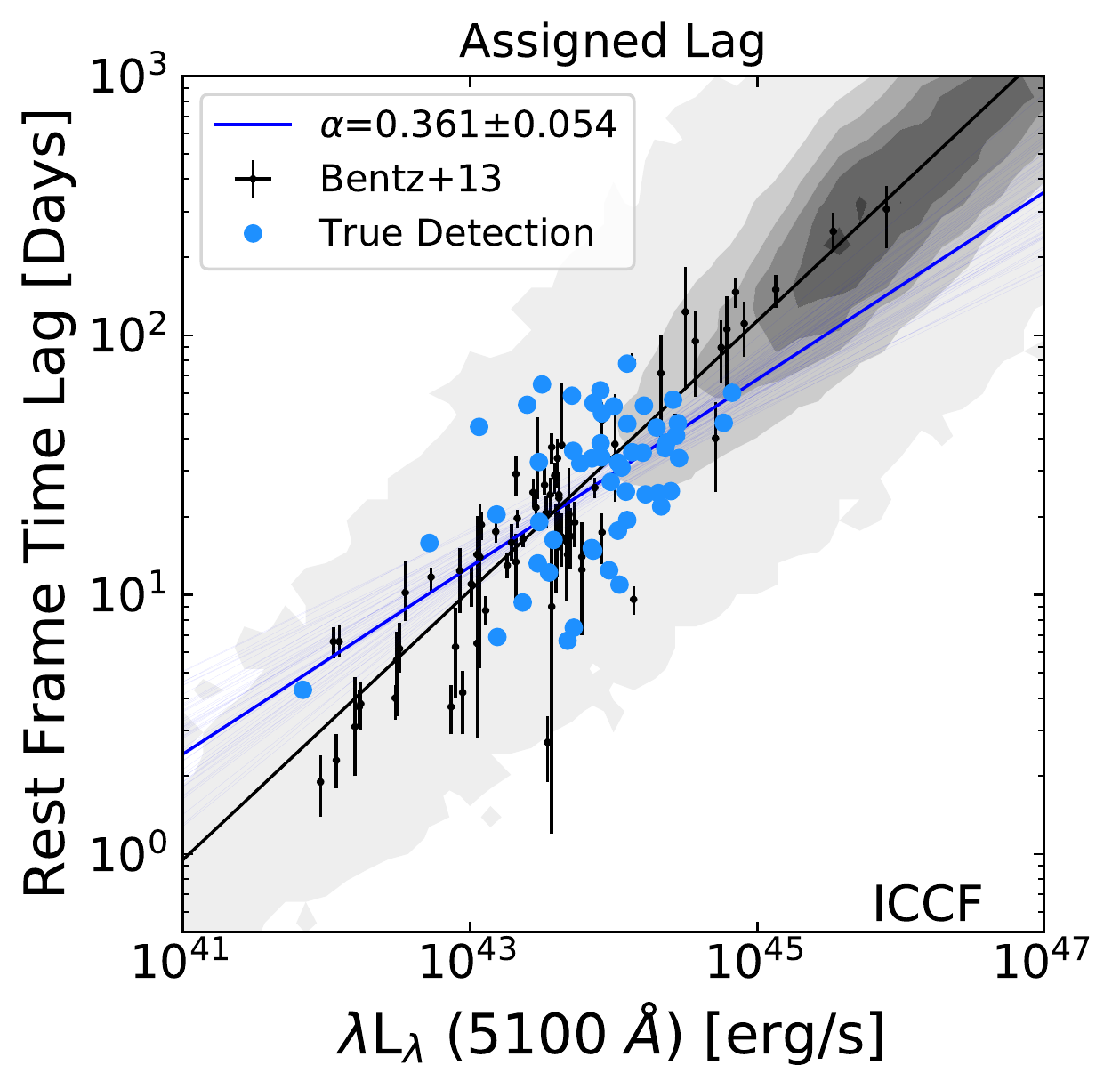}
        \includegraphics[width=0.33\textwidth]{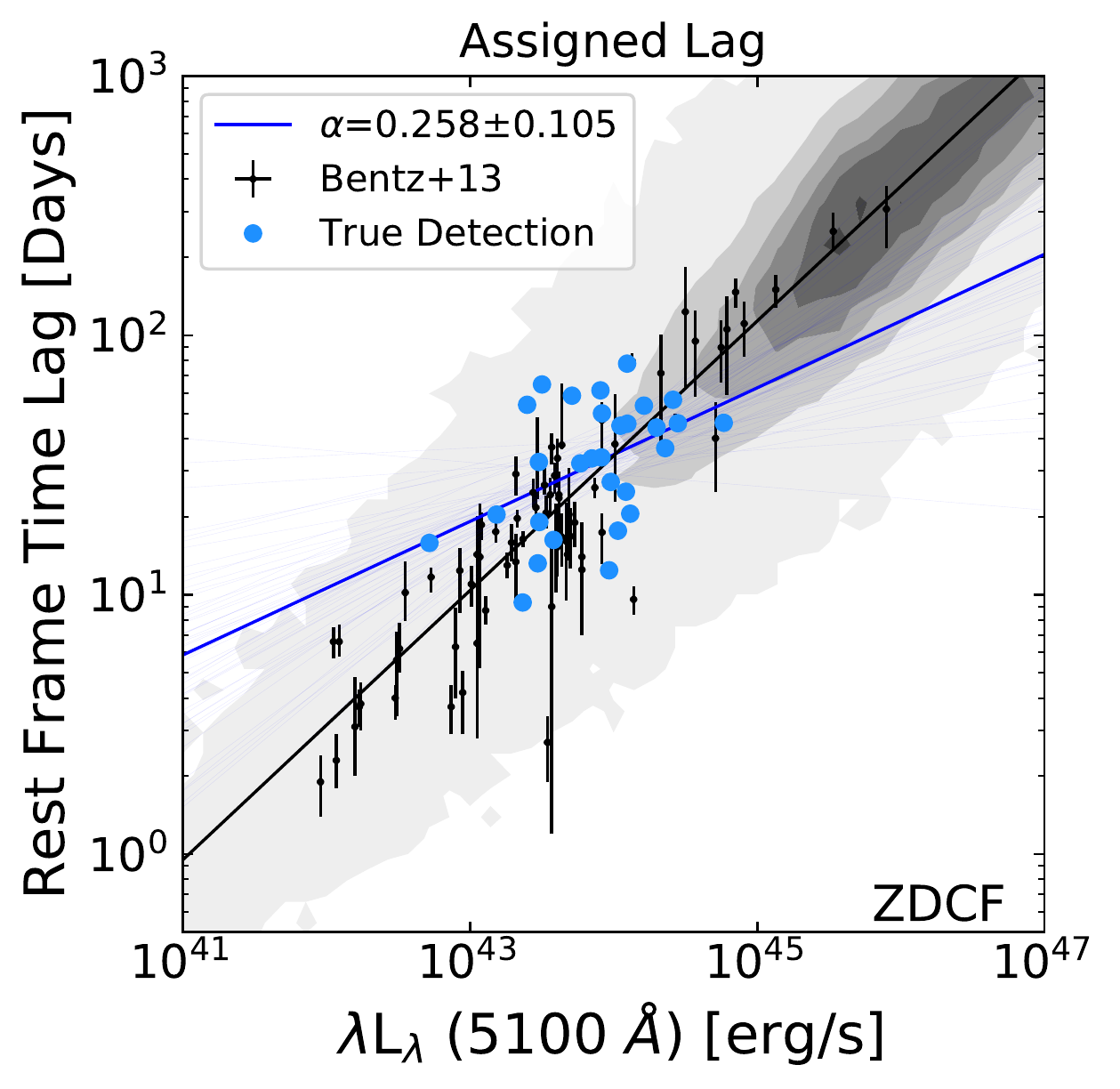} 
        \includegraphics[width=0.33\textwidth]{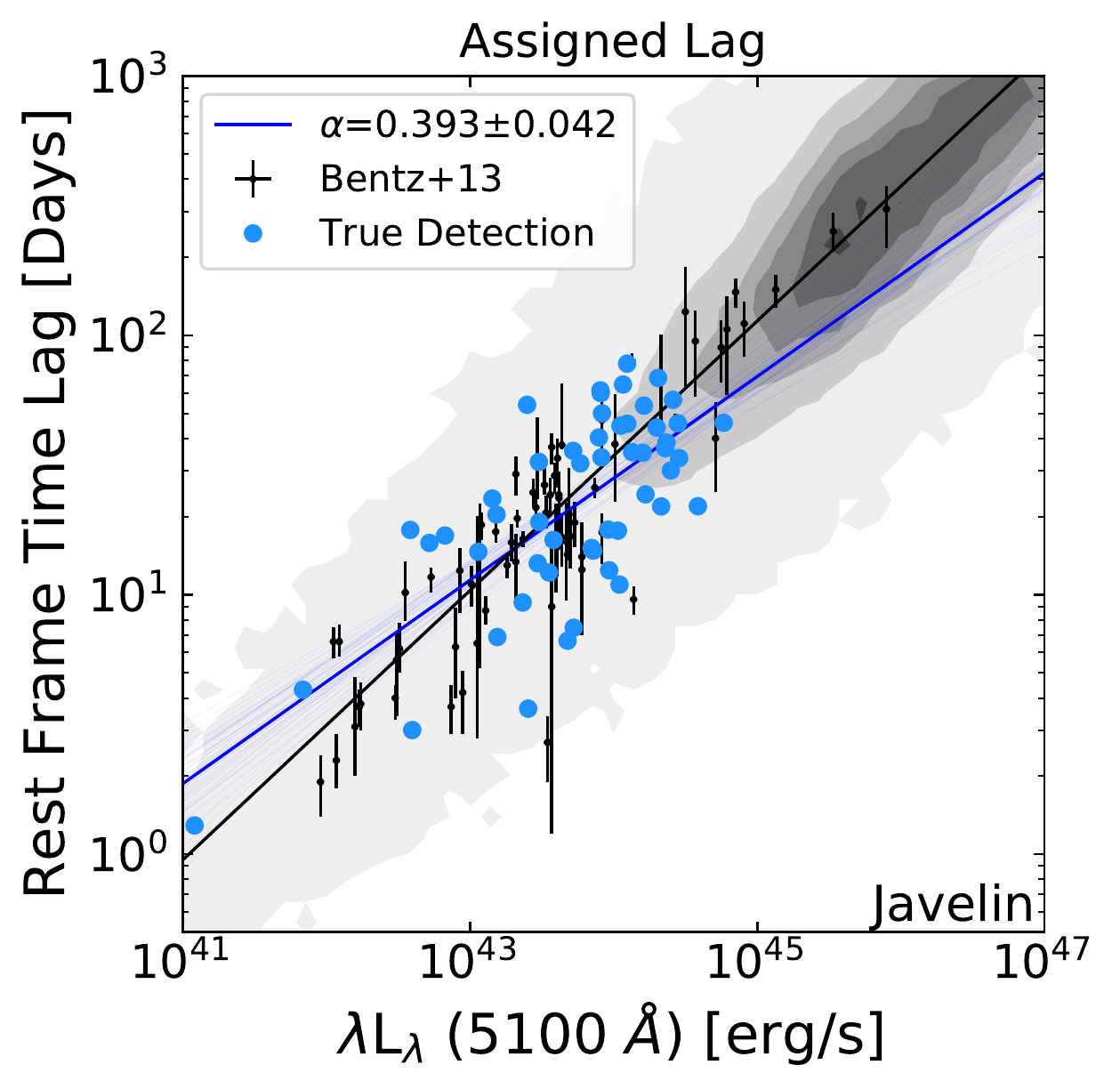}\\
        \includegraphics[width=0.33\textwidth]{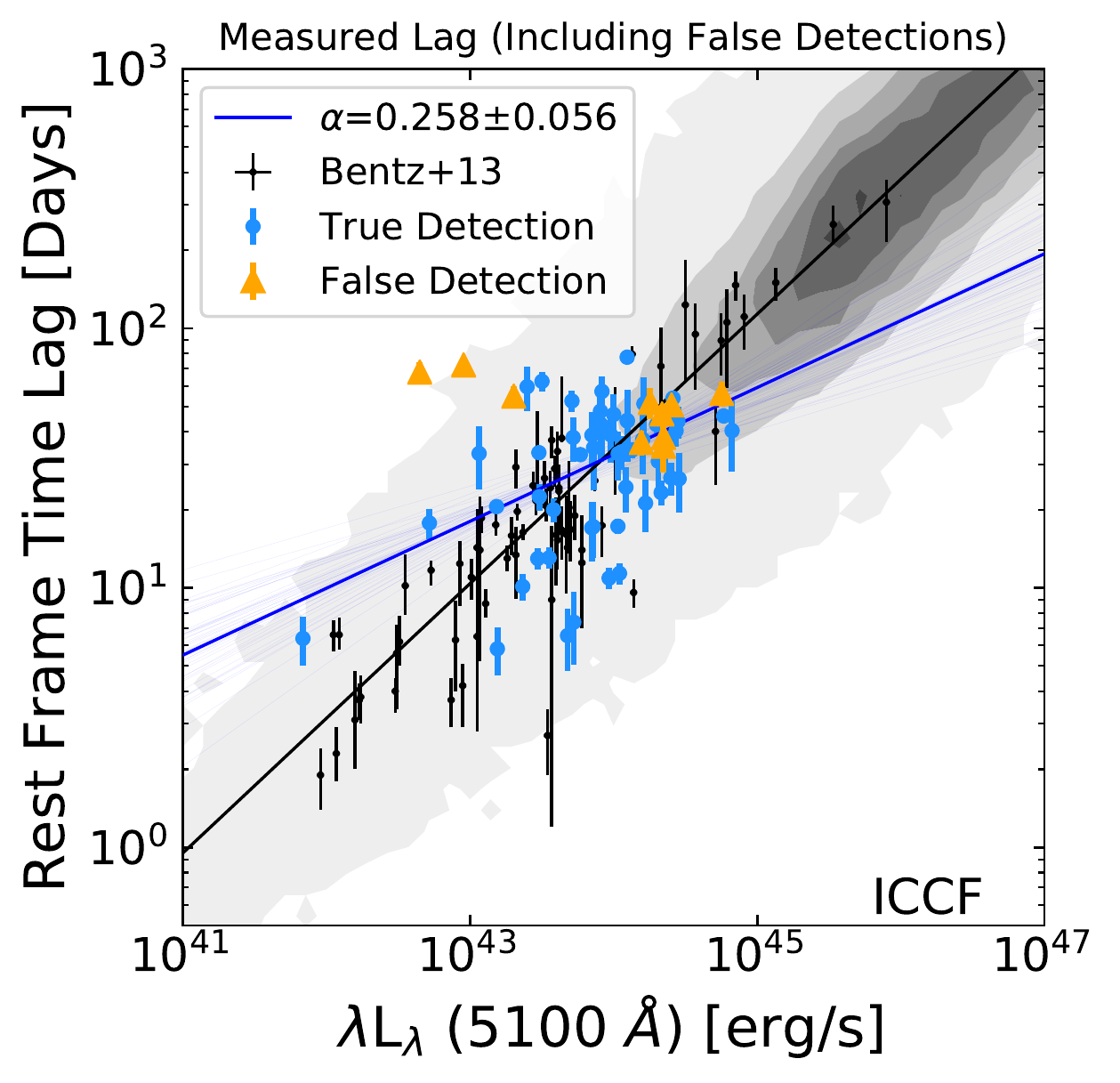}
        \includegraphics[width=0.33\textwidth]{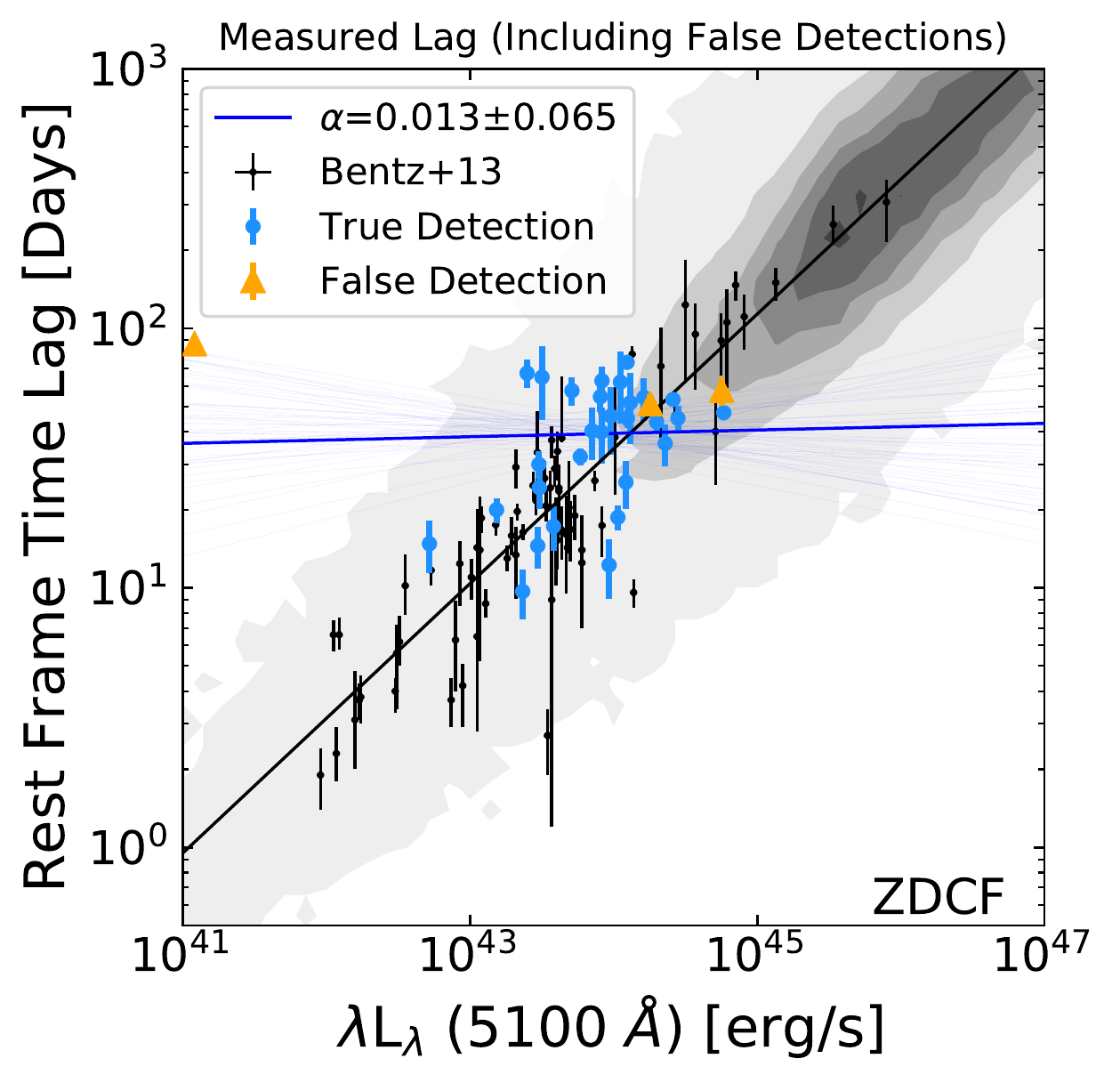} 
        \includegraphics[width=0.33\textwidth]{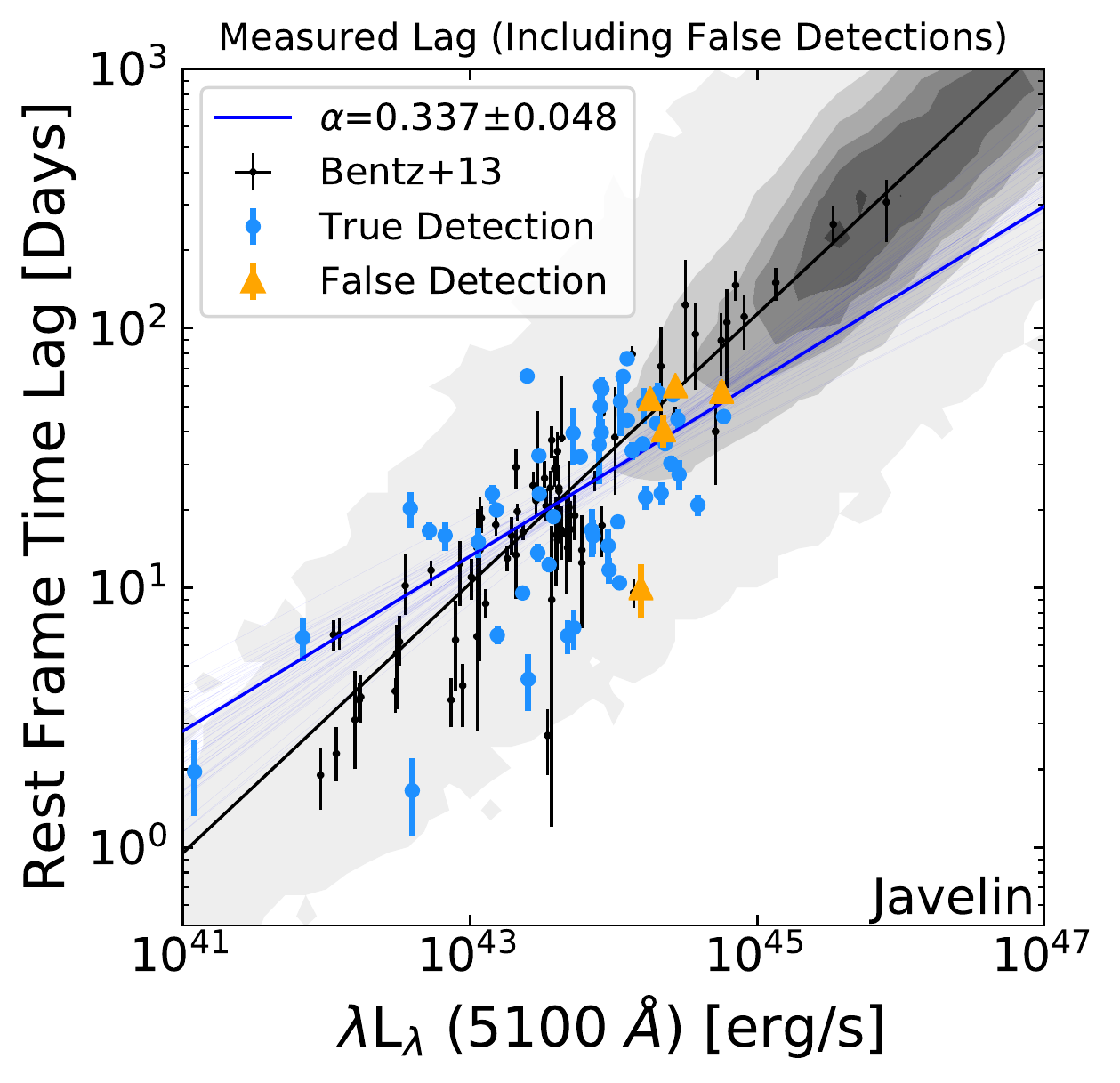}
    	\end{tabular}
    \caption{
    H$\beta$ R-L relation derived from one down-sampling realization in the simulation with more scattered R-L relation. In each panel, the grey contours represent the uniform quasar sample, and the blue and orange points are the true and false detections. The top row shows the R-L relation derived using the assigned lags of the true detections, and the bottom row displays the result using measured lags of both true and false detections. {The black solid line is the input R-L relation \citep{Bentz_2009b}} used to generate the uniform sample and the blue lines are 50 random realizations drawn from the posterior of the Bayesian regression fit to the R-L relation. {The black points are the \cite{Bentz_2013} local RM AGN sample for reference.}}
    \label{fig:rlx2}
    \end{figure*}

\begin{figure}
\centering
\begin{tabular}{@{}cc@{}}
\includegraphics[width=0.5\textwidth]{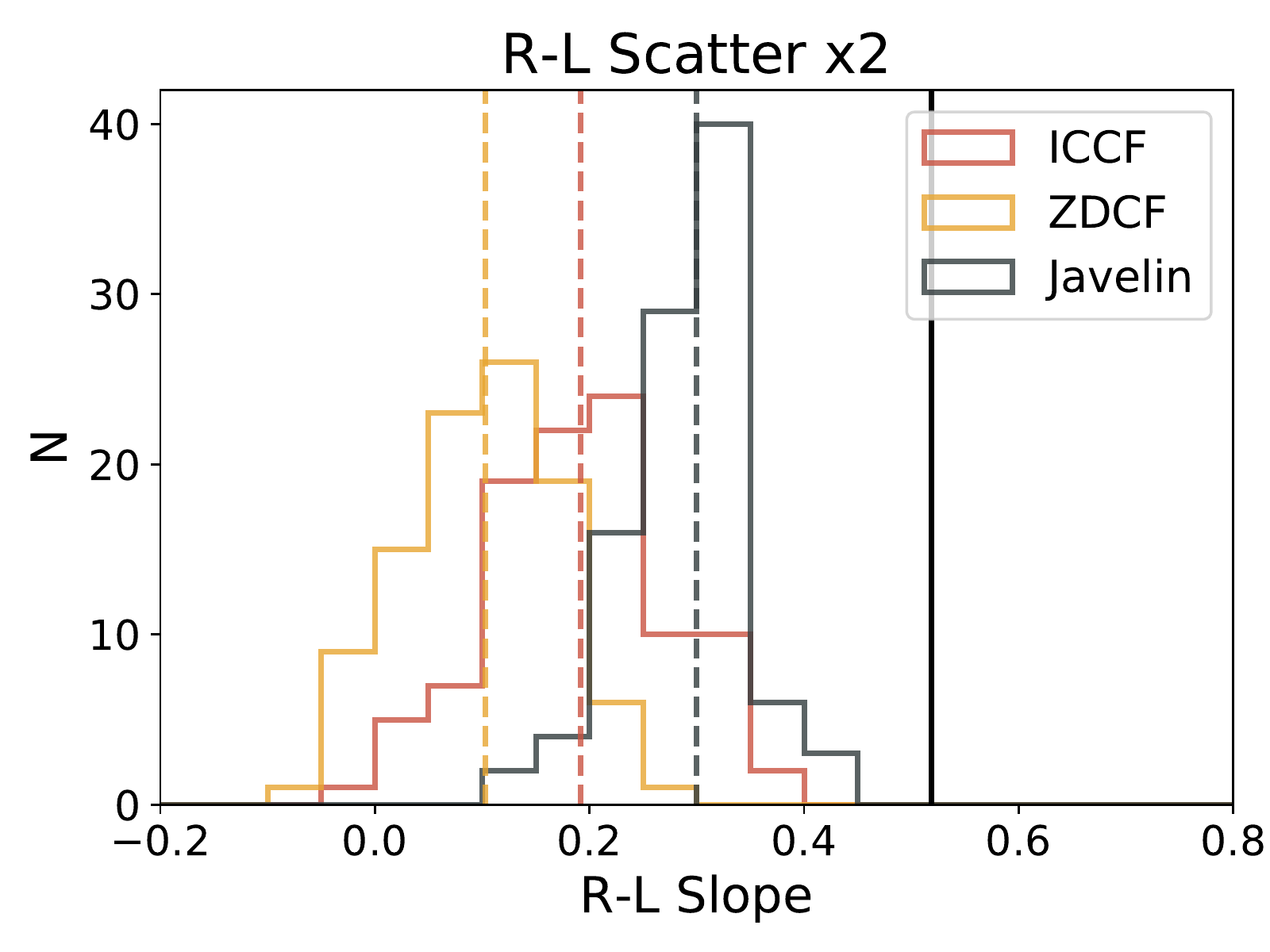}
\end{tabular}
\caption{
Distribution of the best-fit R-L slopes from 100 down-sampling realizations in the simulation with more scattered R-L relation. The vertical dashed lines indicate the median of each distribution and the solid vertical lines mark the slope of the input R-L relation ($\beta$=0.519).}
\label{fig:slope_hist_rlx2}
\end{figure}

With the larger scatter in the input R-L relation, the fitted R-L slope becomes less constrained, as demonstrated in the top rows of Figure \ref{fig:rlx2}. The observed R-L relation slopes are shallower compared to the original simulation for all three methods. In addition, false detection rates increase for all techniques, as it is more difficult {to statistically eliminate false lags by rejecting quasars in certain magnitude and redshift bins} due to the increased scatter in the lags in each bin. These false detections are located near the edge of our search range, mostly in the range of 60--80 days. As a result, the deduced R-L relation is flatter because the fit is skewed by these false detections (see bottom left panel in Figure \ref{fig:rlx2} for an example). The distributions of the fitted R-L relation slopes are presented in Figure \ref{fig:slope_hist_rlx2}: for all three lag measuring methods the slopes are shallower than those in the original simulations {$\sim$0.2 for ICCF, $\sim$0.1 for ZDCF and $\sim$0.3 for {\tt JAVELIN})} and the NMAD increases compared to original simulation {(NMAD $\sim$0.27 for ICCF, $\sim$0.46 for ZDCF and $\sim$0.11 for {\tt JAVELIN})}. The slope from {\tt JAVELIN} lags is the least biased among the three methods.

If the intrinsic scatter in the R-L relation is indeed larger for the SDSS-RM sample than for the local RM sample, then it is likely that we will measure a shallower slope using the measured lags. The shallower measured slopes are primarily a caveat of the limited dynamic range in the measured lags, and should be mitigated with additional lags measured over a broader range in luminosity. 

{The lags from \cite{Grier_2017} also on average fall below the \cite{Bentz_2009b, Bentz_2013} R-L relation \citep[also see, e.g.,][]{Du_2016}. From our simulation, there is no evidence that selection effects can cause a vertical offset from the input R-L relation. However, long lags ($>80$ days) tend to be measured with smaller values than the assigned values using the \cite{Grier_2017} lag-significance criteria, which can partially contribute to the shallower slope in the R-L relation.}

\subsection{Multi-year observations}\label{sec:multi-yr}

Following the SDSS-RM survey design, we ran a 5-year simulation (with 30 observing epochs for the first year, 15 epochs for years 2 and 3, and 6 epochs for years 4 and 5) and examine the lag measurements using ICCF and {\tt JAVELIN} on the flux-limited sample. Since the ZDCF method consistently underperforms over the other two methods, we do not consider ZDCF further in this section.

Similar to our 100-day search range criteria, we set the search range to $\sim$800 days in order to avoid strong CCCD/PDF signals produced with fewer overlapping points. The grid size of the ICCF is set to 15 days, the median of the cadence, which results in smoother ICCFs for light curve pairs with larger lags. The MCMC parameters are set to be the same as for our 180-day simulations (see \S\ref{sec:measurelags}), as this value is sufficient for the results to converge. For the alias removal procedure, the width of the Gaussian smoothing kernel was increased to 7.5 days to improve the ability to capture longer lags. We scale the CCCD/PDFs as a function of the number of overlapping points in each of the 6-month observing seasons. Both ICCF and {\tt JAVELIN} interpolate within the 6-month seasonal gaps, and these lag ranges are down-weighted in the alias removal procedure. Finally, we perform the statistical selection as in \S\ref{sec:reallife} to remove unlikely detections. This approach removes $\sim$10\% false detections and $<$1\% true detections for ICCF, and $\sim$20\% false detections and $\sim$1\% true detections for {\tt JAVELIN}. 

\begin{figure*}
    \centering
    	\includegraphics[width=1.0\textwidth]{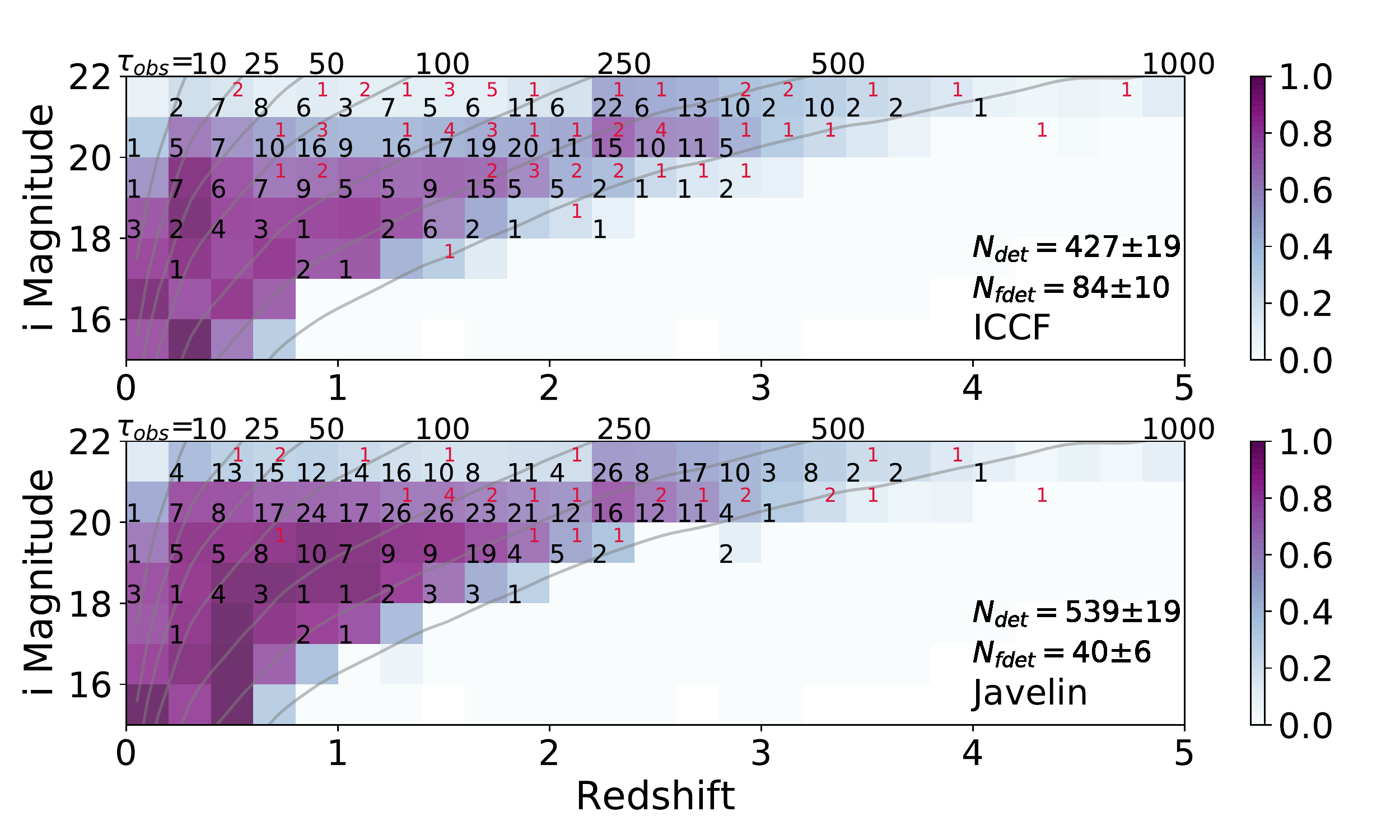}
    \caption{Similar format to Figure \ref{fig:detmap}. Detection maps for the 5-year simulation following the statistical selection described in \S\ref{sec:reallife}. The colormap represents the detection efficiency and the numbers are the detection counts (true detections in black and false detections in red) of a single down-sampling realization. The total numbers of true and false detections shown in the lower-right corner are the median and uncertainties derived from 100 down-sampling realizations. The grey contours show the approximate constant lags from the R-L relation from \cite{Bentz_2009b}. The detection efficiency is calculated for the selected sources in the flux-limited sample, instead of using the uniform sample like in Figure \ref{fig:detmap}.}
    \label{fig:detmap_multiyr}
    \end{figure*}

\begin{figure*}
\centering
	\begin{tabular}{@{}cc@{}}
	\includegraphics[width=0.5\textwidth]{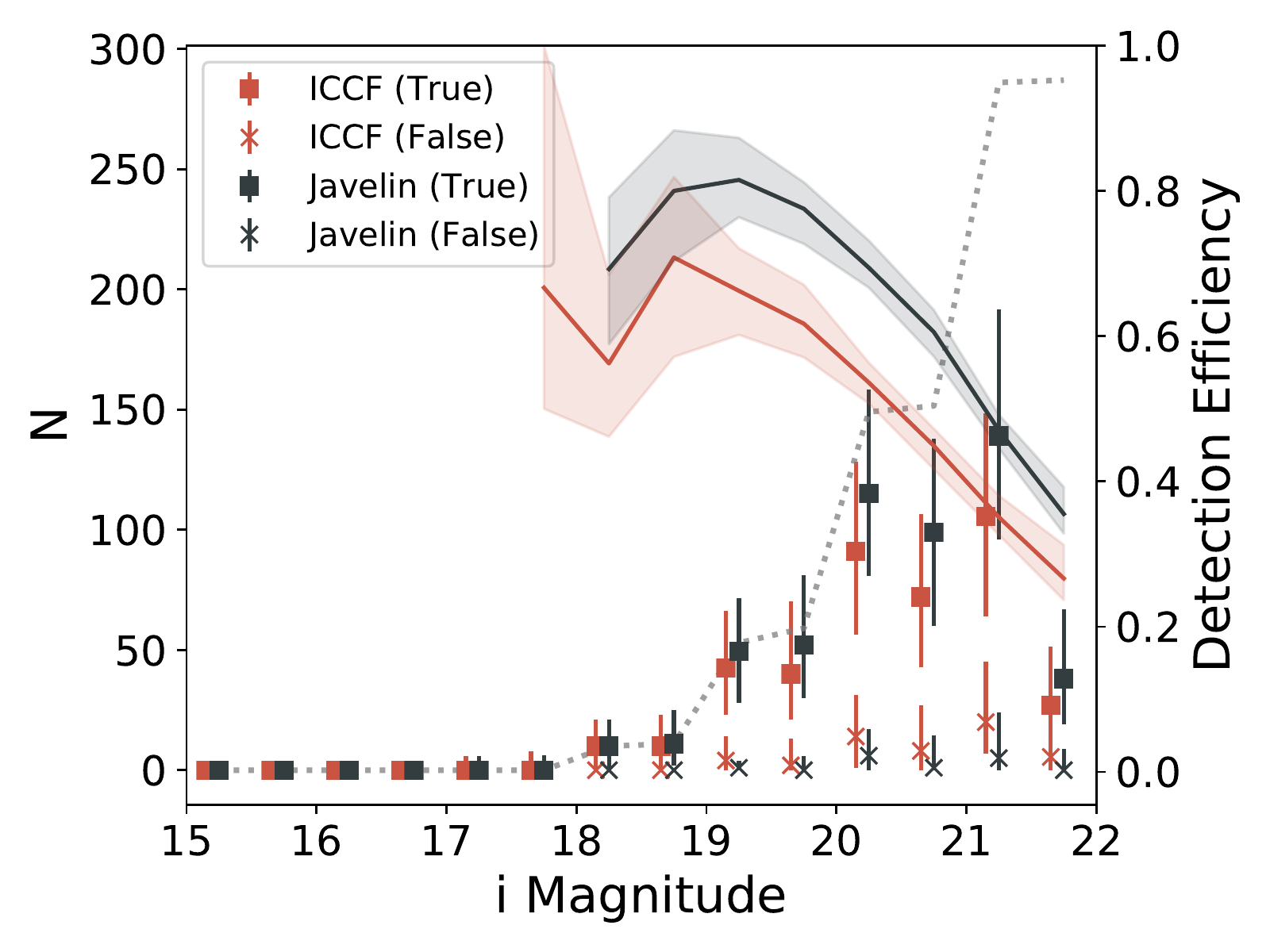}
	\includegraphics[width=0.5\textwidth]{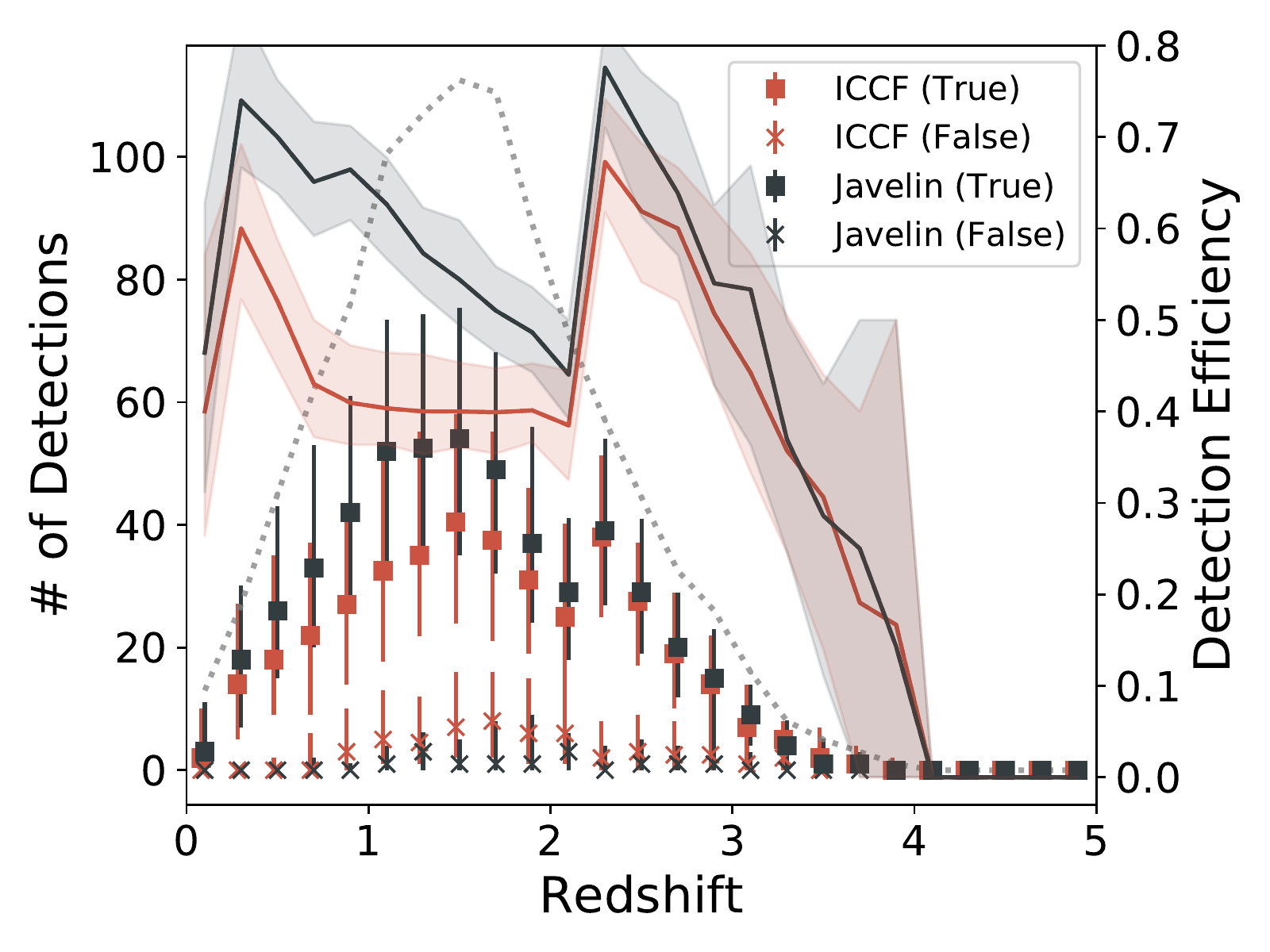}
	\end{tabular}
    \caption{
    Detection efficiency (solid lines and shaded area) and true (square) and false (cross) detection counts of the three methods as functions of $i$-band magnitude (right panel) and redshift (left panel) of the flux-limited sample from the 5-year simulation. The dotted lines show the number of sources with lags shorter than the search range (i.e. 800 days) in each magnitude or redshift bin. For $i<$17 and $z>4$, the detection efficiencies are not shown because there are no quasars selected in more than 95\% of bootstrapping realizations.}
    \label{fig:corr_multiyr}
\end{figure*}

Figure \ref{fig:detmap_multiyr} presents the detection map of the 5-year simulations. The shaded area is the detection efficiency calculated of the flux-limited sample, instead of the uniform quasar sample as in the previous figures (e.g., Figure \ref{fig:detmap}). The overall detection efficiency is $\sim$45\% for ICCF and $\sim$56\% for {\tt JAVELIN} and false detection rates are $\sim$16\% for ICCF and $\sim$6.9\% for {\tt JAVELIN} with the 5-year baseline. Lag detections are limited by the observing baseline, redshift, and light curve S/N. Below redshift $\sim$2, the detection efficiency follows similar trends as the single-season simulations. Compared to the 180-day simulation, the detection efficiency increases for lag of $<100$ days with additional seasons of observation, especially for {\tt JAVELIN}. At longer lags ($>100$ days), lag detection efficiency increases at 2$<z<$2.5 for faint objects. This behavior arises because, with the seasonal gaps and the chosen baseline, our survey will be most sensitive to lags $<$100 days and 250--400 days. These trends are observed in Figure \ref{fig:corr_multiyr}. In our 5-year simulation, detection efficiency peaks at $z\sim$0.5 and $z\sim$2.5 and falls off sharply at $z>$3. Detections mostly fall in the range of 0.5$<z<$2.5 due to the redshift distribution of our sources. 

\begin{figure}
    \centering
        \includegraphics[width=0.5\textwidth]{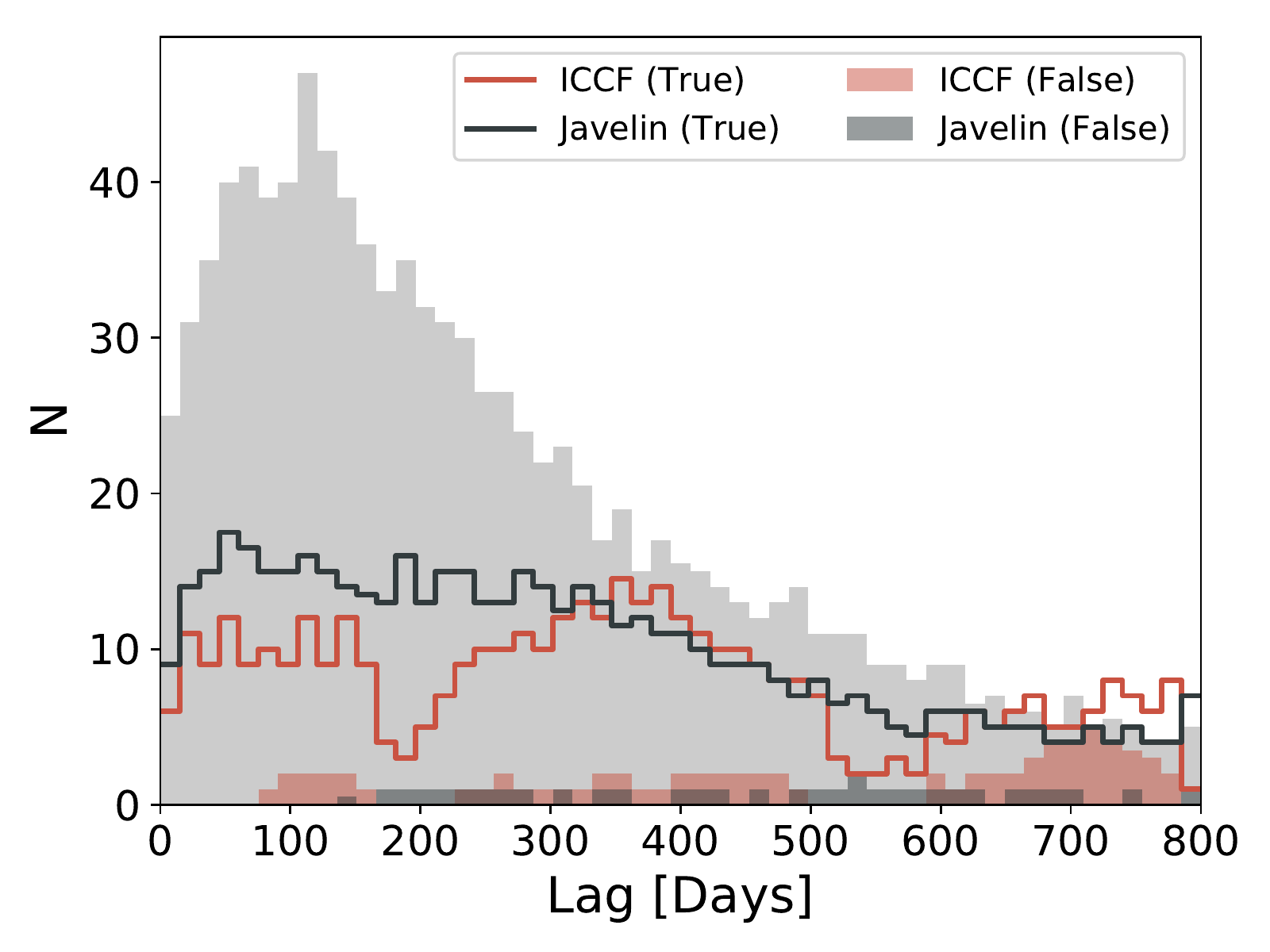}
    \caption{
    Median distribution of the detected lags in the 100 down-sampling realization of the 5-year simulation. The open histograms show the number of true detections and the solid histograms indicate the number of false detections. The grey shaded area represents the median assigned lag distribution.}
    \label{fig:hist_multiyr}
    \end{figure}

The distribution of detected lags (Figure \ref{fig:hist_multiyr}) reveals gaps in the distribution of detected lags with ICCF, which correspond to the seasonal gaps in the observations. For {\tt JAVELIN}, however, these gaps are less obvious, indicating that {\tt JAVELIN} is interpolating reasonably well within long seasonal gaps and measures lags more accurately in multi-year projects than ICCF. In addition, {\tt JAVELIN} has lower and more evenly-distributed false detections throughout the lag ranges. For $>$600 day lags, there are as many false detections as true detections for ICCF, suggesting that it will be very difficult to identify true detections with ICCF in this lag range. 

\begin{figure*}
	\includegraphics[width=0.5\textwidth]{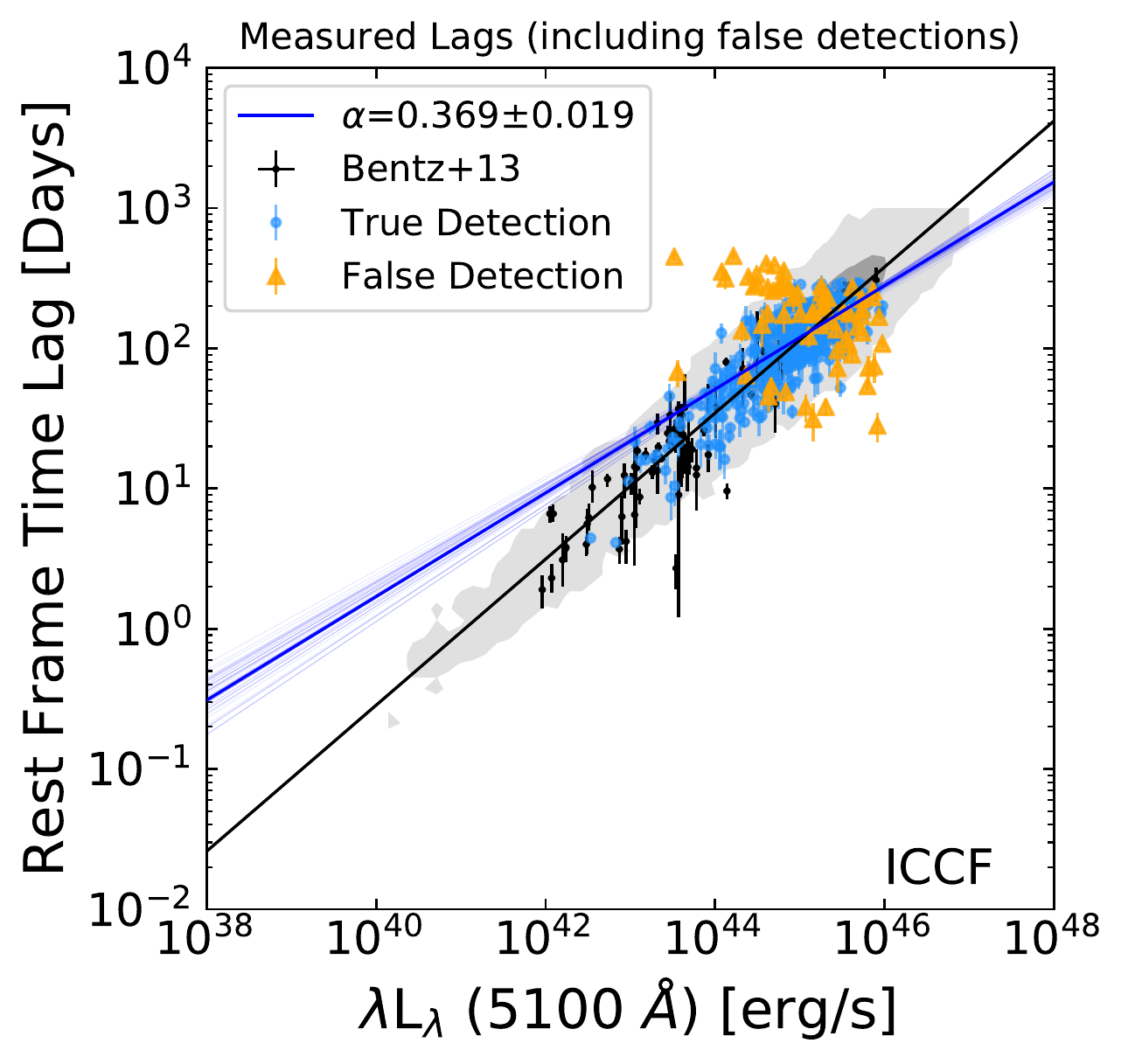}
	\includegraphics[width=0.5\textwidth]{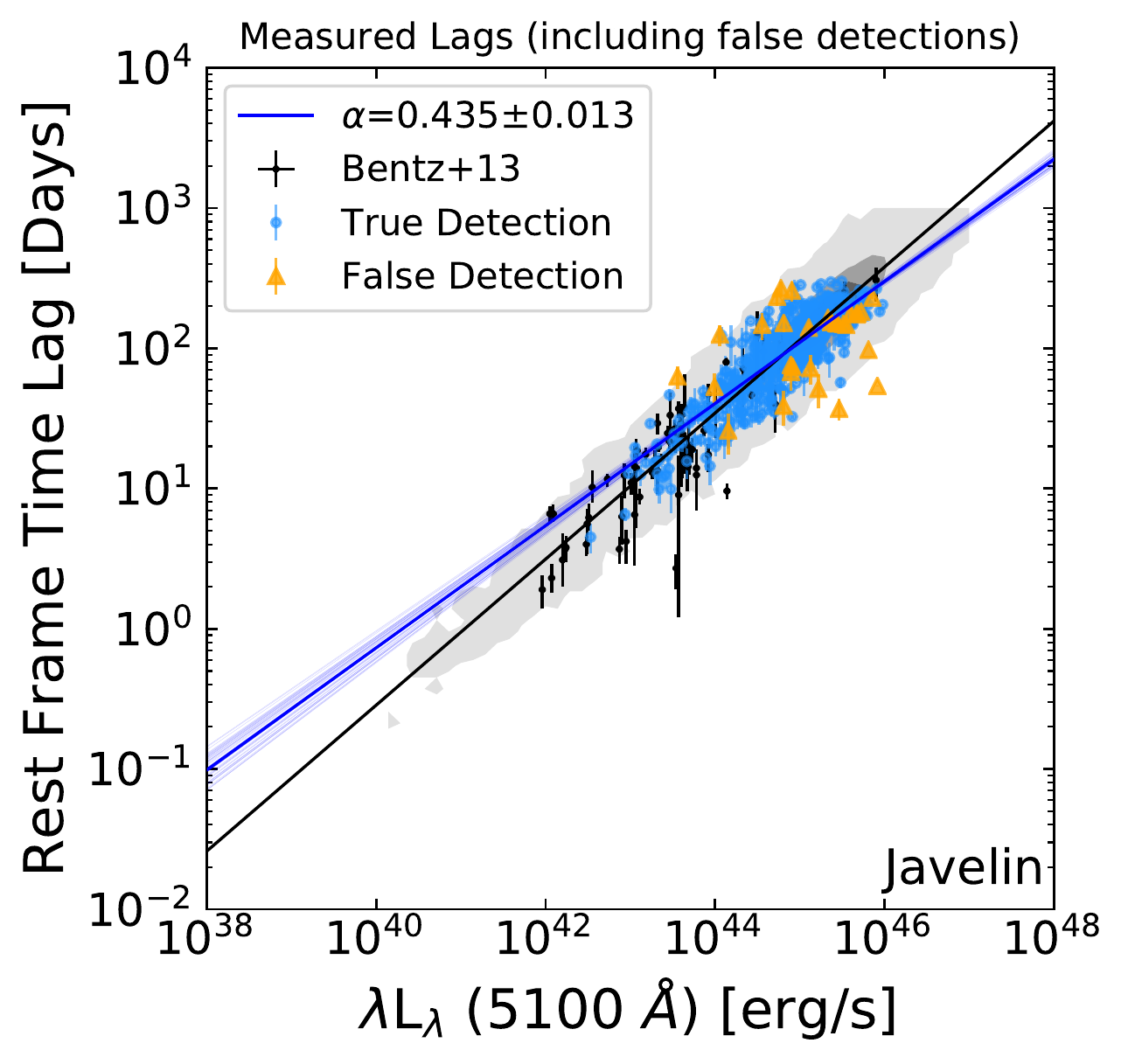}
	\caption{
	H$\beta$ R-L relation derived from one down-sampling realization in the 5-year simulation. The grey contours represent the uniform sample, and the blue and orange points are the true and false detections. The black points are the Bentz et al. (2013) local RM AGN sample for reference. The black solid line is the input R-L relation used to generate the uniform sample and the blue lines are 50 random realizations drawn from the posterior of the Bayesian regression fit to the R-L relation. }
	\label{fig:rl_multiyr}
\end{figure*}

Figure \ref{fig:rl_multiyr} displays the fitting of the R-L relation in one down-sampled realization. Since the detected lags cover a wide range in luminosity, the slopes are less biased by the limited dynamical range than the 180-day simulation. The ICCF R-L relation is still skewed by the false detections clustered at $\sim$600--800 days. Since the distribution of false detections in {\tt JAVELIN} is more uniform over the range of lags, the fitted slope of the R-L relation is more accurate. However, the derived slopes are still somewhat shallower than the assigned value due to the false detections at higher lags and lower detection rate at small lags (in the rest frame). 

\begin{figure}
\centering
\includegraphics[width=0.5\textwidth]{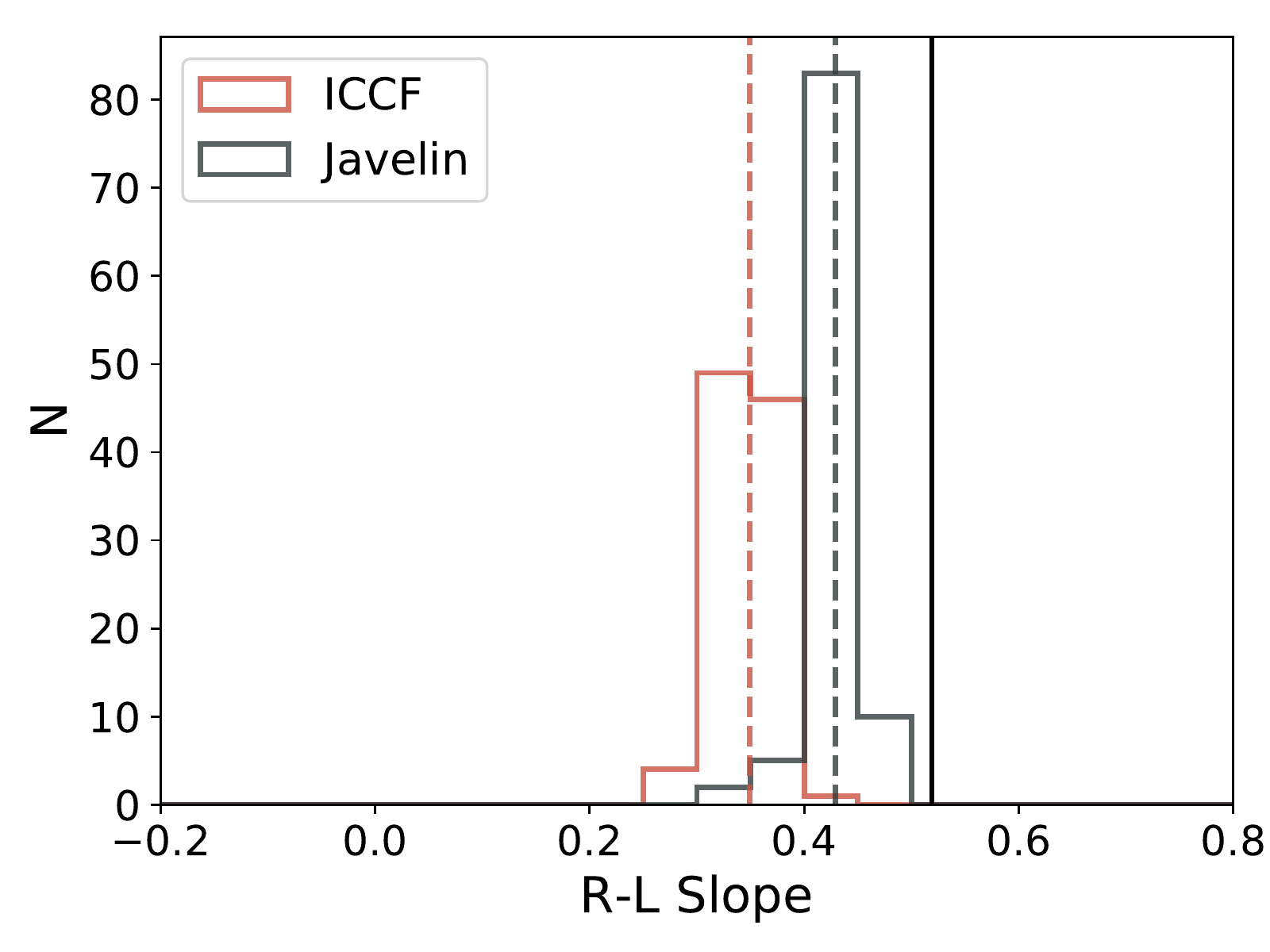}
\caption{
Distribution of the best-fit R-L slopes from 100 down-sampling realizations in the 5-year simulation. The vertical dashed lines indicate the median of each distribution and the solid vertical lines mark the slope of the input R-L relation ($\beta$=0.519).}
\label{fig:slope_hist_multiyr}
\end{figure}

Figure \ref{fig:slope_hist_multiyr} shows the distribution of the fitted slopes from the 100 down-sampling realizations. The measured slopes from the {\tt JAVELIN} lags {(slope $\sim$ 0.42, NMAD $\sim$ 0.03)} are again more consistent with the input slope than those from the ICCF lags {(slope $\sim$ 0.35, NMAD $\sim$ 0.05)}. 

{Expanding the lag sample to a wider AGN luminosity range appears to be necessary to recover the true slope of the R-L relation. The low-luminosity end of the R-L relation can be filled in by measuring short lags in low-luminosity sources at rapid cadence with short baselines. However, the high-luminosity end of the R-L relation is a more difficult problem: it requires continuous monitoring for years or even decades. The longest measurable lags will always be limited by the total baseline. Understanding the lag detection limit and the false detection rate can help understand biases in the high-luminosity end of the R-L relation.}

\section{Conclusions}\label{sec:con}

In this work, we used simulated MOS-RM observations to test the strengths and weaknesses of three popular time lag measuring methods: ICCF, ZDCF and {\tt JAVELIN}. We examined lag detections for a uniform mock quasar sample and down-sample it to mimic flux-limited samples in real surveys. Among the three methods, ZDCF has the lowest detection efficiency and detection quality, indicating that the interpolation between data points in the other two methods enhances the probability of lag detection.  

{\tt JAVELIN} performs better than ICCF in essentially all major benchmarks we tested:
\begin{enumerate}
\item[$\bullet$] {\tt JAVELIN} can recover more lags that are shorter than the cadence, which we ascribe to the more empirically-motivated interpolation scheme based on the DRW model used to describe stochastic quasar continuum variability. 

\item[$\bullet$] Overall, {\tt JAVELIN} produces both more accurate and more precise lag measurements for typical MOS-RM programs. The formal lag errors from {\tt JAVELIN} are also {the most reliable (compared with the deviations from the true lags)} among the three methods. 

\item[$\bullet$] {\tt JAVELIN} in general produces fewer false detections than ICCF, and its detection efficiency and quality are less sensitive to degradation of the S/N of light curves {(Fig. 16)}. 

\item[$\bullet$] {\tt JAVELIN} is less affected by large, seasonal gaps in the light curves, resulting in more lags that are near the seasonal gaps that will otherwise be missed by ICCF. This is again the result of the more physically-motivated interpolation scheme by {\tt JAVELIN} {(Section \ref{sec:multi-yr})}. 

\item[$\bullet$] The advantages of {\tt JAVELIN} in lag measurements lead to less bias in the measured slope in the R-L relation than ICCF {(Section \ref{sec:rl} and Figure \ref{fig:slope_hist})}. 

\item[$\bullet$] {{\tt JAVELIN} performs at least equally well as ICCF in all the aforementioned tests even when the continuum light curves deviate from the DRW model assumed by JAVELIN, in the single power-law PSD models we tested }{(Section \ref{sec:psd})}.

\end{enumerate} 

These results demonstrate the clear preference for {\tt JAVELIN} over the other two methods as the primary method of lag measurements for MOS-RM surveys, where the quality of light curves is generally worse than that achieved for traditional RM programs targeting local low-luminosity AGN. 

We further developed a statistical approach to efficiently eliminate false detections in MOS-RM surveys, without knowing the true lags of the sample. Using this statistical approach, we can recover 90\% of the true (detectable) lags while retaining a reasonably low false detection rate ($\sim 18\%$ for ICCF and $<10\%$ for ZDCF and {\tt JAVELIN}). 

{{\tt JAVELIN} recovers the most accurate R-L relation slope compared to the fiducial slope measured for the low-$z$ RM sample, and the recovered R-L relation slope from ICCF and ZDCF is shallower. When the intrinsic scatter in the R-L relation increases, the recovered R-L relation becomes even shallower.} This is mainly because our 180-day mock observation is not capable of detecting long lags (and lags much shorter than the cadence) and thus limiting the dynamic range in the R-L relation fitting. Indeed, when we include long lags from multi-year observations, this discrepancy in the R-L relation slope is reduced. The deficiency of short lags in the low-luminosity regime still limits the recoverability of the true slope. However, only {\tt JAVELIN} is capable of producing consistent slope measurements when the cadence is reduced or the light curve S/N is degraded. 

Our investigations have not explored the entire parameter space of RM and other less common methods of lag measurements, and it is possible that {\tt JAVELIN} may perform worse than ICCF in special circumstances. However, for large-scale MOS-RM programs, the recently developed, more statistically robust methods (such as {\tt JAVELIN} and \code{CREAM}) convincingly produce superior results than the traditional ICCF in order to utilize the full power of these MOS-RM data. 

\bigskip

{We thank the referee for a thorough report and useful comments, and Zhefu Yu and Brad Peterson for helpful discussions.} JIL and YS acknowledge support from an Alfred P. Sloan Research Fellowship (YS) and NSF grant AST-1715579. LCH was supported by the National Science Foundation of China (11721303) and the National Key R\&D Program of China (2016YFA0400702). JRT and YH acknowledge support from NASA grant HST-GO-15260. WNB and CJG acknowledge support from NSF grants AST-1517113 and AST-1516784. Funding for SDSS-III has been provided by the Alfred P. Sloan Foundation, the Participating Institutions, the National Science Foundation, and the U.S. Department of Energy Office of Science. The SDSS-III web site is http://www.sdss3.org/.

SDSS-III is managed by the Astrophysical Research Consortium for the Participating Institutions of the SDSS-III Collaboration including the University of Arizona, the Brazilian Participation Group, Brookhaven National Laboratory, University of Cambridge, Carnegie Mellon University, University of Florida, the French Participation Group, the German Participation Group, Harvard University, the Instituto de Astrofisica de Canarias, the Michigan State/Notre Dame/JINA Participation Group, Johns Hopkins University, Lawrence Berkeley National Laboratory, Max Planck Institute for Astrophysics, Max Planck Institute for Extraterrestrial Physics, New Mexico State University, New York University, Ohio State University, Pennsylvania State University, University of Portsmouth, Princeton University, the Spanish Participation Group, University of Tokyo, University of Utah, Vanderbilt University, University of Virginia, University of Washington, and Yale University.


\end{document}